\documentclass[11pt]{article}
\usepackage{cite}
\usepackage{amsmath,amsfonts,amssymb}%,calrsfs}
\pdfoutput=1
\usepackage[small,bf,hang]{caption}
\usepackage{slashed}
\input epsf.sty
\usepackage{epsfig}
\usepackage[titletoc,toc]{appendix}
\usepackage{xcolor}

\def\hybrid{
        \topmargin -20pt
        \oddsidemargin 0pt
        \headheight 0pt \headsep 0pt
        \textwidth 6.55in % A4 paper
        \textheight 9.5in % A4 paper
        \marginparwidth .875in
        \parskip 5pt plus 1pt \jot = 1.5ex}

\hybrid

 \csname
@addtoreset\endcsname{equation}{section}

\newcommand{\p}{\partial}

\def\moth{\mathsurround=0pt}
\newdimen\zo \zo=0pt

\def\tick{\leaders\hrule height 0.5ex depth 0pt \hskip 0.5pt}
\def\upboxfill{$\moth \setbox\zo\hbox{\tick}%
  \hskip 3pt\hbox to 0pt{$\tick$\hss}\hrulefill \hbox to 7.5pt{$\tick$\hss}$}

\def\dtick{\leaders\hrule height .34pt depth 0.5ex \hskip 0.5pt}
\def\downboxfill{$\moth \setbox\zo\hbox{\dtick}%
  \hskip 2pt\hbox to 0pt{$\dtick$\hss}\hrulefill \hbox to 2pt{$\dtick$\hss}$}

\def\id{{\mathbb I}}

\def\bec{\begin{center}}
\def\ec{\end{center}}

\def\be{\begin{equation}}
\def\ee{\end{equation}}
\def\bea{\begin{eqnarray}}
\def\eea{\end{eqnarray}}
\def\ba{\begin{array}}
\def\ea{\end{array}}

\usepackage{hyperref}

\graphicspath{ {./images/} }

\begin{document}

\begin{titlepage}
	
	\rightline{\tt MIT-CTP-5753}
	\hfill \today
	\begin{center}
		\vskip 0.5cm
		
		{\Large \bf {Topological recursion for hyperbolic string field theory}
		}
		
		\vskip 0.5cm
		
		\vskip 1.0cm
		{\large {Atakan Hilmi Fırat$^{1,2}$ and Nico Valdes-Meller$^{1}$}}
		
		\vskip 0.5cm
		
		{\em  \hskip -.1truecm
			$^{1}$
			Center for Theoretical Physics \\
			Massachusetts Institute of Technology\\
			Cambridge MA 02139, USA
			\\
			\vskip 0.5cm
			$^{2}$
			Center for Quantum Mathematics and Physics (QMAP) \\
			Department of Physics \& Astronomy, \\
			University of California, Davis, CA 95616, USA
			\\
			\vskip 0.5cm
			\tt \href{mailto:ahfirat@ucdavis.edu}{ahfirat@ucdavis.edu}, \href{mailto:nvaldes@mit.edu}{nvaldes@mit.edu} \vskip 5pt }
		
		\vskip 2.5cm
		{\bf Abstract}
		
	\end{center}
	\vskip 0.5cm
	\noindent
	\begin{narrower}
		\baselineskip15pt
		We derive an analog of Mirzakhani's recursion relation for hyperbolic string vertices and investigate its implications for closed string field theory. Central to our construction are systolic volumes: the Weil-Petersson volumes of regions in moduli spaces of Riemann surfaces whose elements have systoles $L \geq 0$. These volumes can be shown to satisfy a recursion relation through a modification of Mirzakhani's recursion as long as $L \leq 2 \sinh^{-1} 1$. Applying the pants decomposition of Riemann surfaces to off-shell string amplitudes, we promote this recursion to hyperbolic string field theory and demonstrate the higher order vertices are determined by the cubic vertex iteratively for any background. Such structure implies the solutions of closed string field theory obey a quadratic integral equation. We illustrate the utility of our approach in an example of a stubbed scalar theory.

	\end{narrower}
\end{titlepage}

\tableofcontents

\baselineskip15pt

\section{Introduction}

Despite its three decade history, covariant closed string field theory (CSFT) still remains notoriously impenetrable, refer to~\cite{Zwiebach:1992ie,Sen:2024nfd,deLacroix:2017lif,Erler:2019loq,Erbin:2021smf,Maccaferri:2023vns} for reviews. This can be attributed to the seemingly arbitrary structure of its elementary interactions that have been reverse-engineered from string amplitudes. Even though CSFT provides a complete and rigorous definition for closed string perturbation theory, its construction is also responsible for why almost all of its nonperturbative  features remain inaccessible---for example its nonperturbative vacua. Among them the most important one is  arguably the hypothetical tachyon vacuum of the critical bosonic CSFT~\cite{Belopolsky:1994sk,Belopolsky:1994bj,Yang:2005rx,Moeller:2004yy,Moeller:2006cw,Moeller:2007mu,Moeller:2006cv}. Its construction is expected to provide insights on the nonperturbative nature of string theory and its background-independent formulation.

Therefore obtaining any nonperturbative information from CSFT (or the theory and/or principles that govern CSFT perturbatively) appears to require seeking finer structures within its current formulation---beyond those demanded by perturbative quantum field theory, such as amplitude factorization, to tame the infinitely many nonlocal interactions between infinitely many target space fields. The main purpose of this work is to begin unearthing some of these structures and investigate their implications.

This objective immediately brings our attention to~\emph{string vertices} encoding the elementary interactions of CSFT, which is arguably the most nontrivial ingredient of CSFT. Roughly speaking, these are the subsets of appropriate moduli spaces of Riemann surfaces that satisfy the consistency condition known as~\emph{the geometric master equation}
\begin{align} \label{eq:GME}
	\partial \mathcal{V}  + {1 \over 2} \{\mathcal{V}, \mathcal{V} \} + \hbar \Delta \mathcal{V} = 0 \, ,
\end{align}
see section~\ref{sec:Pre} for details. This equation encodes the perturbative consistency of CSFT in the sense of Batalin-Vilkovisky (BV) formalism~\cite{Batalin:1981jr,Schwarz:1992nx,Henneaux:1992ig}. A particular solution to~\eqref{eq:GME} is equivalent to a particular choice of field parametrizations in CSFT~\cite{Hata:1993gf}. Although any allowed field parametrization can be used for a given theory, having one that most efficiently parametrizes the interactions and contains additional structure would be the superior choice.

\emph{Hyperbolic string vertices} have the potential to provide such a canonical choice for all-order investigations in CSFT~\cite{Moosavian:2017qsp,Moosavian:2017sev,Pius:2018pqr,Costello:2019fuh,Cho:2019anu,Erbin:2022rgx,Firat:2021ukc,Firat:2023glo,Firat:2023suh,Firat:2023gfn,Wang:2021aog,Jiang:2024noe,Ishibashi:2022qcz,Ishibashi:2024kdv,Bernardes:2024ncs}. The overall objective of this paper is to demonstrate the existence of a topological recursion among~\emph{the elementary interactions} of hyperbolic CSFT and initiate the investigation of its consequences for the nonperturbative structure of CSFT and its solutions.  

\begin{figure}[t]
	\centering
	\includegraphics[height=4.5in,trim={0.5cm 0cm 1cm 0.75cm}, clip]{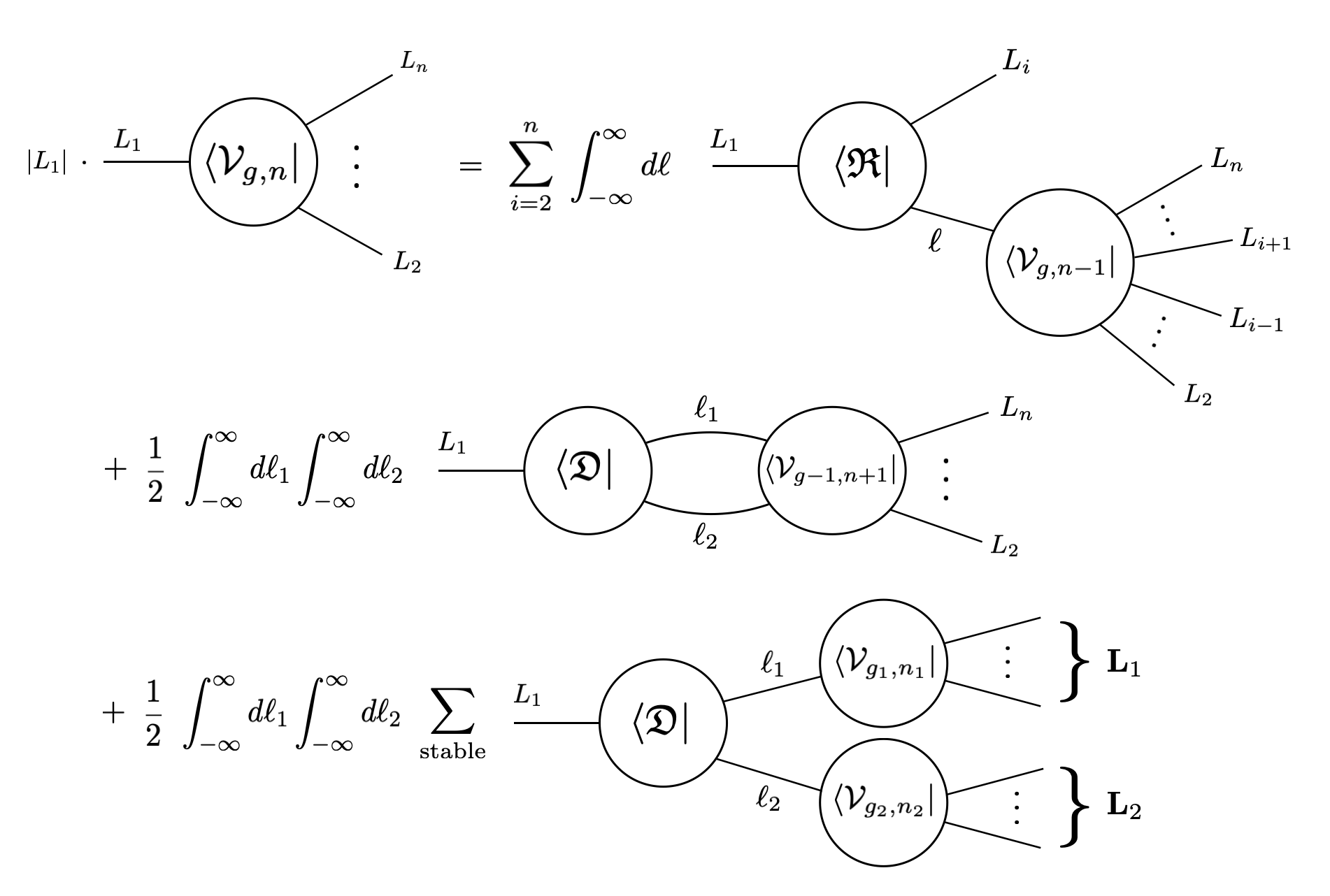}
	\caption{Schematic illustration of the recursion relation~\eqref{eq:A1.2} for hyperbolic vertices. }\label{fig:sft}
\end{figure}

Briefly stated, we show there is an analog of Mirzakhani's recursion for the Weil-Petersson (WP) volumes of moduli spaces~\cite{mirzakhani2007simple,mirzakhani2007weil} for the elementary vertices $\langle \mathcal{V}_{g,n} (L_i) |  = \langle \mathcal{V}_{g,n} (L_1, \cdots, L_n) | $ of hyperbolic CSFT, which are defined using hyperbolic surfaces of genus $g$ with $n$ punctures whose local coordinates are determined up to a phase by grafting semi-infinite flat cylinders to geodesic borders of lengths $|L_1|, \cdots, |L_n|$~\cite{Costello:2019fuh}. It takes the following form (see~\eqref{eq:4.16} and figure~\ref{fig:sft})
\begin{align} \label{eq:A1.2}
	&|L_1| \cdot \langle \mathcal{V}_{g,n} (L_i) | = \sum_{i=2}^n \, \int\limits_{-\infty}^\infty \, d \ell 
	\, \bigg[ \langle \mathfrak{R} (L_1, L_i, \ell) | \otimes \, \langle \mathcal{V}_{g, n-1} (-\ell, \mathbf{L} \setminus \{L_i \}) |
	\bigg] |\omega^{-1} \rangle \\
	&\hspace{0.5in} + {1 \over 2} \int\limits_{-\infty}^\infty  \, d \ell_1  \, \int\limits_{-\infty}^\infty  \, d \ell_2 \,
	\bigg[\langle \mathfrak{D} (L_1, \ell_1, \ell_2) | \otimes \,
	\bigg(\langle \mathcal{V}_{g-1,n+1} (-\ell_1, -\ell_2, \mathbf{L} )| 
	\nonumber \\ 
	&\hspace{2.1in} +
	\sum_{\text{stable}} 
	\langle \mathcal{V}_{g_1,n_1} (-\ell_1, \mathbf{L_1} ) | \otimes \, \langle  \mathcal{V}_{g_2,n_2} (-\ell_2, \mathbf{L_2} ) |
	\bigg)
	\bigg] | \, \omega^{-1} \rangle_1  | \, \omega^{-1}  \rangle_2 \nonumber \, ,
\end{align}
where
\begin{align}
	\mathbf{L} = \{L_2, \cdots, L_n \} \, , \quad  \quad
	\mathbf{L_1} \cup \mathbf{L_2 }= \mathbf{L} \,  , \quad  \quad
	\mathbf{L_1} \cap \mathbf{L_2} = \emptyset \, ,
\end{align}
and the ``stable'' in~\eqref{eq:A1.2}, and henceforth, denotes the sum over all non-negative integers $g_1,g_2,n_1,n_2$ and partitions $\mathbf{L_1}, \mathbf{L_2}$ that satisfy
\begin{align}
	&g_1 + g_2 = g\, , \quad n_1 + n_2 = n + 1 \, , \quad 2g_1 - 2 + n_1 > 0 \, , \quad 2g_2 - 2 + n_2 > 0 \,  ,
\end{align}
for $2g -2 + n > 1$ and $n \geq 1$.\footnote{We comment on the cases $g=1, n=1$ and $g\geq 2, n = 0$ in section~\ref{sec:Sys}.} Here $| \omega^{-1} \rangle$ is the Poisson bivector~\cite{Erler:2019loq} that twist-sews the entries of states whose length is integrated. Its subscript denotes the sewed entry.  The relation~\eqref{eq:A1.2} is a recursion in the negative Euler characteristic $-\chi_{g,n} = 2g-2+n$ of the underlying surfaces, hence we refer it~\emph{topological recursion} in this particular sense.

The amplitudes $\langle \mathcal{V}_{g,n} (L_i) | $ are defined using the hyperbolic surfaces directly when $L_i >0$. However, forming the recursion also requires incorporating the cases when $L_i$ are~\emph{negative} in the sense that they differ by the application of the $ \mathfrak{B}$ ghost operators that result from changing the length of the border
\begin{align}  \label{eq:1.5}
	\langle \mathcal{V}_{g,n} (L_1, \cdots, L_n) | =
	\langle \mathcal{V}_{g,n} (|L_1|, \cdots, |L_n|)  | \, 
	\left(\mathfrak{B}^{\otimes \, \theta(-L_1)} \otimes \cdots \otimes \mathfrak{B}^{\otimes \, \theta(-L_n)} \right) \, ,
\end{align}
where $\theta(x)$ is the Heaviside step function: it is $1$ whenever $x \geq 0$, $0$ otherwise. These cases are recursively determined by~\eqref{eq:A1.2} as well. We particularly highlight that the base case for the recursion is given by the cubic vertex
\begin{align} \label{eq:A1.6}
	\langle \mathcal{V}_{0,3} (L_1, L_2, L_3) |  =  \langle \Sigma_{0,3}( |L_1|, |L_2| , |L_3| )|  
	\, \left(  \mathfrak{B}^{\otimes \, \theta(-L_1)} \otimes \mathfrak{B}^{\otimes \, \theta(-L_2)} \otimes \mathfrak{B}^{\otimes \, \theta(-L_3)} \right) \, ,
\end{align}
where $\langle \Sigma_{0,3}( L_1, L_2 ,L_3)| $ is the generalized hyperbolic three-vertex of~\cite{Firat:2021ukc}, see appendix~\ref{app:H3V}. As an example, a single application of $\mathfrak{B}$ to $\langle \Sigma_{0,3}( L_1, L_2 ,L_3)| $ is given by (see~\eqref{eq:1.7a})
\begin{align} \label{eq:1.7}
	&\langle \Sigma_{0,3}( L_1, L_2 , L_3 )| \left(   \mathfrak{B} \otimes \id \otimes \id \right)
	\\
	&\hspace{0.5in}= 
	\langle \Sigma_{0,3}( L_1, L_2 , L_3 )|
	\left[ {1 \over 2 \pi}  \left(
	{1 \over \rho_1} {\partial \rho_1 \over \partial \lambda_1} \, \left( b_0 + \overline{b}_0 \right)
	+{\lambda_1  \, (\lambda_2^2  - \lambda_3^2) \over (1+ \lambda_1^2)^2} \, \rho_1 
	\left( b_1+ \overline{b}_1 \right)
	+ \cdots
	\right) \otimes \id \otimes \id 
	\right]
	\, .\nonumber
\end{align}
Here $L_i = 2 \pi \lambda_i > 0$ and $\rho_1 = \rho_1(L_1, L_2, L_3)$ is the mapping radius associated with the first puncture~\eqref{eq:R}. Observe that the $ \mathfrak{B}$ insertion produces a dependence on all the border lengths $L_1, L_2, L_3$. One can similarly consider when there are multiple $\mathfrak{B}$ insertions.

The recursion~\eqref{eq:A1.2} also contains~\emph{the string kernels}
\begin{subequations}  \label{eq:1.8}
\begin{align}
	\langle \mathfrak{R}(L_1,L_2,L_3) | &= \widetilde{R}_{|L_1| |L_2| |L_3|} \, \langle \mathcal{V}_{0,3}(L_1,L_2,L_3)| \, ,\\
	\langle \mathfrak{D}(L_1,L_2,L_3) | &= \widetilde{D}_{|L_1| |L_2| |L_3|} \, \langle \mathcal{V}_{0,3}(L_1,L_2,L_3)| \, , 
\end{align}
\end{subequations}
 where $\widetilde{R}_{L_1, L_2, L_3} , \widetilde{D}_{L_1, L_2, L_3} $ are the functions
\begin{subequations}
\begin{align}
	\widetilde{R}_{L_1 L_2 L_3} &= R_{L_1 L_2 L_3} - L_1 \, \theta(L - L_3) \, ,\\
	\widetilde{D}_{L_1 L_2 L_3} &=D_{L_1 L_2 L_3} 
	- R_{L_1 L_2 L_3} \, \theta(L -L_2) - R_{L_1L_3 L_2} \, \theta(L -L_3) + L_1 \, \theta(L-L_2) \, \theta(L -L_3) \, .
\end{align}
\end{subequations}
The kernels depend on~\emph{the threshold length} $L$ that is bounded by
\begin{align}
	0 < L \leq 2 \sinh^{-1} 1 \approx 1.76 \, ,
\end{align}
for the recursion~\eqref{eq:A1.2} to hold true. Finally, the functions $R_{L_1 L_2 L_3}$ and $D_{L_1 L_2 L_3}$ above are the well-known functions that appear in Mirzakhani's recursion for the WP volumes of moduli spaces of Riemann surfaces~\cite{mirzakhani2007simple,mirzakhani2007weil} (also see~\cite{do2011moduli,wright2020tour,Eynard:2007fi,andersen2017geometric,Saad:2019lba,Stanford:2019vob})
\begin{subequations}
\begin{align}
	&R_{L_1 L_2 L_3} = L_1 - \log\left[\frac{\cosh\left(\frac{L_2}{2} \right) + \cosh\left(\frac{L_1 + L_3}{2} \right) } { \cosh\left(\frac{L_2}{2} \right)  + \cosh\left(\frac{L_1 - L_3}{2} \right) }  \right]  \, ,
	\\
	&D_{L_1 L_2 L_3} = 2\log \left[\frac{\exp\left( {\frac{L_1}{2}}\right) + \exp\left({\frac{L_2 + L_3}{2}}\right) }{\exp\left({-\frac{L_1}{2}}\right) + \exp\left({\frac{L_2 + L_3}{2}}\right) } \right] \, .
\end{align}
\end{subequations}

The central idea behind~\eqref{eq:A1.2} is to excise certain pants from surfaces that make up $\langle \mathcal{V}_{g,n} (L_i) |$ (see figure~\ref{Mirz-gluing-figure}) to factorize it in terms of lower-order vertices, then perform the moduli integration over the lengths and twists of the excised seams. We need to make sure that each surface in $\langle \mathcal{V}_{g,n} (L_i) |$ is counted once in the moduli integration after the factorization and this introduces the measure factors $\widetilde{R}_{L_1 L_2 L_3},\widetilde{D}_{L_1 L_2 L_3}$ in the integrand~\eqref{eq:A1.2}. The excisions of pants from the surface further requires insertion of $\mathfrak{B}$, see section~\ref{sec:HSV}.

A careful reader may notice the similarity between \eqref{eq:A1.2} and the recursion proposed by Ishibashi \cite{Ishibashi:2022qcz,Ishibashi:2024kdv}. Despite the structural similarity, there are a few crucial differences between our constructions. First, Ishibashi's off-shell amplitudes are unconventional from the perspective of covariant CSFT. They do not factorize through using the flat propagators. Instead, the simple closed geodesics of hyperbolic surfaces play the role of the propagator and the degenerations occur when the lengths of simple closed geodesics shrink to zero. This obscures its connection to ordinary CSFT, but it provides the straightforward structure to write a recursion relation for off-shell amplitudes as the moduli integration is performed over the entire moduli space.

On the other hand, $\langle \mathcal{V}_{g,n} (L_i) |$ in~\eqref{eq:A1.2} are obtained by performing the moduli integration only over ``systolic subsets'' and the recursion is formed~\emph{exclusively} among them. A priori, it isn't clear that such a recursion has a right to exist---it requires a nontrivial modification of Mirzakhani's method where the systolic subsets are related to each other rather than the entire moduli spaces. But thanks to the ``twisting procedure'' of~\cite{andersen2017geometric} and the collar lemma~\cite{buser2010geometry}, this modification is in fact possible, see section~\ref{sec:Sys} for details. Combining the twisting procedure and the factorization analysis of amplitudes, we derive a recursion relation~\eqref{eq:A1.2} without giving up the connection to CSFT.

In order to see the impact of this modification, first recall the hyperbolic CSFT BV master action
\begin{align} \label{eq:A1.11}
	S_L[\Psi] = S_{0,2} [\Psi] + \sum_{g,n} {1 \over n!} \, \hbar^{g} \kappa^{2g-2+n} 
	\, \langle \mathcal{V}_{g,n} (\underbrace{L, \cdots, L}_{n \text{ times}}) \, | \Psi \rangle^{\otimes n}  \, ,
\end{align}
when $0 < L \leq 2 \sinh^{-1} 1$~\cite{Costello:2019fuh}. Here $S_{0,2}[\Psi]$ is the free part of the action and the sum over the non-negative integers $g,n$ above, and henceforth, denotes the sum over $2g-2+n > 0$. The recursion~\eqref{eq:A1.2} then strikingly shows the different order terms in the hyperbolic CSFT action are related to each other and they are entirely determined by the cubic data~\eqref{eq:A1.6}. This manifestly demonstrates the cubic nature of CSFT in the sense of topological recursion. We point out this result is consistent with the famous no-go theorem for having a cubic level-matched covariant  CSFT~\cite{Sonoda:1989sj}: we don't give such a formulation here, only express a relation between different terms in the action~\eqref{eq:A1.11}.

A similar observation has already been made using the relation between hyperbolic vertices and the classical conformal bootstrap~\cite{Firat:2023glo,Firat:2023suh} by one of the authors. However the equation~\eqref{eq:A1.2} here actually makes this cubic nature even more apparent.\footnote{In fact these features are possibly related to each other, see the discussion on conformal blocks in~\cite{andersen2017geometric}.} We highlight that~\eqref{eq:A1.2}, and its consequences, are background-independent and apply to~\emph{any} bosonic closed string background.

The recursion~\eqref{eq:A1.2} is stronger than the BV structure of CSFT. There, the focus is on how the Feynman diagrams with propagators are related to the elementary vertices---as stated in terms of the geometric master equation~\eqref{eq:GME}. In~\eqref{eq:A1.2}, on the other hand, the elementary vertices themselves relate to each other. Something akin to this feature can be artificially created by deforming a polynomial theory with stubs~\cite{Chiaffrino:2021uyd,Schnabl:2023dbv,Schnabl:2024fdx,Erbin:2023hcs, Erler:2023emp,Maccaferri:2024puc}, for which the interactions are obtained through homotopy transfer of the seed theory using a ``partial propagator'' and the resulting (infinitely many) vertices are related to each other iteratively. This is the primary reason why the stubbed theories are theoretically tractable despite their non-polynomiality. The expectation is that the topological recursion~\eqref{eq:A1.2} would help to achieve similar feats in CSFT.

In order to demonstrate the power of this extra structure we examine its implication for the classical solutions to CSFT~\eqref{eq:A1.11}. Let us call the stationary point of the classical CSFT action $ \Psi_\ast $, keeping its $L$ dependence implicit, and introduce the string field $\Phi(L_1)$
\begin{align}
	\Phi(L_1) = \sum_{n=3}^\infty {\kappa^{n-2} \over (n-1)!} \,
	\bigg(\id \otimes \langle  \mathcal{V}_{0,n} (L_1, \underbrace{L, \cdots, L}_{(n-1) \text{ times}}) | 
	\bigg) \, 
	\left( | \omega^{-1} \rangle \, \otimes | \Psi_\ast \rangle^{\otimes (n-1)} \right) \, ,
\end{align}
associated with the solution $\Psi_\ast $. Observe that we have
\begin{align} \label{eq:A1.13}
	 \Phi(L_1 = L) = - Q_B \,  \Psi_\ast \, ,
\end{align}
by the equation of motion, so $\Phi(L_1)$ should be thought as a certain off-shell generalization of the BRST operator $Q_B$ acting on the solution. It is parameterized by $L_1$. The recursion~\eqref{eq:A1.2} implies the following quadratic integral equation for this string field (see~\eqref{eq:4.24})
\begin{align} \label{eq:A1.15}
	0 = \Phi (L_1)
	- {\kappa \over 2} \, 
	A_{L_1 L L } \big( \beta \, \Phi  (L) , \beta \, \Phi  (L) \big)
	&- \kappa \, \int\limits_{-\infty}^\infty d \ell \, 
	B_{L_1 L \ell}\left(\beta \, \Phi  (L) , \Phi(\ell)\right)
	\\ \nonumber
	&\hspace{0.8in}- {\kappa \over 2} \, \int\limits_{-\infty}^\infty d \ell_1 \, \int\limits_{-\infty}^\infty d \ell_2 \,
	C_{L_1 \ell_1 \ell_2}\left(\Phi(\ell_1), \Phi(\ell_2) \right) \, ,
\end{align}
where the $2$-products $A,B,C$ are constructed using~\eqref{eq:1.8}. Here we fixed the gauge by 
\begin{align}
	\beta(L) \, \Psi_\ast = 0 \, , 
\end{align}
see~\eqref{eq:6.7}. The detailed derivation of this equation is given in section~\ref{sec:6}.

The solutions  $\Phi(L_1)$  to~\eqref{eq:A1.15} (if they exist) are related to the honest CSFT solutions through \eqref{eq:A1.13} and they can be used to obtain the cohomology around them. Even though we haven't made any attempt towards realizing this here, we have investigated the counterpart problem in the stubbed cubic scalar field theory as a proof of principle. We construct the nonperturbative vacuum of~\cite{Erler:2023emp} and read the mass of the linear excitations around it entirely using the analog of~\eqref{eq:A1.15} with some guesswork. In particular we haven't performed a resummation while doing so. This is encouraging for future attempts of solving CSFT with~\eqref{eq:A1.15}: resumming the CSFT action would have been a hopeless endeavor with current technology.

The outline of the paper is as follows. In section~\ref{sec:Pre} we review hyperbolic geometry and CSFT. In section~\ref{sec:Sys} we compute the WP volumes of systolic subsets \`a la Mirzakhani, after reviewing her recursive method for computing WP volumes of the entire moduli spaces. This provides a toy model for the actual case of interest, which is the recursion for hyperbolic string vertices. We derive it in section~\ref{sec:HSV}. We provide more insight on our results by obtaining an analogous recursion relation for the stubbed cubic scalar theory of~\cite{Erler:2023emp} in section~\ref{sec:5} and investigate its implications. This section is mostly self-contained and should be accessible to readers who are not familiar with CSFT. The solutions of the stubbed theory are shown to obey a quadratic integral equation in section~\ref{sec:5}. We also show the solutions to CSFT also obey a similar equation in section~\ref{sec:6} using similar reasoning. We conclude the paper and discuss future prospects in section~\ref{sec:disc}. 

In appendix~\ref{app:H3V} we report the local coordinates for the generalized hyperbolic three-vertex of~\cite{Firat:2021ukc}. Finally we compute certain systolic volumes using the recursion and the bootstrap methods of~\cite{Firat:2023glo,Firat:2023suh} in appendix~\ref{app:B} and~\ref{sec:A} respectively. 

\section{Preliminaries} \label{sec:Pre}

In this section we review hyperbolic geometry and the construction of CSFT. For a mathematically-oriented introduction to hyperbolic geometry refer to~\cite{buser2010geometry} and for the basics of CSFT there are many excellent reviews~\cite{Zwiebach:1992ie,Sen:2024nfd,deLacroix:2017lif,Erler:2019loq,Erbin:2021smf,Maccaferri:2023vns}. We follow~\cite{Costello:2019fuh} for our discussion of hyperbolic vertices.

\subsection{Hyperbolic surfaces and Teichm\"uller spaces}

Let $S_{g,n}$ be an orientable closed two-dimensional marked surface of genus $g$ with $n$ borders that has a negative Euler characteristic\footnote{\emph{The marking} is an orientation-preserving diffeomorphism $f : S_{g,n} \to S_{g,n}$ and two markings are equivalent if they are isotopic to each other. The surface together with the isotopy class of the marking $[f]$ is~\emph{a marked surface}.}
\begin{align}
	-\chi_{g,n} \equiv 2g-2+n > 0 \, .
\end{align}
By the uniformization theorem, such a surface admits metrics with constant negative Gaussian curvature $K = -1$ with geodesic borders of length $L_i \geq 0 $ for $i=1, \cdots, n$. These are hyperbolic metrics with geodesic borders. We denote the space of all such distinct metrics on $S_{g,n}$ by $\mathcal{T}_{g,n}(L_i)$ and call it~\emph{Teichm\"uller space}. This is same as the set of marked (hyperbolic) Riemann surfaces of genus $g$ with $n$ borders. The border is a hyperbolic cusp when its associated length vanishes.

The ordinary moduli spaces of Riemann surfaces $\mathcal{M}_{g,n}(L_i)$ are obtained by forgetting the marking on the surface $S_{g,n}$, i.e., by taking the quotient\footnote{We always consider $\mathcal{M}_{g,n} (L_i) $ to be compactified in the sense of Deligne-Mumford. This also requires using the so-called~\emph{augmented Teichm\"uller space}~\cite{hubbard2014analytic}, however we are not going to be explicit about this.}
\begin{align}
	\mathcal{M}_{g,n} (L_i) = \mathcal{T}_{g,n}(L_i)/ \text{MCG}(S_{g,n}) \, ,
	 \quad \quad
	\text{MCG}(S_{g,n}) = {\text{Diff}^+(S_{g.n}) \over \text{Diff}_0^+(S_{g,n})} \, ,
\end{align}
where $\text{MCG}(S_{g,n})$ is~\emph{the mapping class group} of the surface $S_{g,n}$. Here $\text{Diff}^+(S_{g.n}) $ is the set of orientation-preserving diffeomorphisms and $0$ subscript denotes those  isotopic to the identity. These diffeomorphisms are taken to be restricted to be the identity on the boundary $\partial S_{g.n}$. In other words $\text{MCG}(S_{g,n})$ is the group of large diffeomorphisms of $S_{g,n}$. The action of $\text{MCG}(S_{g,n})$ on $\mathcal{T}_{g,n}(L_i)$ produces the hyperbolic structures on $S_{g,n}$ that are same up to marking.

The Teichm\"uller space is somewhat easier to work with relative to the moduli space $\mathcal{M}_{g,n} (L_i)$ since it is simply-connected~\cite{buser2010geometry}. That is, $\mathcal{T}_{g,n}(L_i)$ is the universal cover of $\mathcal{M}_{g,n}(L_i)$ and there is a covering map
\begin{align} \label{eq:pi}
	\pi : \mathcal{T}_{g,n}(L_i) \to \mathcal{M}_{g,n}(L_i) \, .
\end{align}
In order to better appreciate this feature we remind the reader that marked Riemann surfaces admit~\emph{pants decompositions}, which is done by cutting the surface along a set of $3g-3+n$ disjoint simple closed geodesics~\cite{buser2010geometry}. The pant decompositions are far from unique. There are infinitely many such decompositions and they are related to each other by large diffeomorphisms.

Given a pants decomposition, however, any marked surface can be constructed uniquely through specifying the tuple
\begin{align} \label{eq:FN}
	(\ell_i, \tau_i) \in \mathbb{R}_+^{3g-3+n} \times \mathbb{R}^{3g-3+n} \, , 
\end{align}
where $\ell_i$ are the lengths of the seams of the pants and $\tau_i$ are the relative twists between them. This provides a particularly nice set of coordinates for Teichm\"uller space known as~\emph{Fenchel-Nielsen coordinates}. They descend to the moduli spaces by~\eqref{eq:pi}, but only locally due to the large diffeomorphisms. It is then apparent that the dimensions of Teichm\"uller and moduli spaces are given by
\begin{align}
	\dim_{\mathbb{C}} \mathcal{T}_{g,n}(L_i) = \dim_{\mathbb{C}} \mathcal{M}_{g,n}(L_i) = 3g - 3 + n \equiv d_{g,n} \, .
\end{align}
Note that we present the complex dimensions as these spaces have complex structures~\cite{buser2010geometry}.

Teichm\"uller space further admits a K\"ahler metric known as~\emph{the Weil-Petersson (WP) metric}. The associated symplectic form is the WP form. Its mathematically precise definition is not important for our purposes, but one of its important facets is that the WP form is MCG-invariant so that there is a globally well-defined symplectic form $\omega_{WP}$ over the moduli space $\mathcal{M}_{g,n}(L_i)$. This is reflected in~\emph{Wolpert's magic formula}~\cite{wolpert1983symplectic}
\begin{align}
	\pi^\ast \, \omega_{WP}  = \sum_{i=1}^{3g -3 + n} d \ell_i \wedge d \tau_i \, , 
\end{align}
that presents the pullback of $\omega_{WP}$ under~\eqref{eq:pi} in Fenchel-Nielsen coordinates $(\ell_i, \tau_i)$. Here the ``magic'' refers to the fact that the right-hand side holds for the Fenchel-Nielsen coordinates associated with~\emph{any} pants decomposition (i.e., its invariance under the action of the MCG). Using $\omega_{WP}$, it is possible to construct the WP volume form $\text{vol}_{WP}$ on the moduli space $\mathcal{M}_{g,n}(L_i)$
\begin{align}
	\text{vol}_{WP} = { \omega_{WP}^{3g-3+n} \over (3g-3+n)! } \, , \quad \quad \quad
	\pi^\ast \,  \text{vol}_{WP} = \bigwedge_{i=1}^{3g -3 + n} d \ell_i \wedge d \tau_i  \, , 
\end{align}
and compute integrals over moduli spaces.

We are not going to delve into the proofs of these statements. However, we highlight one of the common ingredients that goes into their proofs:~\emph{the collar lemma}~\cite{buser2010geometry}. This lemma essentially states that two sufficiently short closed geodesics on a hyperbolic surface cannot intersect each other. More precisely, defining a collar $\mathcal{C}_\gamma$ around a simple closed geodesic  $\gamma \in \Sigma_{g,n}(L_i)$ by
\begin{align}
	\mathcal{C}_\gamma = \bigg\{ \,
	p \in \Sigma_{g,n}(L_i)
	\, \big| \,
	\text{dist}(p, \gamma) \leq {w \over 2} \,  , \; \;
	\sinh{w \over 2}  \sinh{\gamma \over 2}  = 1
	\bigg\} \, , 
\end{align}
the collar lemma states that the collars are isometric to hyperbolic cylinders and are pairwise-disjoint for a pairwise-disjoint set of simple closed geodesics. The lengths of the geodesics are always denoted by the same letter and ``$\text{dist}$'' stands for the distance measured by the hyperbolic metric. One can analogously define half-collars around the geodesic borders and they will be pairwise-disjoint from other (half-)collars. Here $w$ is the width of the collar and it increases as $\gamma$ decreases.

The last observation, together with the collar lemma, necessitates the lengths of a simple closed geodesic $\gamma$ and a (not necessarily simple) closed geodesic $\delta$ with $\gamma \cap \delta \neq \emptyset$ to obey
\begin{align} \label{eq:Uncertainty}
	\sinh{\gamma \over 2} \, \sinh{\delta \over 2} > 1 \, .
\end{align}
Defining the maximum threshold length
\begin{align}
	L_\ast = 2 \sinh^{-1} 1 \, ,
\end{align}
it is apparent that a simple closed geodesic $\gamma$ with $\gamma \leq L_\ast$ cannot intersect with a closed (but not necessarily simple) geodesic $\delta$ with $\delta \leq L_\ast$.

\subsection{String vertices}

Now we return our attention to the construction of CSFT. The central geometric ingredient is~\emph{string vertices}. They can be succinctly expressed as the formal sum
\begin{align}
	\mathcal{V} = \sum_{g,n} \hbar^{g} \, \kappa^{2g-2 + n} \, \mathcal{V}_{g,n} \, .
\end{align}
Here the sum runs over non-negative integers $g,n$ with $2g -2 +n > 0$ and $\hbar$ and $\kappa$ (string coupling) are formal variables. The objects $ \mathcal{V}_{g,n} $ are $6g - 6 +2n$ real dimensional singular chains of the bundle $\widehat{\mathcal{P}}_{g,n} \to \mathcal{M}_{g,n}$, where $\mathcal{M}_{g,n} \equiv \mathcal{M}_{g,n}(L_i = 0)$. This bundle encodes a choice of local coordinates without specified global phases around the punctures.

The perturbative consistency of CSFT requires $\mathcal{V}$ to satisfy the geometric master equation
\begin{align} \label{eq:2.2}
	\partial \mathcal{V} = - {1 \over 2} \left\{ \mathcal{V} , \mathcal{V}  \right\} - \hbar \Delta \mathcal{V} \, ,
\end{align}
as stated in the introduction. Expanding in $\hbar, \kappa$, this equation can be seen as a homological recursion relation among $\mathcal{V}_{g,n}$. Here $\p$ is the boundary operator on the chains, while the anti-bracket $\{ \cdot, \cdot \}$ and the Laplacian $\Delta$ are the following multilinear operations:
\begin{itemize}
	\item The chain $\{ \mathcal{V}_{g_1, n_1}, \mathcal{V}_{g_2, n_2}\}$ is the collection of all surfaces constructed by twist-sewing a puncture in a surface belonging to $\mathcal{V}_{g_1, n_1}$ to a puncture in a surface belonging to $\mathcal{V}_{g_2, n_2}$ by
	\begin{align} \label{eq:TS}
		w_1 w_2 = \exp {i \theta} \quad \quad \quad \text{where} \quad \quad \quad 0 \leq \theta < 2 \pi \, .
	\end{align}
	Here $w_i$ are the local coordinates around the sewed punctures. The resulting surfaces are of genus $g_1+g_1$ and have $n_1 + n_2 - 2$ punctures.
	
	\item The chain $\Delta \mathcal{V}_{g,n}$ is the collection of all surfaces constructed by twist-sewing two punctures of the same surface in $\mathcal{V}_{g,n}$. The resulting surfaces are of genus $g+1$ and have $n - 2$ punctures. 
\end{itemize}
These operations, together with an appropriate notion of dot product on the chains over the moduli space of~\emph{disjoint} Riemann surfaces with a choice of local coordinates around their punctures, form a differential-graded Batalin-Vilkovisky (BV) algebra~\cite{Sen:1993kb}.

Any explicit construction of CSFT requires a solution to~\eqref{eq:2.2}. In this paper we are concerned with the one provided by hyperbolic geometry~\cite{Costello:2019fuh}. The primary idea behind its construction is as follows. We first imagine~\textit{bordered} Riemann surfaces endowed with hyperbolic metrics whose borders are geodesics of length $L_i$ and consider
\begin{align} \label{eq:2.14}
	\widetilde{\mathcal{V}}^L_{g,n}(L_i) =
	\big\{ \, \Sigma \in \mathcal{M}_{g,n}(L_i) \; \big| \; \text{sys}(\Sigma) \geq L \,
	\big\} 
	\subseteq \mathcal{M}_{g,n}(L_i)
	\, .
\end{align}
Here $\text{sys}(\Sigma)$ is the~\textit{systole} of the surface $\Sigma$---the length of the shortest non-contractible closed geodesic non-homotopic to any border. We denote these subsets as the~\textit{systolic subsets}.

We remark that the set $\widetilde{\mathcal{V}}^L_{g,n}(L_i) $ may be empty. For example, the maximum value of the systole in $\mathcal{M}_{2,0}$ is $2 \cosh^{-1} ( 1 + \sqrt{2}) \approx 3.06$ and it is realized by the Bolza surface~\cite{parlier2009simple}. Any choice of $L$ greater than this value leads to an empty set. It is clear that taking the threshold length $L$ sufficiently small always leads to a non-empty $\widetilde{\mathcal{V}}^L_{g,n}(L_i)$ since $\widetilde{\mathcal{V}}^L_{g,n}(L_i)$ covers the entire moduli space $\mathcal{M}_{g,n}(L_i)$ as $L \to 0$. In particular the systolic subsets are always non-empty when $L \leq L_\ast$.

The systolic subsets are used to construct hyperbolic string vertices via~\textit{grafting} semi-infinite flat cylinders to each geodesic borders of a surface in $\widetilde{\mathcal{V}}^L_{g,n}(L_i)$\footnote{We use tilde to distinguish systolic subsets before and after grafting. We are not going to make this distinction moving forward unless stated otherwise.}
\begin{align}
	\mathcal{V}^L_{g,n}(L_i) = \text{gr}'_\infty \left( \widetilde{\mathcal{V}}^L_{g,n}(L_i) 
	\right) \,  \quad\quad \text{where} \quad \quad
	\text{gr}'_\infty : \mathcal{M}_{g,n}(L_i) \to \widehat{\mathcal{P}}_{g,n} \, .
\end{align} 
The grafting map $\text{gr}'_\infty$ naturally endows a bordered surface with local coordinates. In fact this map is a homeomorphism~\cite{mondello2011riemann} and $\mathcal{V}^L_{g,n}$ is a piece of a section over the bundle $\widehat{\mathcal{P}}_{g,n} \to \mathcal{M}_{g,n} $ as a result. 

It can be shown that
\begin{align} \label{eq:hyp}
	\mathcal{V}^L = \sum_{g,n} \hbar^{g} \, \kappa^{2g-2 + n} \, \mathcal{V}^L_{g,n} (L_i = L)  \, ,
\end{align}
solves~\eqref{eq:2.2} when $L \leq L_\ast = 2 \sinh^{-1} 1$ by using  the collar lemma~\cite{Costello:2019fuh}. This essentially follows from noticing that $\p \mathcal{V}^L_{g,n} $ consists of surfaces where the length of at least one simple closed geodesic is equal to $L$ and these surfaces are in 1-1 correspondence with those constructed using $\{\cdot, \cdot\}$ and $\Delta$. The crucial insight behind this proof was using the corollary of the collar lemma stated in~\eqref{eq:Uncertainty}, which introduces a nontrivial constraint on the threshold length $L$. Refer to~\cite{Costello:2019fuh} for more details.

\subsection{Differential forms over $\widehat{\mathcal{P}}_{g,n} \to \mathcal{M}_{g,n}$}

String vertices in bosonic CSFT are the chains over which the moduli integration is performed while the integrand is constructed using a 2d matter CFT of central charge $c=26$ together with the $bc$ ghost system whose central charge is $c = -26$. We call their combined Hilbert space $\mathcal{H}$ and consider its subspace $\widehat{\mathcal{H}}$ whose elements are level-matched
\begin{align} \label{eq:2.17}
	b_0^- |\Psi \rangle = ( b_0 - \overline{b}_0) \, |\Psi \rangle  = 0 \, , \quad \quad
	L_0^-  |\Psi \rangle = ( L_0 - \overline{L}_0) \, |\Psi \rangle  = 0 \, .
\end{align}
Here $b_0$ and $L_0$ are the zero modes of the $b$-ghost and stress-energy tensor respectively. In this subsection we often follow the discussion in~\cite{Sen:2014pia}.

The natural objects that can be integrated over the singular $p$-chains are differential $p$-forms. This requires us to introduce a suitable notion of $p$-forms over the bundles $\widehat{\mathcal{P}}_{g,n} \to \mathcal{M}_{g,n}$. There are primarily two ingredients that go into their construction: the surface states and the $b$-ghost insertions. 

The~\emph{surface states} $\langle \Sigma_{g,n}|$ are the elements of the dual Hilbert space $(\widehat{\mathcal{H}}^\ast)^{\otimes n}$ that encode the instruction for the CFT correlator over a given Riemann surface $\Sigma_{g,n}$. That is
\begin{align} \label{eq:2.18}
	\langle \Sigma_{g,n} | \, \Psi_1  \otimes \cdots \otimes  \Psi_n 
	 =
	\big\langle \Psi_1(w_1 = 0) \cdots \Psi_n(w_n = 0) \big\rangle_{\Sigma_{g,n}} \, ,
\end{align}
for any $\Psi_1  , \cdots , \Psi_n  \in \widehat{\mathcal{H}}$. Notice $\langle \Sigma_{g,n} |$ contains the local coordinate data around each puncture. They have the intrinsic ghost number $6g-6$ and have even statistics. Because of this, it is sometimes useful to imagine $\langle \Sigma_{g,n} (L_i) |$ encodes the CFT path integral over~\emph{bordered} surfaces and $\Psi_i$ are inserted through grafting to provide boundary conditions for them in the context of hyperbolic vertices. We usually keep the border length dependence of the hyperbolic string amplitudes to distinguish them from the generic ones.

Often times we apply the surface states to an element
\begin{align}
	\Psi_1 \otimes \cdots  \otimes \Psi_n \in \widehat{\mathcal{H}}^{\otimes n}\, ,
\end{align}
where $\Psi_i$ is understood to be grafted to the $i$-th border. However, the order of the borders in the surface states may come permuted in our expressions, i.e., $\langle \Sigma_{g,n}(L_{\sigma_i}) |$ for $\sigma \in S_n$ is a possibility. In this case we have
\begin{align} \label{eq:2.20}
	\langle \Sigma_{g,n}(L_{\sigma_i}) | \,  \Psi_1  \otimes \cdots \otimes  \Psi_n 
	= \epsilon(\sigma) \, \langle \Sigma_{g,n}(L_{\sigma_i}) | \,
	\Psi_{\sigma_1}  \otimes \cdots \otimes  \Psi_{\sigma_n} \, ,
\end{align}
from~\eqref{eq:2.18}. Here $\epsilon(\sigma)$ is the Koszul sign of the permutation $\sigma$ after commuting string fields. This amounts to replacing $\otimes \to \wedge$, however we keep the tensor product notation.

Now for the $b$-ghost insertions, recall that the defining property of the $p$-forms is that they produce scalars when they are applied to $p$ vectors and are antisymmetric under exchange of these vectors. The surface states don't have this structure---we need to act on them with appropriate $b$-ghosts so that we construct the relevant $(\widehat{\mathcal{H}}^\ast)^{\otimes n}$-valued $p$-forms over $\widehat{\mathcal{P}}_{g,n}  \to \mathcal{M}_{g,n}$. So introduce
\begin{align} \label{eq:2.21}
	B_p (V_1, \cdots V_p) = b (v_1) \cdots b (v_p) \, ,
\end{align}
where $V_1, \cdots, V_p \in T \widehat{\mathcal{P}}_{g,n}$ are vector fields over the bundle $\widehat{\mathcal{P}}_{g,n} \to \mathcal{M}_{g,n}$ while
\begin{align} \label{eq:2.10}
	b(v_q) = \sum_{i = 1}^n \left[
	\oint dw_i \, v_q^{(i)} (w_i) \, b (w_i) + \oint d\overline{w}_i \, \overline{v}_q^{(i)} (\overline{w}_i) \, \overline{b} (\overline{w}_i)
	\right] \, , \quad \quad
	q = 1, \cdots, p \, ,
\end{align}
are given in terms of $b$-ghosts insertions around the punctures. Here 
\begin{align}
	v_q \in T \Sigma_{g,n}\, ,
\end{align}
are the so-called~\textit{Schiffer vector fields} associated with the vector $V_q(\Sigma_{g,n}) \in T_{\Sigma_{g,n}}\widehat{\mathcal{P}}_{g,n} $ and the expressions $v_q^{(i)} (w_i)$ are their presentations in the local coordinates $w_i$ around the punctures. We take 
\begin{align}
	\oint {dz \over z} = \oint { d \overline{z} \over \overline{z} } =1 \, ,
\end{align}
for a contour oriented counterclockwise around $z=0$. Notice the object $B_p$ is antisymmetric under exchanging $V_i$'s by the anticommutation of the $b$-ghosts.

The (generalized) Schiffer vector fields are constructed as follows.\footnote{Strictly speaking Schiffer vectors arise only due to the change of local coordinates around the punctures while keeping the rest of the transition functions of the surface fixed. So it is better to call the objects $v$ described here generalized Schiffer vectors. Unless stated otherwise, we simply call them Schiffer vectors as well.} Suppose that we have two coordinates $z_i$ and $z_j$ on the surface with a holomorphic transition map
\begin{align}
	z_i = f_{i j} (z_j) \, ,
\end{align}
where $z_i$ and $z_j$ can be uniformizing coordinates and/or local coordinates around punctures. Further suppose the vector $V(\Sigma_{g,n}) \in T_{\Sigma_{g,n}} \widehat{\mathcal{P}}_{g,n}$ is associated with some infinitesimal deformation of the transition map corresponding a change of the moduli of the surface. We keep the coordinate $z_i = f_{i j} (z_j)  $ fixed while taking $f_{ij} \to f^{\epsilon}_{ij}$ and $z_j \to z_j^\epsilon$ for this deformation. Then
\begin{align}
	z_i= f^{\epsilon}_{ij}(z^\epsilon_j)
	 \quad \implies \quad
	z^\epsilon_j = (f^{\epsilon}_{ij})^{-1} (z_i) = (f^{\epsilon}_{ij})^{-1} (f_{i j} (z_j) ) 
	= z_j + \epsilon \, v^{(i)} (z_j) \, .
\end{align}
Here $ v^{(i)} (z_j) $ defines the Schiffer vector field on the surface associated with $V$. It is regular on the intersection of coordinate patches $z_i, z_j$ but it can develop singularities away from this region. We further have
\begin{subequations} \label{eq:2.12}
	\begin{align}
		&f_{ij}^\epsilon (z_j) =  f_{ij}^\epsilon (z^\epsilon_j - \epsilon \, v^{(i)} (z_j)  )  
		= f_{ij}^\epsilon (z^\epsilon_j ) - \epsilon \, { \p f_{ij} (z_j) \over \p z_j} \, v^{(i)} (z_j)
		= z_i -  \epsilon  \, v^{(i)} (z_i)  \, , \\
		&v^{(i)} (z_i)  =  { \p f_{ij} (z_j) \over \p z_j} \, v^{(i)} (z_j)
		\, ,
	\end{align}
\end{subequations}
after pushing the vector forward to $z_i$ coordinates. Here we abuse the notation and use the arguments of the Schiffer vectors to implicitly remind us the coordinates in which they are expressed. The superscript on $v^{(i)} (z_j)$ informs which patch is held fixed.

Given the local coordinates $t^a$ on $\widehat{\mathcal{P}}_{g,n}$ we have the dependence $f_{ij} (z_j) = f_{ij}(z_j ; t^a)$. The variation associated with $t^a \to t^a + \delta t^a$ (which is also associated with the vectors $V_{a}$) is then given by 
\begin{align}
	f_{ij} (z_j; t^a + {\delta t^a}) = f_{ij} (z_j; t^a) + \delta t^a \, {\p f_{ij} (z_j; t^a) \over \p t^a}
	= z_i + \delta t^a \, {\p f_{ij} (z_j; t^a) \over \p t^a} \, .
\end{align}
Using~\eqref{eq:2.12} we find
\begin{align} \label{eq:2.15}
	&v^{(i)}_a (z_i) = -  {\p f_{ij} (z_j; t^a) \over \p t^a} = -  {\p f_{ij} (f_{i j}^{-1} (z_i); t^a) \over \p t^a} 
	\\ \nonumber
	&\hspace{1in}\implies
	v^{(i)}_a (z_j) = - \left(  { \p f_{ij} (z_j; t^a) \over \p z_j} \right)^{-1}  {\p f_{ij} (z_j; t^a) \over \p t^a} \, .
\end{align}
From here it is natural to take Schiffer vectors as 1-forms over $\widehat{\mathcal{P}}_{g,n}$ and express the operator-valued $p$-form over $\widehat{\mathcal{P}}_{g,n}$~\eqref{eq:2.21} as
\begin{align}
	B_p = b(v_{ a_1}) \, d t^{a_1} \wedge \cdots \wedge b(v_{a_p}) \, d t^{a_p}\, ,
\end{align}
in the coordinates $t^a$. The $b$-ghost insertions here can be chosen to act either on coordinates $z_i$ or $z_j$ using the first or second expressions in~\eqref{eq:2.15}
\begin{align} \label{eq:2.29}
	b(v_{a_q}) = \left[
	\oint d u \, v^{(i)}_{a_q} (u) \, b (u) + 
	\oint d\overline{u} \, 
	\overline{v}_{a_q}^{(i)} (\overline{u}) \, \overline{b} (\overline{u})
	\right]  \, , 
	\quad \quad u = z_i, z_j \, ,
\end{align}
since this is conformally invariant. We point out the Schiffer vectors $v^{(i)} (u)$ and $\overline{v}^{(i)} (\overline{u})$ above may not be complex conjugates of each other.

Given these ingredients, we define
\begin{align}
	\langle \Omega^{(p)}_{g,n} | = \left(- 2\pi i\right)^{- d_{g,n}}
	\langle \Sigma_{g,n} | \, B_p  \, .
\end{align}
The prefactor is inserted to generate the correct factorization of the amplitudes. It carries an intrinsic ghost number $6g-6-p$ and has the statistics determined by $(-1)^p$. The object $\langle \Omega^{(p)}_{g,n} | $ is indeed a $(\widehat{\mathcal{H}}^\ast)^{\otimes n}$-valued $p$-form over the bundle $\widehat{\mathcal{P}}_{g,n} \to \mathcal{M}_{g,n}$ due to the antisymmetry of exchanging $b$-ghosts. We call $\langle \Omega^{(6g-6+2n)}_{g,n} | = \langle \Omega_{g,n} |$~\textit{the string measure}. This is the correct integrand for the moduli integrations in string amplitudes.

\subsection{Closed string field theory action}

Given $p$-forms over $\widehat{\mathcal{P}}_{g,n} \to \mathcal{M}_{g,n}$ we are now ready to construct the CSFT action and discuss its perturbative consistency~\cite{Sen:1993kb}. The free part of it is given by
\begin{align} \label{eq:2.32}
	S_{0,2} [\Psi] = {1 \over 2} \, \langle \Psi | c_0^- Q_B | \Psi \rangle \, ,
\end{align}
where $c_0^- = ( c_0 - \overline{c}_0 ) /2$ is the zero mode of the $c$-ghosts, $Q_B$ is the BRST operator of matter + ghost CFT,  and $\langle \cdot | \cdot \rangle$ is the ordinary BPZ inner product
\begin{align} \label{eq:2.22}
	\langle \Sigma_{0,2} | \, \Psi_1 \otimes \Psi_2 
	= \langle \Psi_1|\Psi_2 \rangle =
	\langle \Psi_1(\widetilde{z} = 0) \, \Psi_2(z = 0) \rangle
	\quad \quad \text{where} \quad\quad
	 \widetilde{z}(z) = {1 \over z} \, .
\end{align}

Above we have defined $\langle \Sigma_{0,2} |$ encoding the BPZ product, which carries intrinsic ghost number $-6$ and is even.  In a given basis $\phi_\alpha$ of $\mathcal{H}$ it can be expressed as
\begin{align} \label{eq:2.35}
	\langle \Sigma_{0,2} | = \sum_\alpha \,  \langle \phi_\alpha | \otimes \langle \phi_\alpha^c | \, , 
	\quad \quad
	| \Sigma_{0,2} \rangle = \sum_{\alpha} | \phi_\alpha \rangle \otimes | \phi_\alpha^c \rangle \, ,
\end{align}
and the conjugate states $\phi_\alpha^c \in \mathcal{H}$ are given by
\begin{align}
	\langle \phi_\alpha^c | \phi_\beta \rangle = 
	(-1)^{\phi_\alpha} \langle \phi_\alpha | \phi_\beta^c \rangle =
	\delta_{\alpha \beta} \, ,
\end{align}
where $(-1)^{\phi_\alpha}$ denotes the statistics of the state $\phi_\alpha $. Notice the conjugate state has the same statistics as $\phi_\alpha$ and it has ghost number $6-\text{gh}(\phi_\alpha)$. The associated partitions of the identity operator $\id$ acting on $\mathcal{H}$ are
\begin{align}
	\id = \sum_\alpha | \phi_\alpha \rangle \langle \phi_\alpha^c | 
	= \sum_\alpha (-1)^{\phi_\alpha} | \phi_\alpha^c \rangle \langle \phi_\alpha | \, .
\end{align}

By further introducing the symplectic form~\cite{Erler:2019loq}
\begin{align}
	\langle \omega | = \langle \Sigma_{0,2} | \, \id \otimes c_0^- \in (\mathcal{H}^{\ast})^{\otimes 2} \, ,
\end{align}
the quadratic part of the action~\eqref{eq:2.32} can be expressed as
\begin{align}
	S_{0,2} [\Psi] = {1 \over 2} \, \langle \omega|  \Psi   \otimes Q_B \Psi  \, ,
\end{align}
since $Q_B$ is cyclic under the symplectic form and the string field $\Psi$ is even in the CSFT master action. The object $\langle \omega|$ has odd statistics and carries intrinsic ghost number $-5$.

Relatedly, we also introduce~\emph{the Poisson bivector}
\begin{align} \label{eq:2.40}
	|\omega^{-1} \rangle &=  \id  \otimes b_0^- \delta(L_0^-)  \, | \Sigma_{0,2} \rangle  
	=
	 \sum_{\alpha} (-1)^{\phi_\alpha } | \phi_\alpha \rangle \, \otimes \, b_0^- \, \delta(L_0^-) \, | \phi_\alpha^c \rangle  \, ,
\end{align}
which inverts the symplectic form in the following sense~\cite{Erler:2019loq}
\begin{align}
	\big( \langle \omega | \otimes \id \big) \, \big( \id \otimes  |\omega^{-1} \rangle \big)
	= b_0^- \, \delta(L_0^-) \, c_0^- \, .
\end{align}
Note that $b_0^- \delta(L_0^-) c_0^- $  is the projector onto $\widehat{\mathcal{H}}$ so it restricts to the identity $\id$ there. That is,
\begin{align} \label{eq:2.42a}
	b_0^- \delta(L_0^-) c_0^- \, | \Psi \rangle = | \Psi \rangle \, , 
	\quad \quad \quad
	\langle \Psi | \, c_0^- \delta(L_0^-) b_0^- = \langle \Psi |  \, .
\end{align}
for every $\Psi \in \widehat{\mathcal{H}}$. Furthermore the Poisson bivector is annihilated by $b_0^{-}$ and $L_0^{-}$ at both entries, hence
\begin{align}
	|\omega^{-1} \rangle \in \widehat{\mathcal{H}}^{\otimes 2} \, .
\end{align}
Like $\langle \omega|$, it has odd statistics and carries intrinsic ghost number $5$. The Poisson bivector essentially implements the entirety of the twist-sewing~\eqref{eq:TS}, thanks to the combination of these properties~\cite{Erler:2019loq}.

Finally, it is also useful to introduce a (odd, ghost number 5) bivector that implements the sewing with a particular value of twist $\tau$
\begin{align} \label{eq:2.42}
	| \mathfrak{S} (\tau) \rangle &\equiv
	\bigg[ \id \otimes b_0^- \exp\left( {2 \pi i \tau L_0^- \over \ell} \right) \bigg] | \Sigma_{0,2} \rangle
	\\ \nonumber
	&= \sum_{\alpha} (-1)^{\phi_\alpha} | \phi_\alpha \rangle \otimes b_0^- \exp\left({2 \pi i \tau L_0^- \over \ell} \right)
	| \phi_\alpha^c \rangle \in \mathcal{H}^{\otimes 2} \, .
\end{align}
Note that this bivector does not belong to the space $ \widehat{\mathcal{H}}^{\otimes 2}$. Only upon integrating over all twists do we obtain the Poisson bivector
\begin{align} \label{eq:2.43}
	| \omega^{-1} \rangle = \int\limits_0^\ell \, {d \tau \over \ell} \, | \mathfrak{S} (\tau) \rangle \in \widehat{\mathcal{H}}^{\otimes 2} \, .
\end{align}
and we restrict to the level-matched states.

We now include the interactions to the action~\eqref{eq:2.32}. These are defined by integrating the string measure over the string vertices
\begin{align} \label{eq:2.25}
	\langle \mathcal{V}_{g,n} | = \int\limits_{\mathcal{V}_{g,n}} \langle \Omega_{g,n} | = 
	\left(- 2\pi i \right)^{-d_{g,n}} \int\limits_{\mathcal{V}_{g,n}}  \langle \Sigma_{g,n} | \, B_{6g-6+2n} \, ,
\end{align}
We denote the resulting bras by the same letter as the integration chain. Combining them for all $g,n$ produces the interacting CSFT action
\begin{align}  \label{eq:2.37}
	S[\Psi] = S_{0, 2}[\Psi] + \sum_{g,n} {1 \over n!} \, \hbar^{g} \, \kappa^{2g-2+n} 
	\, \langle \mathcal{V}_{g,n} | \Psi \rangle^{\otimes n} \, .
\end{align}
Observe that each term in the action is finite individually since the moduli integration~\eqref{eq:2.25} doesn't get contributions from the degenerating surfaces. The action $S[\Psi]$ can be shown to be BV quantizable as a result of string vertices $\mathcal{V}$ solving the geometric master equation~\eqref{eq:2.2}, refer to~\cite{Sen:1993kb} for details.

A somewhat convenient way of writing the action $S[\Psi]$ is
\begin{align} \label{eq:46}
	S[\Psi] = 
	S_{0,2}[\Psi] 
	+ \sum_{g,n} {1 \over n!} \, \hbar^g \kappa^{2g-2+n} \,
	\langle \omega | \, \Psi \otimes L_{g,n-1} \left(\Psi^{(n-1) } \right) \, ,
\end{align}
for which we have defined~\emph{the string products} $L_{g,n-1} : \widehat{\mathcal{H}}^{\otimes (n-1)} \to \widehat{\mathcal{H}} $
\begin{align} \label{eq:2.47}
	L_{g,n-1} = \bigg( \, \id \otimes \langle \mathcal{V}_{g,n} \, | \, \bigg) \,
	\bigg( | \omega^{-1} \rangle \otimes \id^{\otimes (n-1)} \bigg) \, .
\end{align}
Using them and their cyclicity under the symplectic form, the classical equation of motion and the gauge transformations can be expressed as
\begin{subequations} \label{eq:2.48}
\begin{align}
	& Q_B \, \Psi + \sum_{n=2}^\infty {\kappa^{n-1} \over n!} \, L_{0,n} \left(\Psi^n\right) = 0 \, , \label{eq:2.48a}
	\\
	&\delta_\Lambda \Psi = Q_B \, \Lambda 
	+ \sum_{n=2}^\infty {\kappa^{n-1} \over (n-1)!} \, L_{0,n} (\Lambda, \Psi^{n-1}) \label{eq:2.48b}\, .
\end{align}
\end{subequations}
Here $\Psi$ and $\Lambda $ should be restricted to have ghost numbers $2$ and $1$ respectively. The latter is also taken to have odd statistics.

We see CSFT is simply a result of  reverse engineering string amplitudes as far as generic string vertices $\mathcal{V}$ are concerned. It is a natural anticipation that a smart choice of string vertices can manifest extra information about the theory however, as we have argued before. This situation already presents itself for open strings: Witten's vertex is superior relative to other string vertices since it makes the theory's underlying associative algebra manifest and the analytic solution becomes accessible as a result---even though any other choice would have accomplished the BV consistency. 

As we shall argue, using hyperbolic vertices $\mathcal{V}^L$~\eqref{eq:hyp} for the action~\eqref{eq:2.37} manifests a peculiar recursive structure among its elementary interactions. This structure is topological in nature and somewhat resembles the recursive structure of the stubbed theories~\cite{Erler:2023emp,Chiaffrino:2021uyd,Schnabl:2023dbv,Erbin:2023hcs,Schnabl:2024fdx,Maccaferri:2024puc}. On the other hand, some of its features are quite distinct due to the presence of closed Riemann surfaces and their modularity properties.

\section{The recursion for the volumes of systolic subsets} \label{sec:Sys}

Before we discuss the recursion in hyperbolic CSFT it is beneficial to investigate a simpler analog to communicate some of the central ideas behind its construction. In this section, after reviewing Mirzakhani's method for computing the WP volumes of the moduli spaces of bordered Riemann surfaces $\mathcal{M}_{g,n}(L_i)$~\cite{mirzakhani2007simple,mirzakhani2007weil}, we form a recursion among the WP volumes of the systolic subsets $\mathcal{V}_{g,n}^L(L_i)$ for $L \leq L_\ast$ using~\emph{the twisting procedure} developed in~\cite{andersen2017geometric}. The twisting allows one to restrict the considerations to the vertex regions of CSFT, which is one of the most crucial ingredients for deriving the recursion for hyperbolic vertices in the next section.

\subsection{Mirzakhani's recursion for the volumes of $\mathcal{M}_{g,n}(L_i)$}

We begin by reviewing Mirzakhani's method for recursively computing the WP volumes of the moduli spaces $\mathcal{M}_{g,n}(L_i)$ in order to set the stage for our discussion~\cite{mirzakhani2007simple,mirzakhani2007weil}. The central idea behind her method is to replace the original volume integration with another integration over a suitable covering space---at the expense of introducing measure factors in the integrand that compensate the overcounting when working on the covering space.

\begin{figure}[t]
	\centering
	\includegraphics[height=3in]{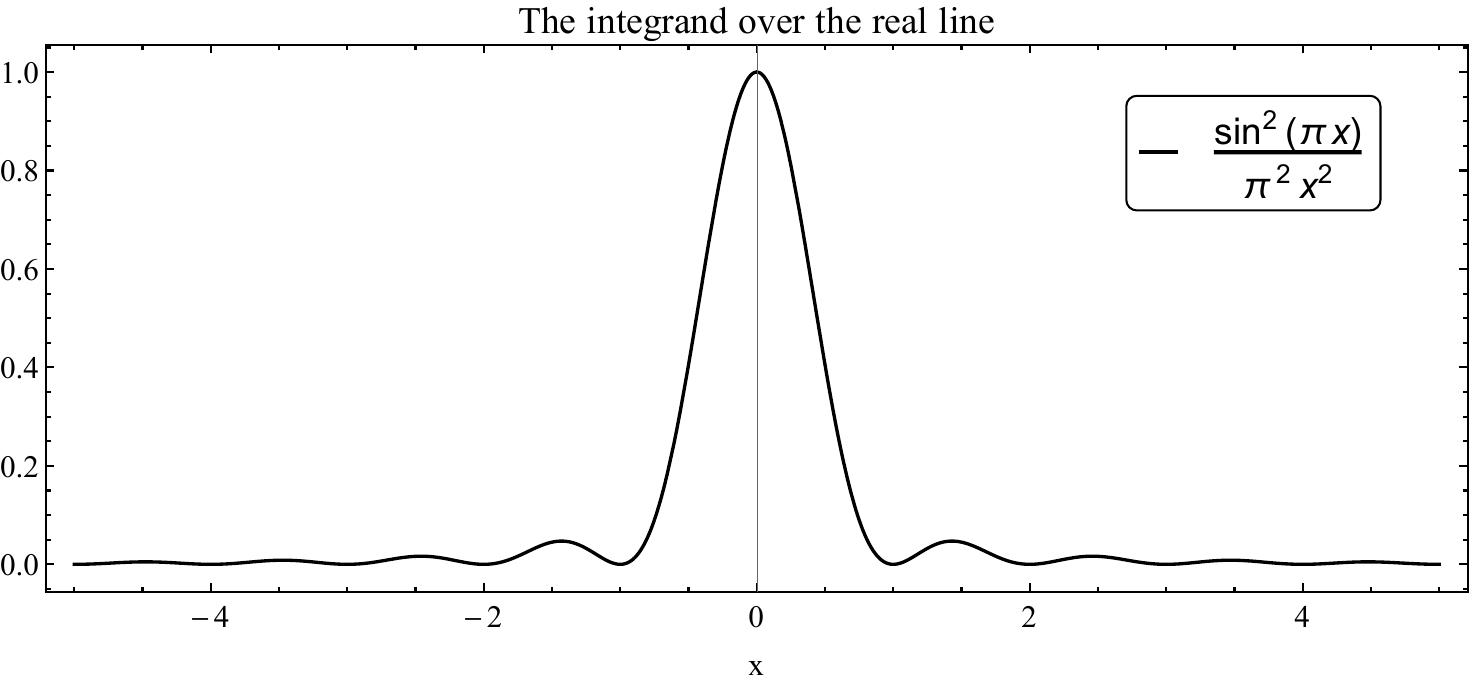}
	\caption{The measure factor of~\eqref{eq:example}. }\label{fig:integrand}
\end{figure}

Let us demonstrate this procedure by considering a toy example provided in~\cite{Moosavian:2017sev}. Imagine we have a circle $S^1$ described by the interval $[0,1]$ with identified endpoints. We would like to evaluate
\begin{align}
	I = \int\limits_{0}^1 f(x) \, dx \, ,
\end{align}
where $f(x)$ is a periodic integrable function $f(x) = f(x + 1)$ on $S^1$. This integral can be replaced with another integral over the universal cover of the circle $S^1$, the real line $\mathbb{R}$, by noticing the partition of unity
\begin{align} \label{eq:3.2}
	1 = \sum_{k \in \mathbb{Z}}  {\sin^2 \left( \pi (x-k) \right) \over \pi^2 \, (x-k)^2} \, ,
\end{align}
and inserting it into the integrand of $I$
\begin{align} \label{eq:example}
	I = \int\limits_{0}^1 1 \cdot f(x) dx 
	&= \int\limits_{0}^1  \left( \,\sum_{k \in \mathbb{Z}}  {\sin^2 \left( \pi (x-k) \right) \over \pi^2 (x-k)^2} \right) f(x) \, dx 
	 \\
	&= \int \limits_{0}^1 \sum_{k \in \mathbb{Z}}  {\sin^2 \left( \pi (x-k) \right) \over \pi^2 (x-k)^2} f(x-k) \, dx 
	\nonumber \\ 
	&=  \sum_{k \in \mathbb{Z}} \, \int \limits_{0}^1  {\sin^2 \left( \pi (x-k) \right) \over \pi^2 (x-k)^2} f(x-k) \, dx 
	= \int \limits_{-\infty}^\infty  \, {\sin^2 \left( \pi x\right) \over \pi^2 x^2} f(x) \, dx \, . \nonumber
\end{align}
Above we have used the periodicity of $f$ in the second line and commuted the sum out of the integral in the third line. Then we replaced the sum and integral with a single integration over $\mathbb{R}$ using the translation property of the integrand.  For $f(x)=1$ we plot the integrand of the last integral of~\eqref{eq:example} in figure~\eqref{fig:integrand}. As one can see, not only $[0,1]$ contributes to the integral, but there are contributions from all of its images in $\mathbb{R}$ accompanied by a suitable measure factor. Summing all of them adds up to integrating $1$ over $[0,1]$. 

We can formalize this result. Consider a space $M$ and its covering space $N$ with the projection $\pi: N \to M$. Imagine the top form $\varepsilon$ on $M$ and its pullback $\pi^\ast \varepsilon$ on $N$. Also imagine a function $h$ on $N$. This can be pushed forward to $M$ by
\begin{align} \label{eq:3.4}
	\pi_\ast h (x) = \sum_{y \in \pi^{-1}\{ x \} } h(y) \, ,
\end{align}
assuming the sum converges. Then we see
\begin{align} \label{eq:main}
	\int\limits_{M} \pi_\ast h \cdot \varepsilon = \int\limits_{N}  h \cdot \pi^\ast \varepsilon \, ,
\end{align}
following the reasoning in~\eqref{eq:example}. There, for instance, $M=S^1$, $N = \mathbb{R}$, $\varepsilon = dx$ and $h(x) = \sin^2 \left( \pi x\right) f(x) / \pi^2 x^2$. We note the nontrivial step of this method is coming up with the function $h$ in the cover $N$ from a seed function $\pi_\ast h$ on the original manifold $M$, i.e., identifying the appropriate partition of unity like in~\eqref{eq:3.2}.

Remarkably, Mirzakhani found a way to calculate the volumes of $\mathcal{M}_{g,n}(L_i)$ by identifying the appropriate partitions of unity via the hyperbolic geometry of pair of pants~\cite{mirzakhani2007simple} and using~\eqref{eq:main} recursively, refer to appendix D of~\cite{Stanford:2019vob} for a derivation accessible to a physicist. Before we illustrate this procedure let us introduce the functions which we call~\textit{the Mirzakhani kernels}
\begin{subequations} \label{eq:3.6}
	\begin{align}
		D_{L_1 L_2 L_3} &= 2\log \left[\frac{\exp\left( {\frac{L_1}{2}}\right) + \exp\left({\frac{L_2 + L_3}{2}}\right) }{\exp\left({-\frac{L_1}{2}}\right) + \exp\left({\frac{L_2 + L_3}{2}}\right) } \right] \, , \\
		T_{L_1 L_2 L_3}  &=
		\log \left[{\cosh\left({L_3 \over 2}\right)+\cosh \left({L_1 + L_2 \over 2}\right) 
		\over \cosh\left({L_3 \over 2}\right) + \cosh\left({L_1 - L_2 \over 2}\right) } \right] \, , \\
		R_{L_1 L_2 L_3} &= L_1 - \log\left[\frac{\cosh\left(\frac{L_2}{2} \right) + \cosh\left(\frac{L_1 + L_3}{2} \right) } { \cosh\left(\frac{L_2}{2} \right)  + \cosh\left(\frac{L_1 - L_3}{2} \right) }  \right]  \, .
	\end{align}
\end{subequations}
They have the symmetry properties
\begin{subequations} \label{eq:27}
	\begin{align}
		&D_{L_1 L_2 L_3} = D_{L_1 L_3 L_2} \, , \\
		&T_{L_1 L_2 L_3} = T_{L_2 L_1 L_3} \, ,  \\
		&D_{L_1 L_2 L_3} + T_{L_1 L_2 L_3} = L_1 - T_{L_1 L_3 L_2}
		= R_{L_1 L_2 L_3} 
	\end{align}
\end{subequations}
and satisfy
\begin{align} \label{eq:38}
	&{\p D_{L_1 L_2 L_3} \over \p L_1} = H_{ L_2+L_3, L_1 }\, ,
	\quad \quad {\p R_{L_1 L_2 L_3} \over \p L_1} = {1 \over 2} \, (H_{L_3, L_1+L_2}+ H_{L_3, L_1-L_2} ) \, ,
\end{align}
where
\begin{align}
	H_{L_1,L_2} = \left[ {1 + \exp\left({L_1+L_2\over 2}\right)} \right]^{-1}
	+  \left[ {1 + \exp\left({L_1-L_2 \over 2}\right) } \right]^{-1} \, .
\end{align}
We can express these kernels compactly as
\begin{subequations} \label{eq:310}
	\begin{align}
		&D_{L_1 L_2 L_3} = F_{L_1 - L_2 - L_3} - F_{-L_1 - L_2 - L_3} \\
		&R_{L_1 L_2 L_3} = {1 \over 2} \left(
		F_{L_1 + L_2 - L_3} + F_{L_1 - L_2 - L_3} - F_{-L_1 + L_2 - L_3} - F_{-L_1 - L_2 - L_3}
		\right) \, ,
	\end{align}
\end{subequations}
by introducing
\begin{align}
	F_L = 2 \log \left[ {1 + \exp\left({L \over 2}\right)} \right] \, .
\end{align}
Some of these identities are going to be useful later.

\begin{figure}[t]
	\centering
	\includegraphics[height=1.8in]{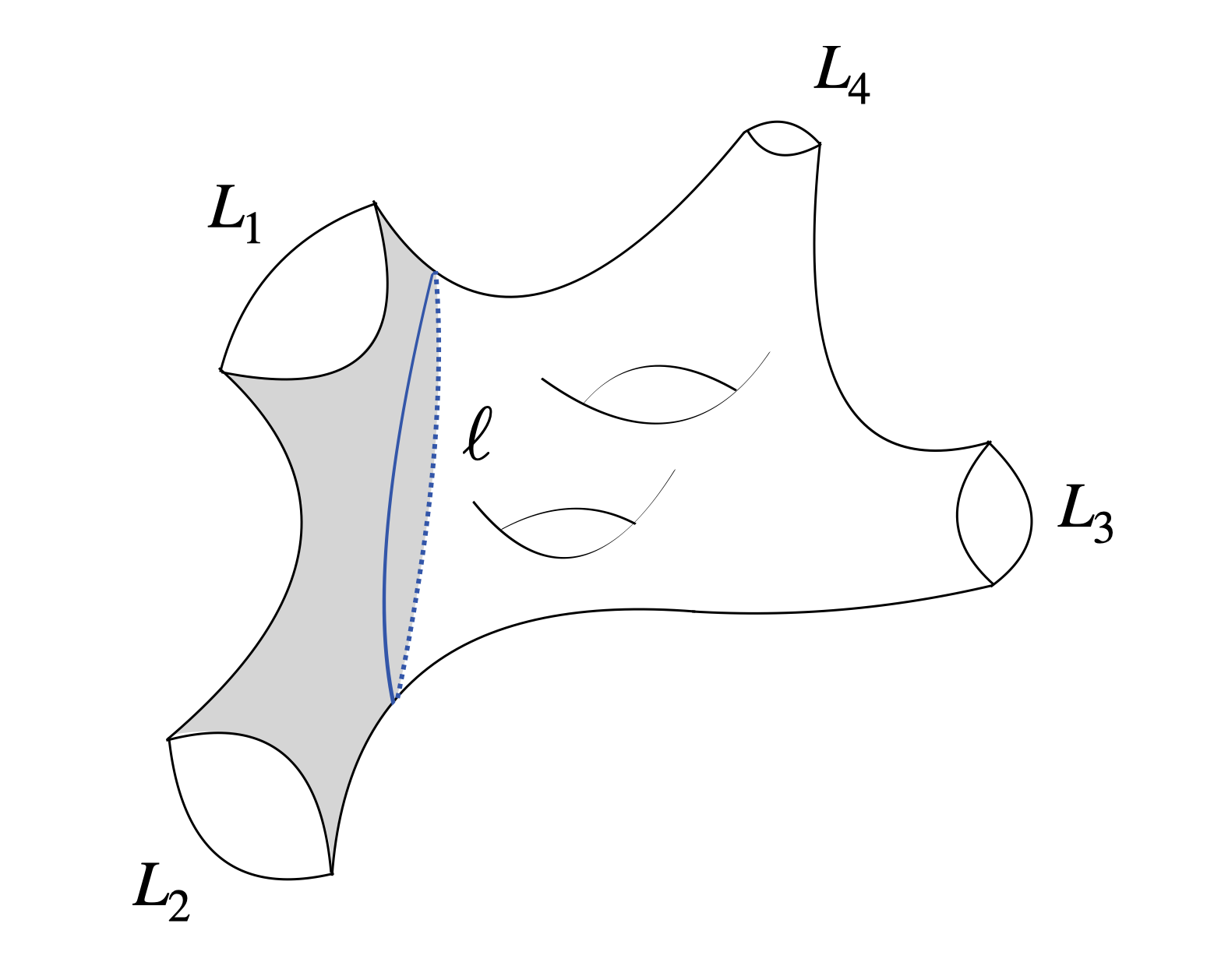}
	\includegraphics[height=1.8in]{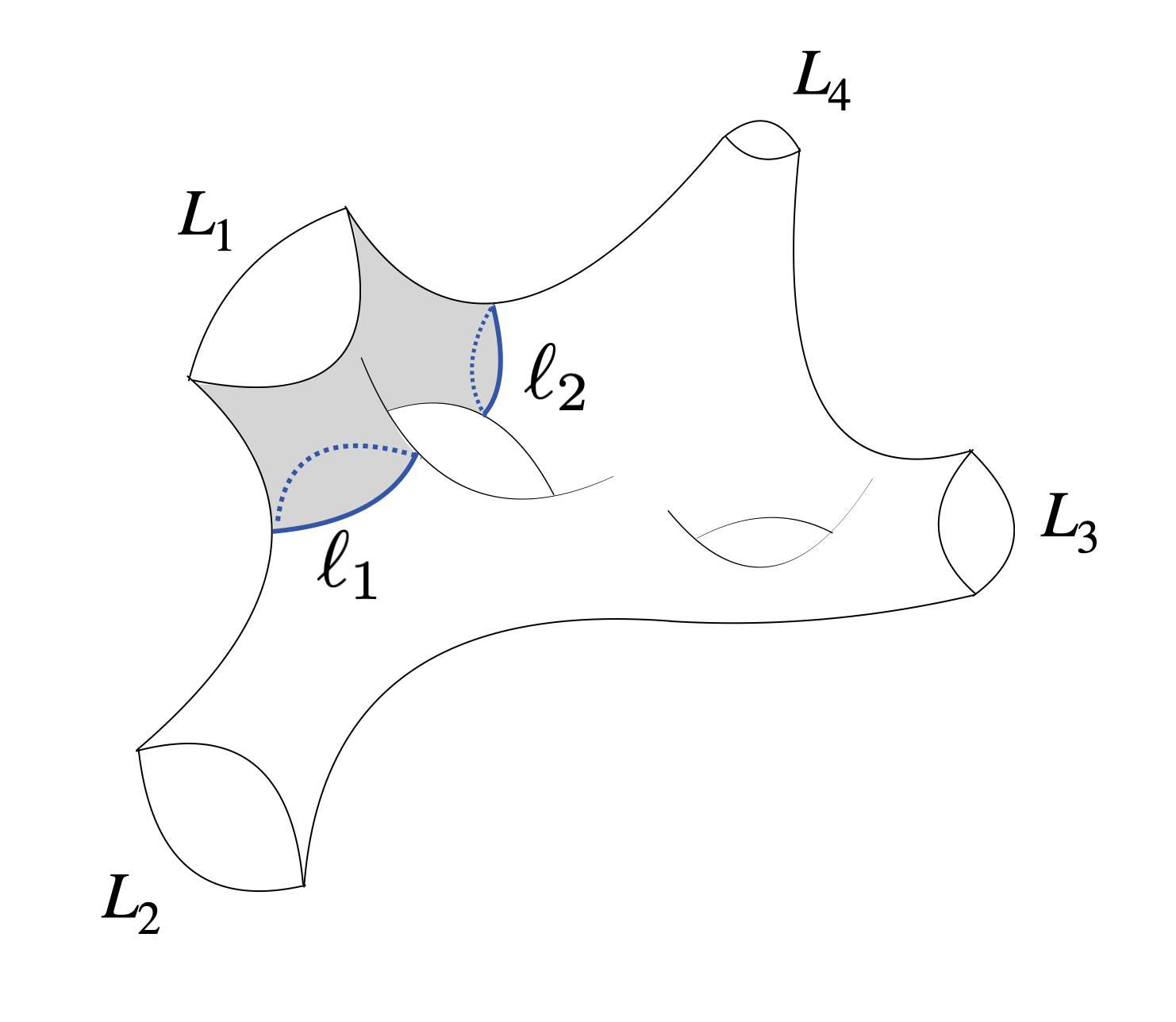}
	\includegraphics[height=1.8in]{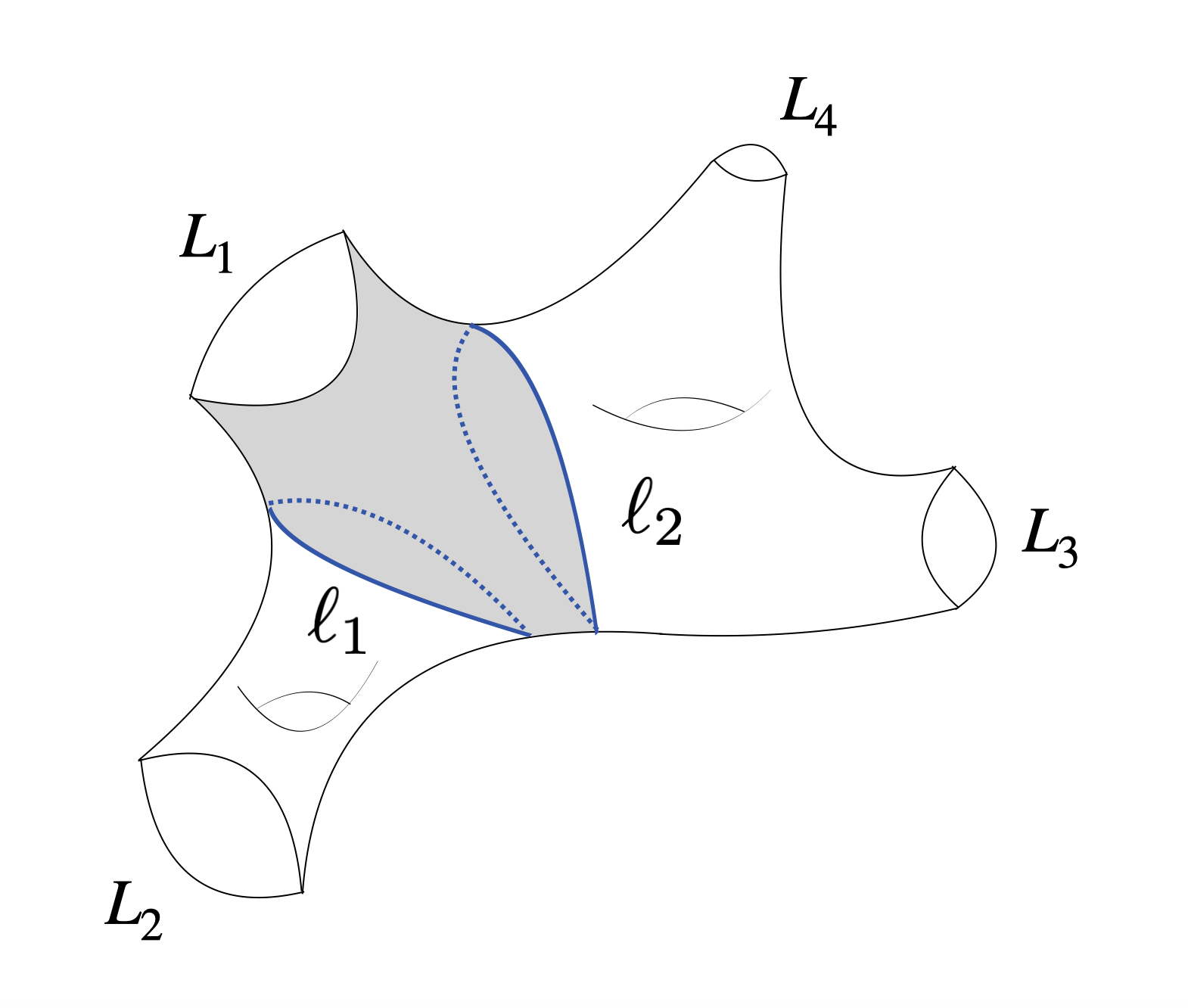}
	\caption{Different ways of excising a pair of pants and their corresponding terms in the expression~\eqref{eq:110}. From left to right: $R$-term, nonseparating $D$-term and separating $D$-term excisions.}\label{Mirz-gluing-figure}
\end{figure}

In terms of the kernels $D$ and $R$, the relevant partition of the length $L_1$ for a Riemann surface with $n$ borders of length $L_i$ is given by~\emph{the Mirzakhani–McShane identity}~\cite{mirzakhani2007simple}
\begin{align} \label{eq:17}
	L_1 = \sum_{\{\gamma, \delta\} \in \mathcal{P}_1 }  \hspace{-0.12in} D_{L_1\gamma \delta} 
	+\sum_{i = 2}^n \sum_{\gamma \in \mathcal{P}_i} R_{L_1L_i \gamma} 
	\, .
\end{align}
Here the sets $\mathcal{P}_i$ are certain collections of simple geodesics. For $i=1$, it is the set of pairs of simple closed internal geodesics that bound a pair of pants with the first border $L_1$, and for $i = 2, \cdots, n$, it is the set of simple closed internal geodesics that bound a pair of pants with the first and $i$-th borders $L_1, L_i$. We often imagine these pants are ``excised'' from the surface as shown in figure~\ref{Mirz-gluing-figure} to interpret various different terms in the expressions below.

The volumes of the moduli spaces $\mathcal{M}_{g,n}(L_i)$ are given by the integrals
\begin{align} \label{eq:18}
	V \mathcal{M}_{g,n}(L_i) = \int\limits_{\mathcal{M}_{g,n}(L_i)} \text{vol}_{WP}  = 
	\int\limits_{\mathcal{M}_{g,n}(L_i)} {\omega_{WP}^{3g-3+n} \over (3g-3+n)!}\, ,
\end{align}
for which we take $V \mathcal{M}_{0,3}(L_i)= 1$ to set the units. This definition is implicitly taken to be divided by $2$ for $(g,n) = (1,1)$ due to the presence of a nontrivial $\mathbb{Z}_2$ symmetry~\cite{Stanford:2019vob}. This simplifies the expressions below.

It is possible to derive the following recursion relation by substituting the partition of identity~\eqref{eq:17} to the integrand of~\eqref{eq:18} for $2g -2 + n > 1$ and $n \geq 1$ to obtain
\begin{align} \label{eq:110}
	L_1 \cdot V \mathcal{M}_{g,n} (L_i) &= 
	\sum_{i=2}^n \int\limits_{0}^\infty \ell d \ell \, R_{L_1L_i\ell} \, V \mathcal{M}_{g,n-1}\left(\ell, \mathbf{L} \setminus \{L_i \} \right) 
	\\
	&\hspace{-0.8in} + {1 \over 2} \int\limits_{0}^\infty \ell_1 d \ell_1 \, \int\limits_{0}^\infty \ell_2 d \ell_2 \,
	D_{L_1 \ell_1 \ell_2} \bigg[
	V \mathcal{M}_{g-1, n+1} (\ell_1, \ell_2, \mathbf{L}) 
	+ \sum_{\text{stable}} V \mathcal{M}_{g_1, n_1} (\ell_1,  \mathbf{L}_1) \cdot V \mathcal{M}_{g_2, n_2} (\ell_2,  \mathbf{L}_2)
	\bigg] \, , \nonumber
\end{align}
by 	``unrolling'' the integral over the relevant fibers in the covering space, see~\cite{mirzakhani2007simple} for mathematical details. Here
\begin{align}
	\mathbf{L} = \{L_2, \cdots, L_n \} \, , \quad \quad
	\mathbf{L}_1 \cup \mathbf{L}_2 = \mathbf{L} \,  , \quad  \quad
	\mathbf{L}_1 \cap \mathbf{L}_2 = \emptyset \, ,
\end{align}
and the ``stable'' in~\eqref{eq:110} denotes the sum over all $g_1,g_2,n_1,n_2,\mathbf{L}_1, \mathbf{L}_2$ that satisfy
\begin{align}
	&g_1 + g_2 = g\, , \quad n_1 + n_2 = n + 1 \, , \quad 2g_1 - 2 + n_1 > 0 \, , \quad 2g_2 - 2 + n_2 > 0 \,  .
\end{align}
The pictorial representations of various terms in the relation~\eqref{eq:110} along with their associated pants excisions are given in figure~\ref{Mirz-gluing-figure}. The first few of the volumes are given in table~\ref{tab:M}. Observe that it is possible to ignore the first term in the second line of~\eqref{eq:110} (i.e., nonseparating $D$ term) as long as surfaces of genus zero are concerned.

\begin{table}[t]
	\begin{center}
		\bgroup
		\def\arraystretch{1.75}
		\Large
		\begin{tabular}{ |c | c | }
			\hline
			$(g,n)$& $V \mathcal{M}_{g,n} (L_i) $  \\ 
			\hline
			$(0,3)$ & 1  \\ 
			\hline
			$(0,4)$ & $ 2\pi^2 + {1 \over 2} \sum\limits_{i=1}^4 L_i^2 $  \\ 
			\hline
			$(1,1)$ & $ {1 \over 12} \pi^2  + {1 \over 48} L_1^2$  \\
			\hline
			$(0,5)$ & $10 \pi^4 + {1 \over 8} \sum\limits_{i=1}^5 L_i^4 + {1 \over 2} \sum\limits_{1 \leq i<j \leq 5} L_i^2 L_j^2 + 3 \pi^2 \sum\limits_{i=1}^5 L_i^2$ \\ 
			\hline
			$(1,2)$ & ${1 \over 192} \left(4\pi^2 + L_1^2 +L_2^2 \right)\left(12\pi^2 + L_1^2 +L_2^2 \right)$ \\
			\hline
		\end{tabular}
		\normalfont
		\egroup
	\end{center}
	\caption{\label{tab:M}The WP volumes of the moduli spaces $V \mathcal{M}_{g,n} (L_i) $, see~\cite{mirzakhani2007simple,do2011moduli}. Despite the fact that border $L_1$ is singled out in the recursion, the resulting expressions are symmetric under the exchange of borders.}
\end{table}

We note the formula~\eqref{eq:110} doesn't apply to $1$-bordered tori directly and this case requires special treatment. Nevertheless, it can be still found using the partition of unity~\eqref{eq:17}
\begin{align} \label{eq:3.17}
	V \mathcal{M}_{1,1} (L_1) = {1 \over 2} {1 \over L_1} \int\limits_{0}^\infty \ell \, d \ell \,  {D_{L_1 \ell\ell}} \cdot
	V \mathcal{M}_{0,3} \left(L_1, \ell, \ell \right) 
	= {\pi^2 \over 12} + {L_1^2 \over 48}
	\, ,
\end{align}
and together with $V\mathcal{M}_{0,3}(L_i) = 1$, the recursion~\eqref{eq:110} can be used to find the volumes for the rest of the moduli spaces of surfaces with at least one border. We highlight the presence of the extra $1/2$ in front of this equation: this comes from the $\mathbb{Z}_2$ symmetry of one-bordered tori.

Finally, we point out the formula, known as~\emph{the dilaton equation},
\begin{align} \label{eq:3.18}
	(2g-2+n) \, V \mathcal{M}_{g,n}(L_i) =  {1 \over 2 \pi i} V \mathcal{M}_{g,n+1}' (L_i, 2 \pi i) \, ,
\end{align}
can be used for the volumes of the moduli spaces of surfaces without borders~\cite{Eynard:2007fi,do2009weil}. Here the prime stands for the derivative with respect to the border length $L_{n+1}$.

\subsection{The recursion for the volumes of $\mathcal{V}_{g,n}^L(L_i)$}

In this subsection we generalize Mirzakhani's recursion discussed in the previous subsection to the recursion for the WP volumes of the systolic subsets $\mathcal{V}_{g,n}^L(L_i)$. Only the cases with $L_i = L$ are relevant for CSFT, but as we shall see, it will be necessary to consider $L_i \neq L$ to construct the desired recursion. We always take $L \leq L_\ast \equiv 2 \sinh^{-1} 1$ for reasons that are going to be apparent soon, unless stated otherwise. 

The volumes of the systolic subsets are given by
\begin{align} \label{eq:V}
	V \mathcal{V}_{g,n}^L(L_i) = \int\limits_{\mathcal{V}_{g,n}^L(L_i) } \text{vol}_{WP}
	=\int\limits_{\mathcal{M}_{g,n}(L_i) } \mathbf{1}_{\mathcal{V}_{g,n}^L(L_i) } \cdot \text{vol}_{WP} 
	\, .
\end{align}
Here $\mathbf{1}_{\mathcal{V}_{g,n}^L(L_i) }$ is the indicator function for the region $\mathcal{V}_{g,n}^L(L_i)$, that is
\begin{align} \label{eq;3.20}
	\mathbf{1}_{\mathcal{V}_{g,n}^L(L_i) }(\Sigma) =
	\begin{cases}
		0 & \text{for} \quad \text{sys}(\Sigma) < L \\
		1 & \text{for} \quad  \text{sys}(\Sigma) \geq L
	\end{cases} \, .
\end{align}
The volume $V \mathcal{V}_{g,n}^L(L_i) $ can be understood as integrating the indicator function over the moduli space with respect to the WP volume form $\text{vol}_{WP}$. The definition~\eqref{eq:V} is implicitly divided by $2$ for $(g,n) = (1,1)$ like before.

It is beneficial to have a ``regularized'' representation for the indicator function $\mathbf{1}_{\mathcal{V}_{g,n}^L(L_i) }$  by considering its slight generalization~\cite{andersen2017geometric}. Let us denote the set $S(\Sigma)$ to be the set of simple closed ``short'' geodesics $\gamma$ of $\Sigma \in \mathcal{M}_{g,n}(L_i)$ with $\gamma < L$. For $L \leq L_\ast$ the curves in the set $S(\Sigma)$ cannot intersect by the collar lemma~\eqref{eq:Uncertainty}, hence
\begin{align}
	|S(\Sigma)| \leq 3g - 3 + n < \infty \, ,
\end{align}
and the following holds
\begin{align} \label{eq:3.20}
	\Omega_t(\Sigma) \equiv (1+t)^{|S(\Sigma)|} 
	&= \sum_{k=0}^{|S(\Sigma)|} \binom{|S(\Sigma)|}{k} \, t^k
	= \sum_{\mu \subseteq S(\Sigma)} t^{|\mu|}
	= \sum_{c \in M(\Sigma) } \prod_{\gamma \in \pi_0(c)} t \, \theta(L - \gamma ) 
	\, , 
\end{align}
for $t \in \mathbb{R}$, as each sum and product is finite. Here $\mu \subseteq S(\Sigma)$ is a collection of short geodesics on $\Sigma$, which we obtained by binomial theorem. It can be empty.

We also define $M(\Sigma)$ to be the set of~\emph{primitive multicurves} on the surface $\Sigma$. A~\emph{multicurve} $c$ is defined as the homotopy class of one-dimensional submanifolds on $\Sigma$ whose components are not homotopic to borders. Primitive refers to no disjoint component of $c \in M(\Sigma)$ is homotopic to each other. The primitive multicurve can be thought of the union of pairwise-disjoint simple closed geodesics on the surface $\Sigma$. From this definition~\eqref{eq:3.20} directly follows after employing the step functions. Finally $\pi_0(c)$ stands for the set of components of the multicurve $c$.

The equation~\eqref{eq:3.20} then shows
\begin{align} \label{eq:3.22}
	\mathbf{1}_{\mathcal{V}_{g,n}^L(L_i) } (\Sigma) = \lim_{t \to -1} \Omega_t(\Sigma)
	= \Omega_{-1}(\Sigma)
	= \sum_{c \in M(\Sigma) } \prod_{\gamma \in \pi_0(c)} \, (-1) \, \theta(L - \gamma ) \, ,
\end{align}
since the product and sum are finite. The indicator function, and its regularized version, are MCG-invariant because they only depend on the set $S(\Sigma)$. This is also manifest in~\eqref{eq:3.22} since the terms in the sum map to each other under large diffeomorphisms. Based on this analysis the volumes~\eqref{eq:V} take the form
\begin{align} \label{eq:3.23}
	V \mathcal{V}_{g,n}^L(L_i)  
	= \int\limits_{\mathcal{M}_{g,n}(L_i) }  \Omega_{-1} \cdot \text{vol}_{WP} 
	= \int\limits_{\mathcal{M}_{g,n}(L_i) }  \sum_{c \in M(\Sigma) } \prod_{\gamma \in \pi_0(c)} (-1) \, \theta(L - \gamma ) \cdot \text{vol}_{WP}  \, .
\end{align}

As before, we need to find an appropriate partition of the function $\Omega_{-1}(\Sigma)$~\eqref{eq:3.4} and apply~\eqref{eq:main}. However we now have a somewhat complicated integrand that involves a sum over the multicurves, rather than a mere constant, and using the Mirzakhani-McShane identity~\eqref{eq:17} wouldn't be sufficient as it is. We can handle this situation by classifying the multicurves $c$ appropriately and refining~\eqref{eq:17} with respect to this classification. This would lead to the desired recursion \`a la Mirzakhani.

\begin{figure}[t]
	\centering
	\includegraphics[height=1.95in]{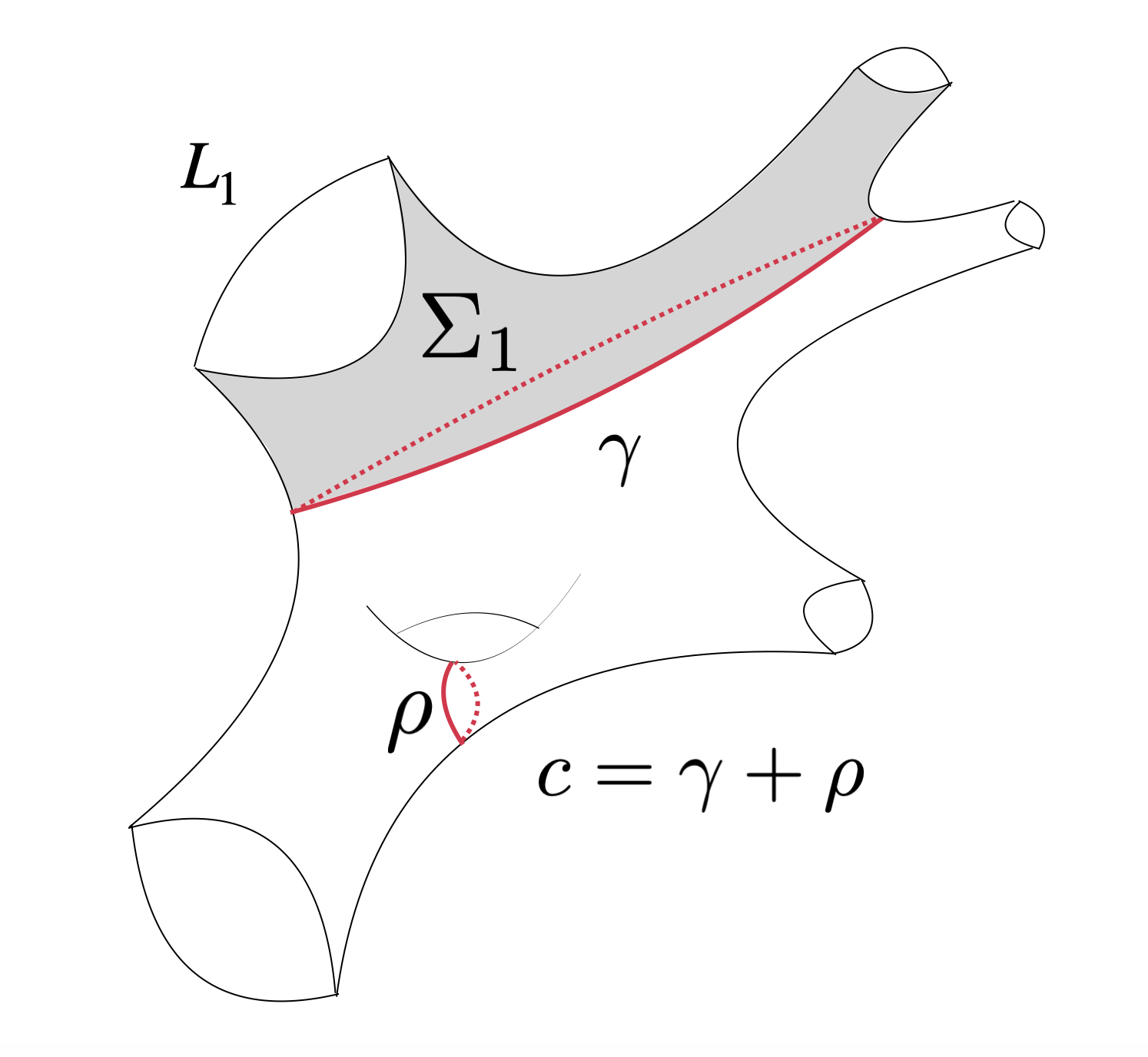}
	\includegraphics[height=1.95in]{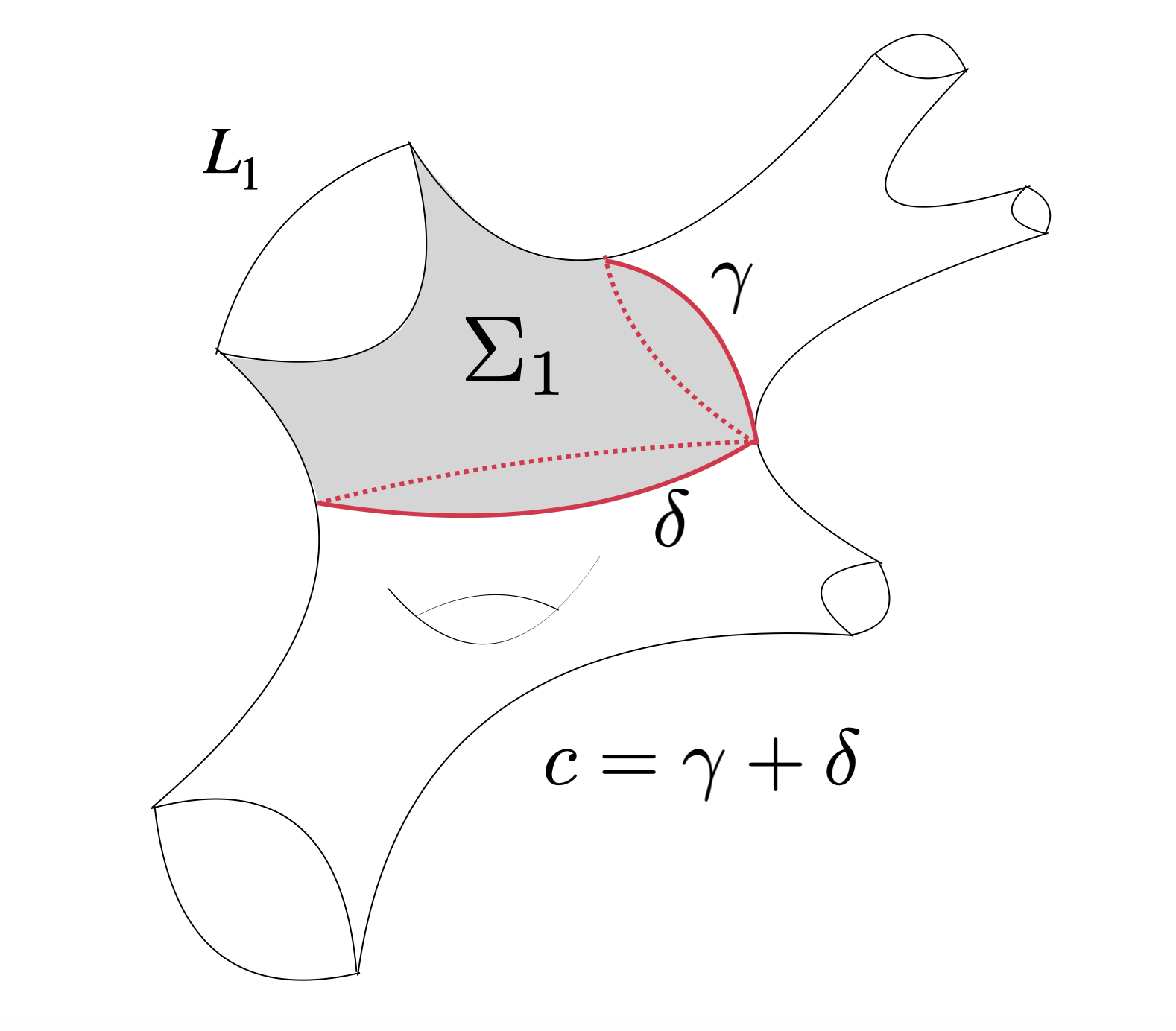}
	\includegraphics[height=1.95in]{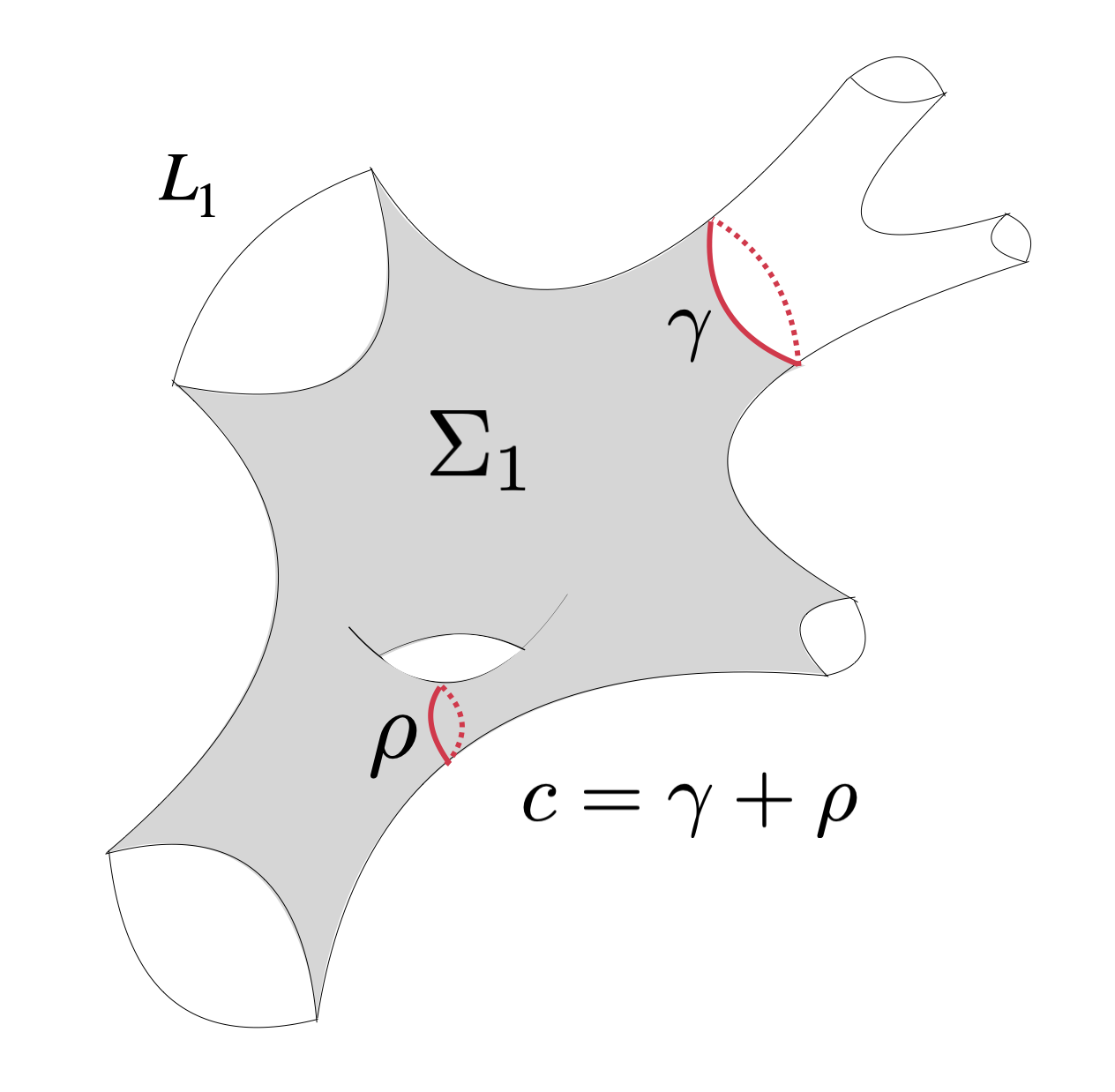}
	\caption{The categorization of multicurves in the order described in the main text. The components of the multicurve $c$ are shown in red.}\label{fig:sys1}
\end{figure}

Let us investigate this in more detail. We can think of the multicurves $c$ as decomposition of the surface $\Sigma$ into a disjoint collection of connected surfaces. Having done this, let us isolate the connected surface $\Sigma_1$ which contains the border $L_1$. Then we have the following categorization of multicurves:
\begin{enumerate}
	\item Those having a component $\gamma$ that bounds a pair of pants with the border $L_1$ and $L_i$. The surface $\Sigma_1 = P_{\gamma i}$ is a pair of pants in this case.
	\item Those having two components $\gamma$ and $\delta$ that together with the border $L_1$ bound a pair of pants.  The surface $\Sigma_1 = P_{\gamma \delta}$ is a pair of pants in this case as well.
	\item Those such that $\Sigma_1$ is not a pair of pants. 
\end{enumerate}
We remind that each multicurve $c$ may contain any number of components on the remaining surface $\Sigma \setminus \Sigma_1$. These distinct situations are shown in figure~\ref{fig:sys1}.

Now we evaluate the sum over $M(\Sigma)$ in~\eqref{eq:3.22} for each of these cases. For the first case we find the contribution to the sum over the multicurves is
\begin{align} \label{first-contribution-indicator}
	\Omega_{-1}(\Sigma)
	&\supset - \sum_{i=2}^n \sum_{\gamma \in \mathcal{P}_{i}} \theta(L-\gamma) \sum_{c\in M(\Sigma \setminus P_{\gamma i})} \prod_{\delta \in \pi_0(c)} (-1) \, \theta(L-\delta) \\
	&=-\sum_{i=2}^n \sum_{\gamma \in \mathcal{P}_{i}} \theta(L - \gamma) \, \Omega_{-1}(\Sigma \setminus P_{\gamma i}) \nonumber \, ,
\end{align}
after singling out the geodesic $\gamma\in \mathcal{P}_{i}$ that bounds $\Sigma_1=P_{\gamma i}$ in the decomposition~\eqref{eq:3.20}. Similarly we find the contribution
\begin{align}  \label{second-contribution-indicator} 
	\Omega_{-1}(\Sigma)  \supset \sum_{\{\gamma,\delta\} \in \mathcal{P}_1} \theta(L- \gamma) \, \theta(L- \delta) \, \Omega_{-1}(\Sigma \setminus P_{\gamma \delta}) \, ,  
\end{align}
for the second case, after singling both $\gamma$ and $\delta$. Note that the surface $\Sigma \setminus P_{\gamma \delta}$ can be disjoint in this case. For those cases we implicitly consider the function $\Omega_{-1}(\Sigma \setminus P_{\gamma \delta})$ to be the product of $\Omega_{-1}$ associated with these two disjoint surfaces.

\begin{figure}[t]
	\centering
	\includegraphics[height=2in]{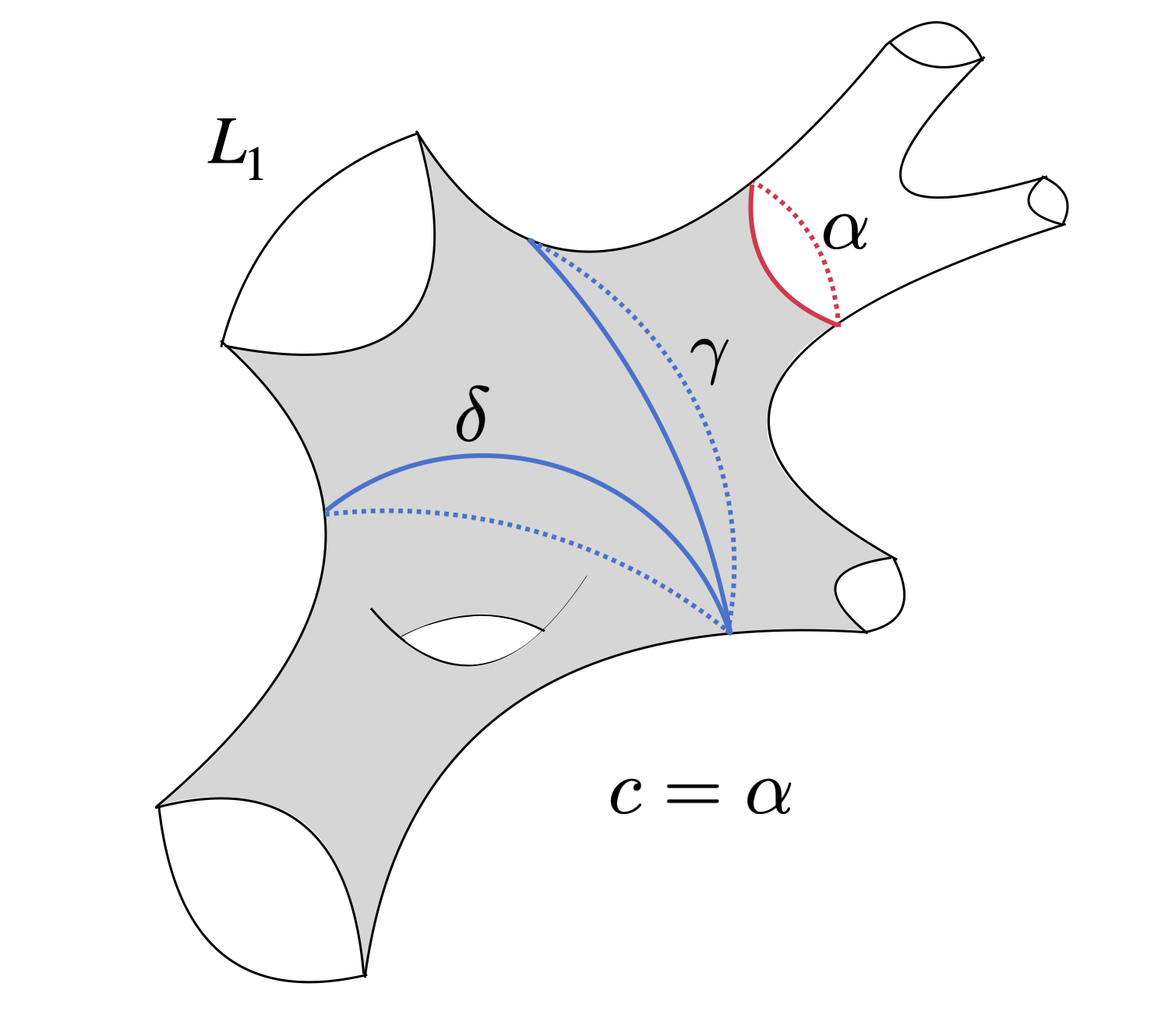}
	\includegraphics[height=2in]{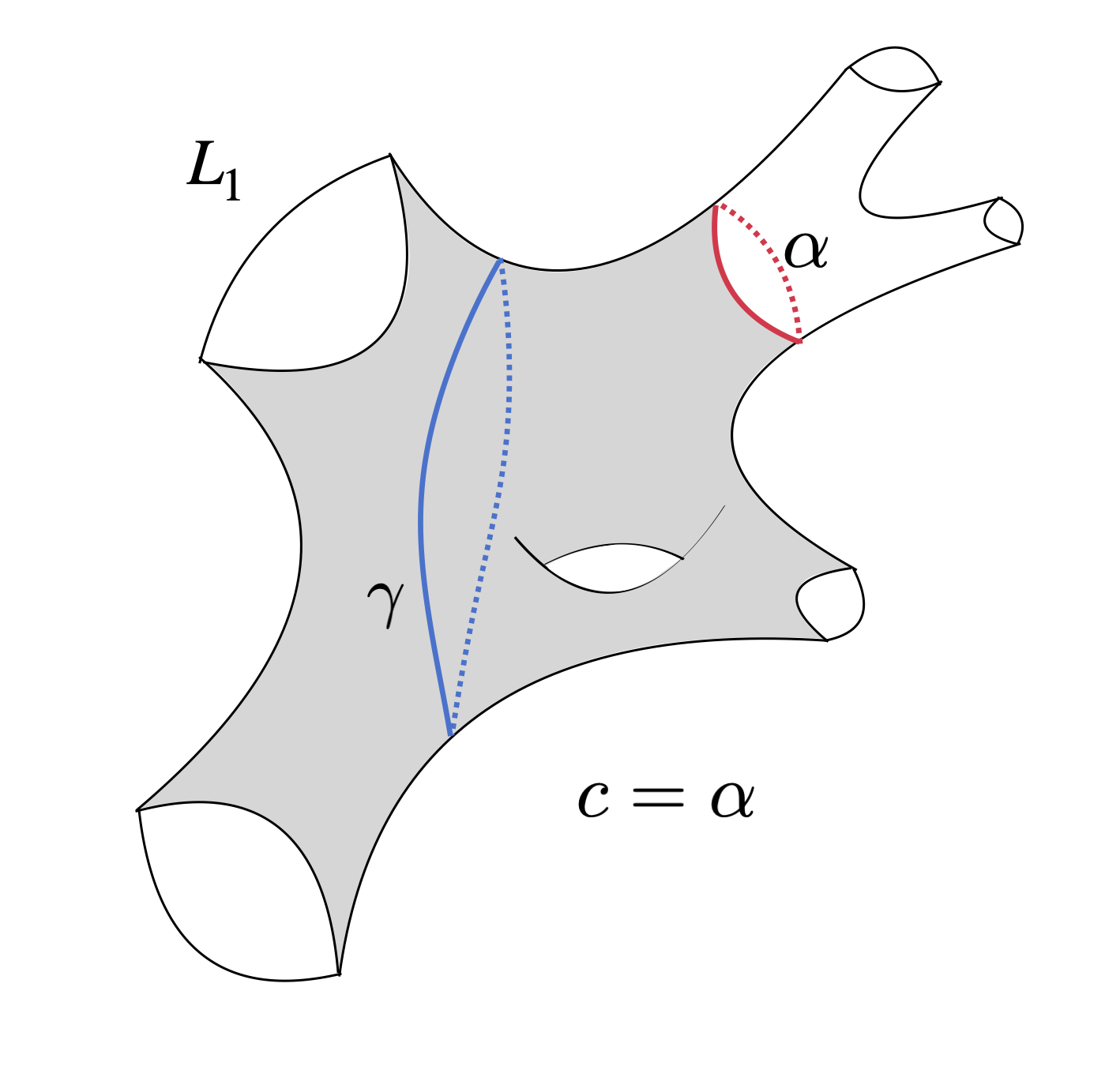}
	\includegraphics[height=2in]{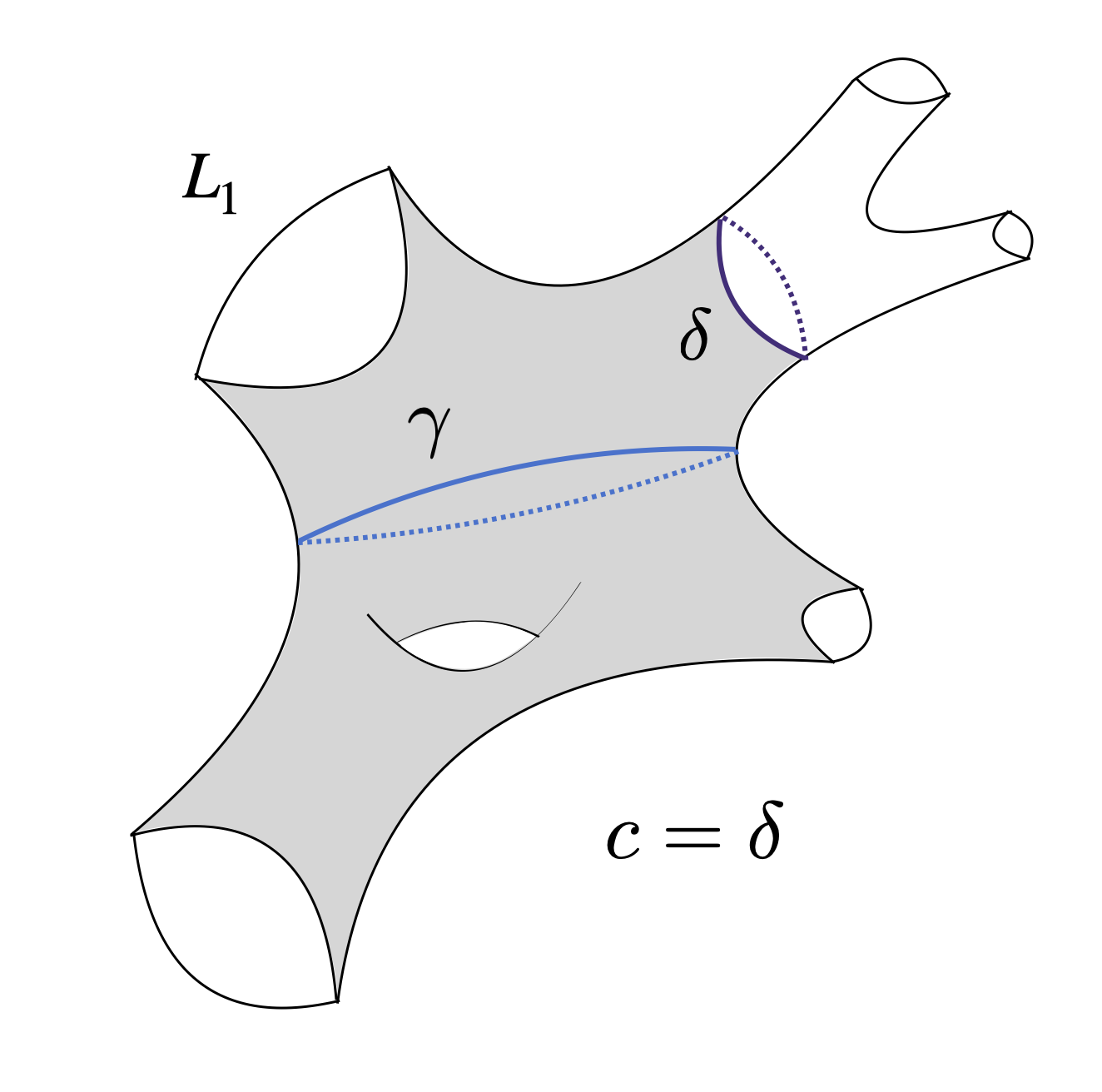}
	\caption{The finer categorization of the third case in the order described in the main text. The components of the multicurve $c$ are shown in red, while the geodesics used in the pants excision are shown in blue. If the geodesic in question is also a part of the multicurve it is shown in purple.}\label{fig:sys2}
\end{figure}

The third case is somewhat more involved. We begin by rewriting their total contribution as
\begin{align} \label{eq:3}
	\Omega_{-1}(\Sigma) \supset \frac{1}{L_1} \sum_{c\in M'(\Sigma)} L_1 \prod_{\gamma \in \pi_0(c)} (-1) \, \theta(L-\gamma) ,
\end{align}
where $M'(\Sigma)$ is the set of primitive multicurves that excludes the components bounding a pair of pants with the border $L_1$ whose contributions have already been considered in~\eqref{first-contribution-indicator} and~\eqref{second-contribution-indicator}. This form allows us to use the Mirzakhani–McShane identity~\eqref{eq:17} for 
\begin{align} \label{eq:4}
	L_1 &=  
	\sum_{\{\gamma,\delta\} \in \mathcal{P}_1^{(1)}} \hspace{-0.12in} D_{L_1 \gamma \delta} +
	\sum_{i=2}^{m_1} \sum_{\gamma \in \mathcal{P}_{i}^{(1)}} R_{L_1 L_i \gamma} + 
	\sum_{i=m_1+1}^m \sum_{\gamma \in \mathcal{P}_{i}^{(1)}} R_{L_1 L_i \gamma} \, ,
\end{align}
in the summand~\eqref{eq:3}. This partition is associated with the surface $\Sigma_1$ determined by $c \in M' (\Sigma)$ and the superscript on the set $\mathcal{P}_i^{(1)}$ is there to remind us that we only consider the relevant pair of pants over the surface $\Sigma_1$, which is assumed to have $m$ borders. We split the contribution of the $R$-terms into two: the first $m_1$ of the borders are assumed to be borders of $\Sigma$, while the rest are the components of the relevant multicurve $c$, see the second and third surfaces in figure~\ref{fig:sys2}.

After this substitution we commute the sums over $\mathcal{P}_i^{(1)}$ in~\eqref{eq:4} with the sum over $M'(\Sigma)$ in ~\eqref{eq:3}. Note that this is allowed since $M'(\Sigma)$ is a finite set. To do this, we point out that there are three independent cases for a simple closed geodesic $\gamma$ to bound a pant with the border $L_1$ in $\Sigma_1$:
\begin{enumerate}
	\item The geodesic $\gamma$ bounds a pant $P_{\gamma  \delta}$ with the border $L_1$ and some other internal simple closed geodesic $\delta$ of $\Sigma_1$. These are not components of the multicurve $c \in M'(\Sigma)$ by design.
	
	\item The geodesic $\gamma$ bounds a pant $P_{\gamma i}$ with the border $L_1$ and $L_i$ of $\Sigma_1$, which is also a border of $\Sigma$. This border is not a component of a multicurve $c \in M'(\Sigma)$ by design.
	
	\item The geodesic $\gamma$ bounds a pant $P_{\gamma \delta}$ with the border $L_1$ and $\delta$ of $\Sigma_1$, which is~\emph{not} a border of $\Sigma$. This border is a component of $c \in M'(\Sigma)$ by design.
\end{enumerate}
These can be associated with the three sums in~\eqref{eq:4}, also refer to figure~\ref{fig:sys2}. Note that the geodesic $\gamma$ above is not a component of a multicurve by the construction. These cases yield the total contributions
\begin{subequations} \label{eq:third}
	\begin{align}
		\Omega_{-1}(\Sigma) &\supset
		{1 \over L_1} \sum_{\{\gamma,\delta\} \in \mathcal{P}_1} D_{L_1 \gamma \delta} \,  \Omega_{-1}(\Sigma \setminus P_{\gamma \delta}) \, , \\
		\Omega_{-1}(\Sigma)  &\supset
		{1 \over L_1} \sum_{i=2}^n \sum_{\gamma\in \mathcal{P}_i} R_{L_1 L_i \gamma}  \,  \Omega_{-1}(\Sigma \setminus P_{\gamma i}) \, , \\
		\Omega_{-1}(\Sigma) &\supset
		- {1 \over L_1} \sum_{\{\gamma, \delta\}\in \mathcal{P}_1} \big[R_{L_1 \gamma \delta} \, \theta(L-\gamma) +
		R_{L_1 \delta \gamma} \, \theta(L-\delta)\big] \Omega_{-1}(\Sigma \setminus P_{\gamma \delta}) \, ,
	\end{align}
\end{subequations}
respectively after summing over the multicurves $c \in M'(\Sigma)$ following the reasoning around~\eqref{first-contribution-indicator}. Notice  the symmetrization over $\gamma$ and $\delta$ in the last equation after the sum is performed. This is due to $\{\gamma, \delta\}\in \mathcal{P}_1$  being an unordered set, while both of the situations where only one of them belongs to $c \in M'(\Sigma)$ contributes.

Combining the exhaustive contributions~\eqref{first-contribution-indicator}~\eqref{second-contribution-indicator} and~\eqref{eq:third} we find the desired partition
\begin{align} \label{eq:3.29}
	L_1 \cdot \Omega_{-1}(\Sigma) = 
	\sum_{i=2}^n \sum_{\gamma \in \mathcal{P}_i} \widetilde{R}_{L_1 L_i \gamma} \, \Omega_{-1}(\Sigma \setminus P_{\gamma i})
	+\sum_{\{\gamma, \delta\}\in \mathcal{P}_1} \widetilde{D}_{L_1\gamma \delta} \, \Omega_{-1}(\Sigma \setminus P_{\gamma \delta}) \, ,
\end{align}
for which we defined~\emph{the twisted Mirzakhani kernels}
\begin{subequations} \label{eq:218}
\begin{align} 
	& \widetilde{R}_{L_1 L_2 L_3} = R_{L_1 L_2 L_3} - L_1 \, \theta(L - L_3) \, , \\
	&\widetilde{D}_{L_1 L_2 L_3}  = D_{L_1 L_2 L_3}  - R_{L_1 L_2 L_3 } \, \theta(L - L_2 )  
	- R_{L_1 L_3 L_2 }  \, \theta(L - L_3 )  + L_1 \,  \theta(L - L_2 ) \,  \theta(L - L_3 )  
	 \, .
\end{align}
\end{subequations}
We highlight that the symmetries in~\eqref{eq:27} are also satisfied and taking $L\to 0$ reduces them to the Mirzakhani kernels. Similarly the partition~\eqref{eq:3.29} reduces to~\eqref{eq:17} in this limit.

Upon inserting the partition~\eqref{eq:3.29} to~\eqref{eq:3.23} and exactly repeating Mirzakhani's analysis in~\cite{mirzakhani2007simple} we conclude for $2g -2 + n > 1$ and $n \geq 1$ the systolic volumes satisfy a recursion relation among themselves
\begin{align}  \label{eq:218A}
	L_1 \cdot V \mathcal{V}^L_{g,n} (L_i) &= 
	\sum_{i=2}^n \int\limits_{0}^\infty \ell d \ell \, \widetilde{R}_{L_1L_i\ell} \, V \mathcal{V}^L_{g,n-1}\left(\ell, \mathbf{L} \setminus \{L_i \} \right) 
	\\
	&\hspace{-0.5in} + {1 \over 2} \int\limits_{0}^\infty \ell_1 d \ell_1 \, \int\limits_{0}^\infty \ell_2 d \ell_2 \,
	\widetilde{D}_{L_1 \ell_1 \ell_2} \bigg[
	V \mathcal{V}^L_{g-1, n+1} (\ell_1, \ell_2, \mathbf{L}) 
	+ \sum_{\text{stable}} V \mathcal{V}^L_{g_1, n_1} (\ell_1,  \mathbf{L}_1) \cdot V \mathcal{V}^L_{g_2, n_2} (\ell_2,  \mathbf{L}_2)
	\bigg] \, , \nonumber
\end{align}
whenever $L \leq L_\ast$. Like before, we take $V \mathcal{V}^L_{0,3} (L_i) = V \mathcal{M}_{0,3} (L_i) = 1$ to set the units. We are going to discuss the case $(g,n)=(1,1)$ shortly.

The first few volumes $V \mathcal{V}^L_{g,n} (L_i)$ are listed in table~\ref{tab:V}. The results of this systolic recursion can be shown to be independent of the choice of the border $L_1$ following the arguments for the counterpart statement for~\eqref{eq:110}~\cite{mirzakhani2007simple}. The dilaton equation~\eqref{eq:3.18} applies to the systolic volumes as well and can be used to find the systolic volumes for the surfaces without borders. We remark in passing that this particular twisting procedure has been used in~\cite{andersen2017geometric} to establish that hyperbolic vertices~\eqref{eq:hyp} satisfy the geometric master equation~\eqref{eq:2.2}, providing an alternative to the argument of~\cite{Costello:2019fuh}.
\begin{table}[t]
	\begin{center}
		\bgroup
		\def\arraystretch{1.75}
		\Large
		\begin{tabular}{ |c | c | }
			\hline
			$(g,n)$& $V \mathcal{V}^L_{g,n} (L_i) $  \\ 
			\hline
			$(0,3)$ & 1  \\ 
			\hline
			$(0,4)$ & $ V \mathcal{M}_{0,4} (L_i) - {3 \over 2}  L^2  $  \\ 
			\hline
			$(1,1)$ & $ V \mathcal{M}_{1,1} (L_1) - {1 \over 4} L^2 $  \\
			\hline
			$(0,5)$ & $V \mathcal{M}_{0,5} (L_i)  -{3 \over 2} L^2 \sum_{i=1}^5 L_i^2 + {5 \over 2} L^4 -10 \pi^2 L^2$ \\ 
			\hline
			$(1,2)$ & $ V \mathcal{M}_{1,2} (L_i) - {1 \over 8} L^2 \sum_{i=1}^2 L_i^2  + {23 \over 192} L^4 - {13 \pi^2 \over 24} L^2 $ \\
			\hline
		\end{tabular}
		\normalfont
		\egroup
	\end{center}
	\caption{\label{tab:V}The WP volumes $V \mathcal{V}^L_{g,n} (L_i) $, see appendix~\ref{app:B}. Even though the border $L_1$ is singled out in the recursion, the resulting volumes are symmetric under exchanging the borders. We also see the dilaton equation~\eqref{eq:3.18} holds for them.}
\end{table}

Take note that using step functions $\theta(L-L_i)$ wasn't essential for the twisting procedure itself. For instance, we could have formed a recursion similar to~\eqref{eq:218A} for the integrals of the functions~\cite{andersen2017geometric} 
\begin{align}
	\zeta (\Sigma) = \sum_{c \in M(\Sigma) } \prod_{\gamma \in \pi_0(c)} \, f(\gamma) \, ,
\end{align} 
provided $f$ is a sufficiently well-behaved real integral function. In the twisted Mirzakhani kernels~\eqref{eq:218} this simply amounts to replacing $-\theta(L-\gamma)$ with $f(\gamma)$. In particular we can use a suitable family of functions $f(\gamma)$ that limits to $-\theta(L-\gamma)$, which may be particularly helpful to evaluate~\eqref{eq:218A} efficiently.

In order to get an intuition for the systolic volumes it is instructive to evaluate~\eqref{eq:218A} for the first nontrivial case, $(g,n)=(0,4)$,
\begin{align} \label{eq:321}
	V \mathcal{V}^L_{0,4} (L_i) &= 
	{1 \over L_1} \sum_{i=2}^4 \int\limits_{0}^\infty \ell d \ell \, \widetilde{R}_{L_1L_i\ell}  
	= V \mathcal{M}_{0,4} (L_i) - \sum_{i=2}^4 \int\limits_{0}^L  \ell d \ell  \\
	&
	= V \mathcal{M}_{0,4} (L_i) - {3 \over 2} { L^2} 
	= 2 \pi^2 + {1 \over 2} \sum_{i=1}^4 L_i^2 - {3 \over 2} { L^2 }  \, . \nonumber
\end{align}
Note that for all $L_i \geq 0$
\begin{align} \label{eq:3.34}
	V \mathcal{V}^L_{0,4} (L_i)  \geq 2 \pi^2 - {3 \over 2}  L^2 > 0\, \quad \quad
	\text{when} \quad \quad 0 \leq L \leq  L_\ast = 2 \sinh^{-1} 1 \, .
\end{align}
which shows this systolic volume is always positive in the allowed regime, as it should be. 

Let us illustrate another way to perform the same computation. Recall that we exclude surfaces whose systoles are smaller than $L$ from the systolic subset $\mathcal{V}^L_{0,4} (L_i)$. This means we need to subtract the WP volume associated with the region 
\begin{align} \label{eq:reggion}
	0 <  \ell \leq L \leq L_\ast \, , \quad \quad 0 \leq \tau < \ell \, ,
\end{align}
from the WP volume $V \mathcal{M}_{0,4} (L_i)$ as part of evaluating $\mathcal{V}^L_{0,4} (L_i)$.  Here $(\ell, \tau)$ are the Fenchel-Nielsen coordinates determined by a particular way to split a four-bordered sphere into two pairs of pants with an internal simple closed geodesic $\ell$.

The volume of this excluded region is $L^2/2$ since the MCG acts freely there, i.e., it always maps surfaces outside of~\eqref{eq:reggion}---there isn't any multiple counting of surfaces. Recall that the effect of MCG is to exchange geodesics on the surface with possible twists. The restriction on the twist $\tau$ has already eliminated the possibility of the Dehn twists. Further, two short geodesics never intersect on the surface by the collar lemma~\eqref{eq:Uncertainty}, so short geodesics always get exchanged with the longer ones, i.e., the MCG takes us outside of the region~\eqref{eq:reggion}. For the additional multiplication by $3$, notice the decomposition~\eqref{eq:reggion} is just one way to split a four-bordered sphere: there are two other distinct channels to perform a similar decomposition. So one should subtract $3L^2/2$ from the total volume to obtain~\eqref{eq:321}. We emphasize again that these regions can't intersect by the collar lemma and there was no oversubtraction in our results.

\begin{figure}[t]
	\centering
	\includegraphics[height=1.8in]{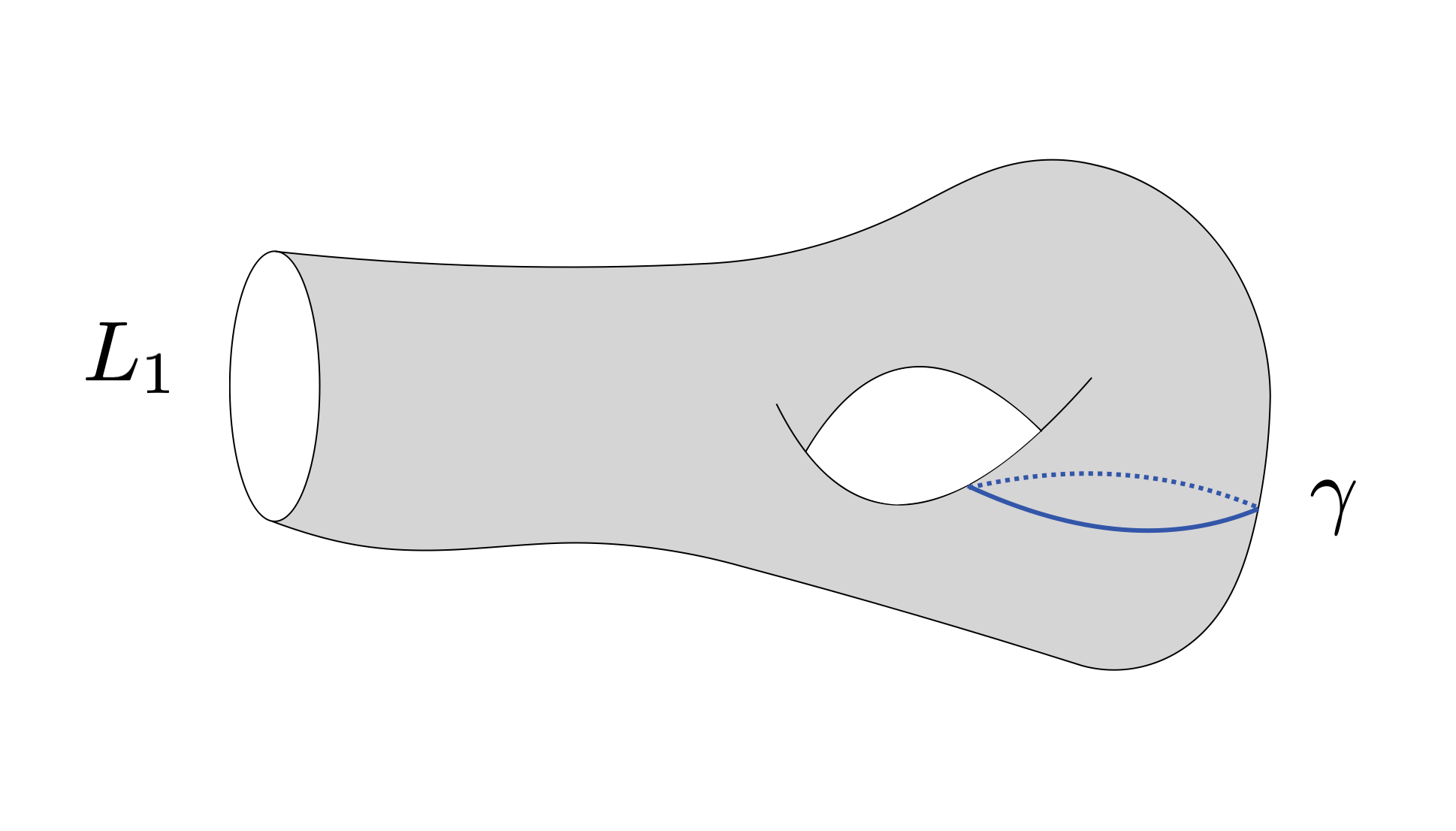}
	\includegraphics[height=1.8in]{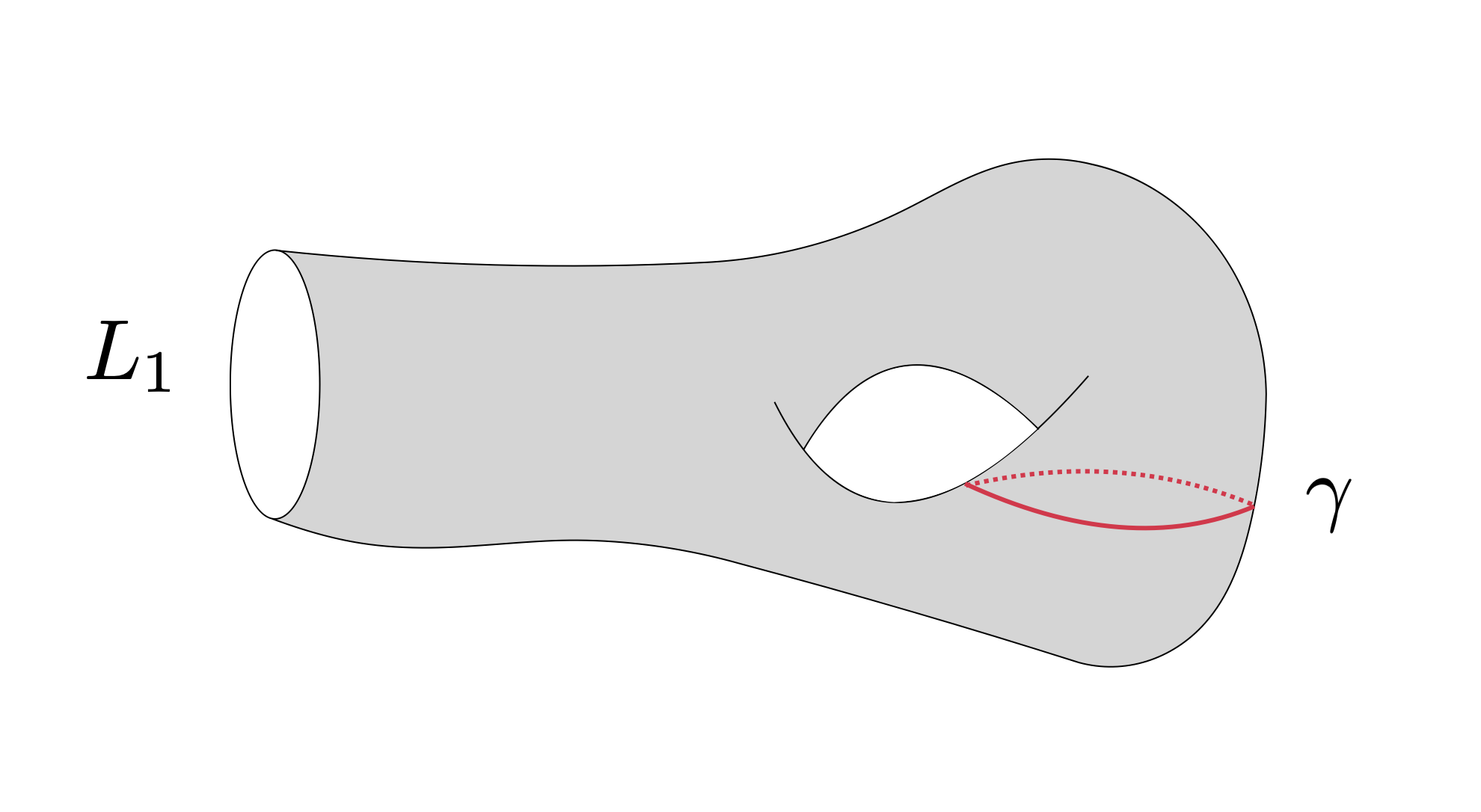}
	\caption{The one-bordered torus relevant for the partition~\eqref{eq:3.35} in the order described in the main text. The color conventions from previous figures apply.}\label{fig:tori}
\end{figure} 
Now we turn our attention to $(g,n)=(1,1)$ required to run~\eqref{eq:218A}. This case requires special attention like~\eqref{eq:3.17}. The partition of $ \Omega_{-1} $ for a one-bordered torus $\Sigma$ is
\begin{align} \label{eq:3.35}
	L_1 \cdot \Omega_{-1}(\Sigma) = \sum_{\gamma \in \mathcal{P}_1} 
	\left[ D_{L \gamma \gamma} - L_1 \, \theta(L-\gamma) \right]
	\, \Omega_{-1}(\Sigma \setminus P_{\gamma \gamma}) \, .
\end{align}
This can be argued similarly to~\eqref{eq:3.29} after observing there is only one geodesic $\gamma$ that has been cut, which is either part of a primitive multicurve (the second term) or isn't (the first term), refer to figure~\ref{fig:tori}. Using this partition we then find
\begin{align} \label{eq:322}
	V \mathcal{V}^L_{1,1} (L_1) &= {1 \over 2} {1 \over L_1} \, \int\limits_{0}^\infty \ell d \ell \,  
	\left[ D_{L\ell\ell} - L_1 \theta(L- \ell) \right]
	\cdot V \mathcal{V}^L_{0,3} \left(L, \ell, \ell \right) 
	= V \mathcal{M}_{1,1}(L_1) - {1 \over 4} L^2 
	\, .
\end{align}
This result can be alternatively argued from the geometric perspective similar to~$V \mathcal{V}^L_{0,4} (L_i)$. We also have
\begin{align} \label{eq:3.39a}
	V \mathcal{V}^L_{1,1} (L_1) \geq {\pi^2 \over 12} -{1 \over 4} L^2 > 0  \, 
	\quad \quad \text{when} \quad \quad
	0 \leq L \leq L_\ast = 2 \sinh^{-1} 1 \, ,
\end{align}
for all $L_1 \geq 0$ and this volume is positive in the allowed range.

For the higher-order cases the systolic volumes are somewhat more complicated. We subtract the volumes of the regions with short geodesics from the total volume of the moduli space, however the MCG action is not necessarily free in these subtracted regions and the evaluations of their volumes requires suitable weighting of the integrand by Mirzakhani kernels as a result. The computations for some sample systolic volumes are given in appendix~\ref{app:B}.

\begin{figure}[t]
	\centering
	\hspace{-0.3in}
	\includegraphics[width=0.50\textwidth]{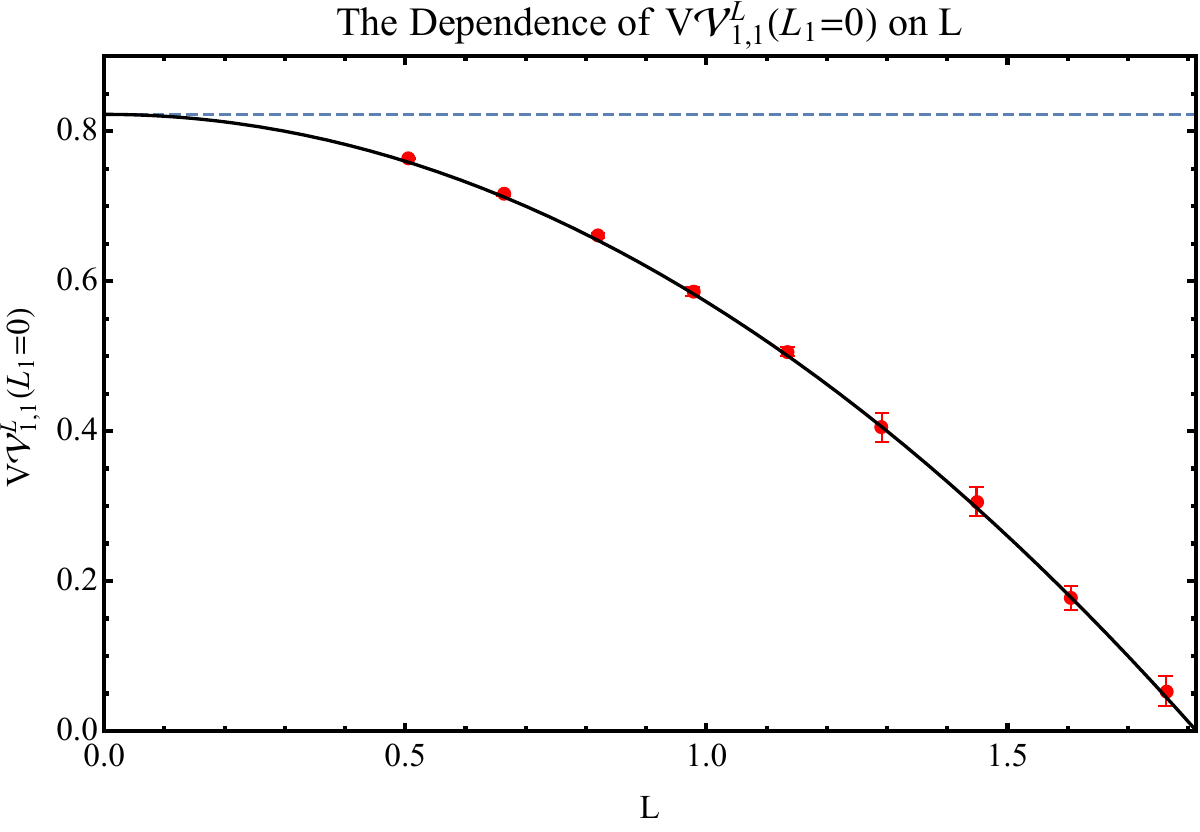}
	\includegraphics[width=0.50\textwidth]{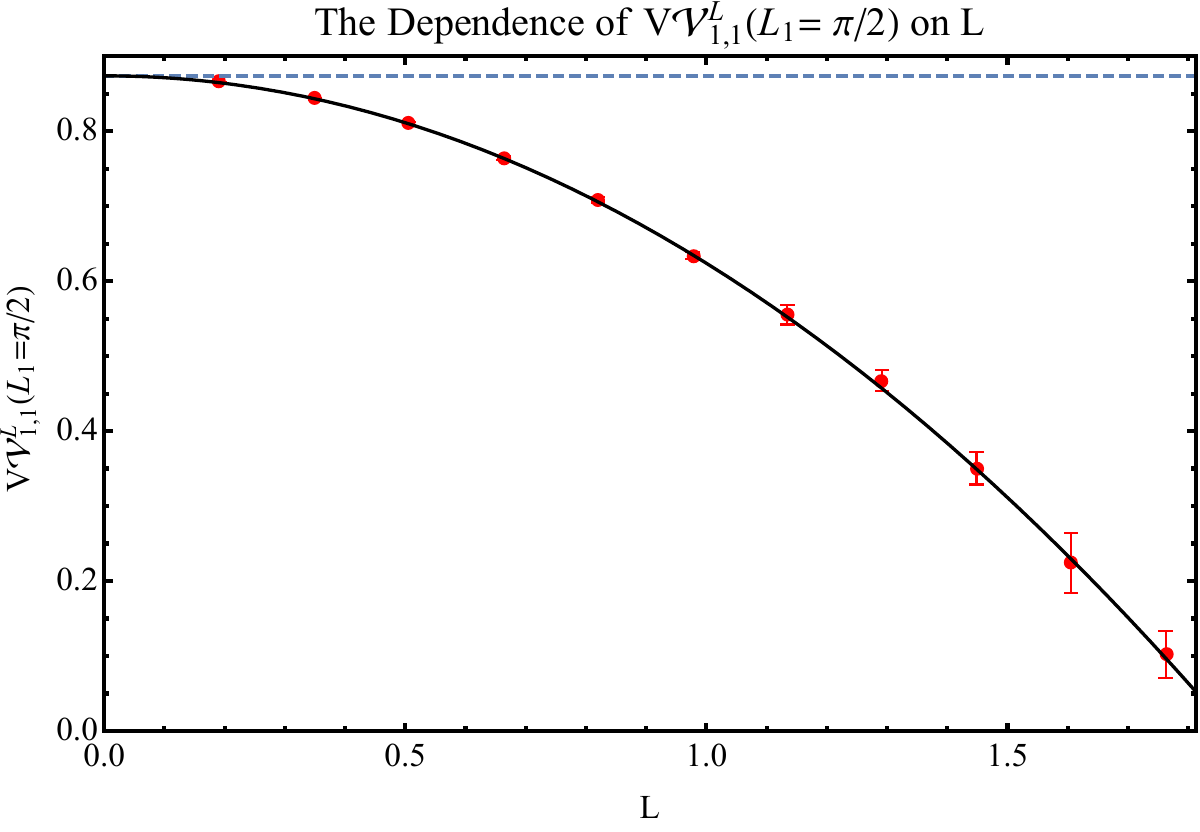}
	\begin{tabular}{ |c|c|c|c|c|c|c| }  
		\hline
		Border Length  & $c_1$ & $c_2$  & $c_1  \text{  (true) }$ & $c_2 \text{  (true) }$ \\ 
		\hline
		$0$  & $0.8275 \pm 0.0004$ & $-0.2502 \pm 0.0012$ & $\pi^2/12 \approx 0.8225$ & $-0.25$ \\ 
		\hline
		$\pi/4$ & $0.87541 \pm 0.00004$ & $-0.25172 \pm 0.00095$ &$17 \pi^2/192 \approx 0.87387$ & $-0.25$  \\ 
		\hline
	\end{tabular}
	\caption{\label{fig:vol}The systolic volumes $V \mathcal{V}^L_{1,1}(L_1)$ as a function of  $L$ when $L_1 = 0$~\textit{(left)} and $L_1 = \pi/2$~\textit{(right)}. The red points are obtained by integrating the WP metric of~\cite{Firat:2023suh} and the black curve is the function~\eqref{eq:322}. The blue dashed line is $V \mathcal{M}_{1,1}(L_1)$. Refer to appendix~\ref{sec:A} for details.} 
\end{figure}

We can further crosscheck the results for $V \mathcal{V}_{1,1}(L_1)$ and  $V \mathcal{V}_{0,4}(L_i)$ by numerically integrating the WP metric derived using the Polyakov conjecture~\cite{zograf1987liouville, zograf1988uniformization, Artemev:2023bqj,Hadasz:2003kp, Piatek:2013ifa, Firat:2023glo, Firat:2023suh}. The dependence of $V \mathcal{V}^L_{1,1}(L_1)$ on $L$ for $L_1 = 0$ and $L_1 = \pi/2$, along with their fits to the quadratic polynomial
\begin{align}
	f(L) = c_1 + c_2 \, L \, ,
\end{align}
are shown in figure~\ref{fig:vol} for instance. It is apparent that we obtain consistent results. For the details of this computation refer to appendix~\ref{sec:A}.

As a final remark in this section, we highlight the systolic recursion~\eqref{eq:218A} simplifies upon restricting to genus 0 surfaces. Like in Mirzakhani's recursion, we don't need to consider the first term in the second line of~\eqref{eq:218A}. This means for $n > 3$
\begin{align}  \label{eq:3.25}
	L_1 \cdot V \mathcal{V}^L_{0,n} (L_i) &= 
	\sum_{i=2}^n \int\limits_{0}^\infty \ell d \ell \, \widetilde{R}_{L_1L_i\ell} \, V \mathcal{V}^L_{0,n-1}\left(\ell, \mathbf{L} \setminus \{L_i \} \right) 
	\\
	&\hspace{0.5in} + {1 \over 2} \int\limits_{0}^\infty \ell_1 d \ell_1 \, \int\limits_{0}^\infty \ell_2 d \ell_2 \,
	\widetilde{D}_{L_1 \ell_1 \ell_2} 
	\sum_{\text{stable}} V \mathcal{V}^L_{0, n_1} (\ell_1,  \mathbf{L}_1) \cdot V \mathcal{V}^L_{0, n_2} (\ell_2,  \mathbf{L}_2)
	\, . \nonumber
\end{align}
This again corresponds to considering only the first and second types of pants excisions in figure~\ref{Mirz-gluing-figure}. 

Remember it is possible to take $0 < L \leq \infty$ for string vertices in the classical CSFT~\cite{Costello:2019fuh,Firat:2023glo}. It is then natural to ask whether the condition on the threshold length, $L \leq L_\ast$, can also be eliminated for~\eqref{eq:3.25}. However, this doesn't follow from our construction. Clearly we can't take $0 < L \leq \infty$ for the general border lengths: taking $L$ too large while keeping $L_i$ small would eventually make systolic volumes negative, see table~\ref{tab:V}. So the bound for them should persist in general. Although we can argue it can be relaxed to
\begin{align} \label{eq:3.39}
	0 < L \leq 2 L_\ast = 4 \sinh^{-1}1 \approx 3.53 \, ,
\end{align}
since two intersecting closed geodesics always traverse each other at least twice. We point out~\eqref{eq:3.34} remains positive with this weaker bound, but~\eqref{eq:3.39a} doesn't apply anymore. It is still in the realm of possibilities~\eqref{eq:3.25} to hold when the border lengths $L_i$ scale with $L$ while taking some of them the same however. We comment on this possibility more in the discussion section~\ref{sec:disc}. 

\section{The recursion for hyperbolic string vertices} \label{sec:HSV}

We now turn our attention to deriving the recursion relation satisfied by the elementary vertices of hyperbolic CSFT. We remind that they are given by integrating the string measure over the systolic subsets\footnote{Strictly speaking, only the amplitudes $\langle \mathcal{V}_{g,n}(L_i) | $ with $L_i = L$ for $0<L\leq L_\ast$ form string vertices. However we also denote the cases with generic $L_i$ and $L$ as string vertices and/or elementary interactions for brevity.}
\begin{align} \label{eq:4.1}
	\langle \mathcal{V}_{g,n}(L_i) | =
	\int\limits_{\mathcal{V}^L_{g,n} (L_i) } \langle \Omega_{g,n} (L_i)  | 
	=  \int\limits_{\mathcal{M}_{g,n} (L_i)} \mathbf{1}_{\mathcal{V}_{g,n}^L(L_i) } \,
	  \cdot \langle \Omega_{g,n} (L_i)  |  \  
\end{align}
and the resulting recursion would be among $\langle \mathcal{V}_{g,n} (L_i)  | \in (\widehat{\mathcal{H}}^\ast)^{\otimes n}$.  It is useful to introduce
\begin{align} \label{eq:4.2}
	\langle \omega_{g,n} (L_i) | = \mathbf{1}_{\mathcal{V}_{g,n}^L(L_i) } \,
	\cdot \langle \Omega_{g,n} (L_i)  | \, ,
\end{align}
in this section. The indicator function $\mathbf{1}_{\mathcal{V}_{g,n}^L(L_i) }$ is already given in~\eqref{eq;3.20}.

The integrations~\eqref{eq:4.1} above take place in the moduli space of Riemann surfaces with geodesic borders of length $L_i$, $\mathcal{M}_{g,n} (L_i)$.  In the context of CSFT, on the other hand, it is often considered that such integrations take place in the sections of moduli space of punctured Riemann surfaces endowed with suitable local coordinates around the punctures $\widehat{\mathcal{P}}_{g,n}$.  In hyperbolic CSFT, however, they are equivalent since both the string measure (and the local coordinate data it contains) and the regions of integrations are defined by the bordered Riemann surfaces through grafting. More precisely, $\text{gr}_\infty' \mathcal{M}_{g,n} (L_i)$ is a section of $\widehat{\mathcal{P}}_{g,n}$ by $\text{gr}_\infty' $ being a homeomorphism~\cite{mondello2011riemann}. We are performing the integration over this section of $\widehat{\mathcal{P}}_{g,n}$.

\subsection{$b$-ghost insertions for the Fenchel-Nielsen deformations} \label{sec:4.1}

In our arguments we are going to need to excise pairs of pants from the surface states. For this we ought to understand the Schiffer vectors corresponding to the Fenchel-Nielsen deformations. So consider a Riemann surface $\Sigma_{g,n}(L_i)$ together with its pants decomposition. The lengths $\ell_i$ and the twists $\tau_i$ of the seams of the pants define local coordinates over the moduli space~\eqref{eq:FN}.  Let us denote the Schiffer vectors corresponding to the coordinate vectors $\p/ \p \tau_i$ and $\p/ \p \ell_i$ by $u_i$ and $v_i$ respectively. In order to derive their expression, and subsequently the associated $b$-ghost insertions, we need to find the transition functions between two coordinate patches for pants after twisting and increasing the length of the seams. Similar analysis has been considered with different levels of detail in~\cite{Moosavian:2017qsp,Ishibashi:2022qcz}, however we are going to be more explicit.

\begin{figure}[t]
	\centering
	\includegraphics[height=3.75in,trim={2cm 0.5cm 3.25cm 2.5cm}, clip]{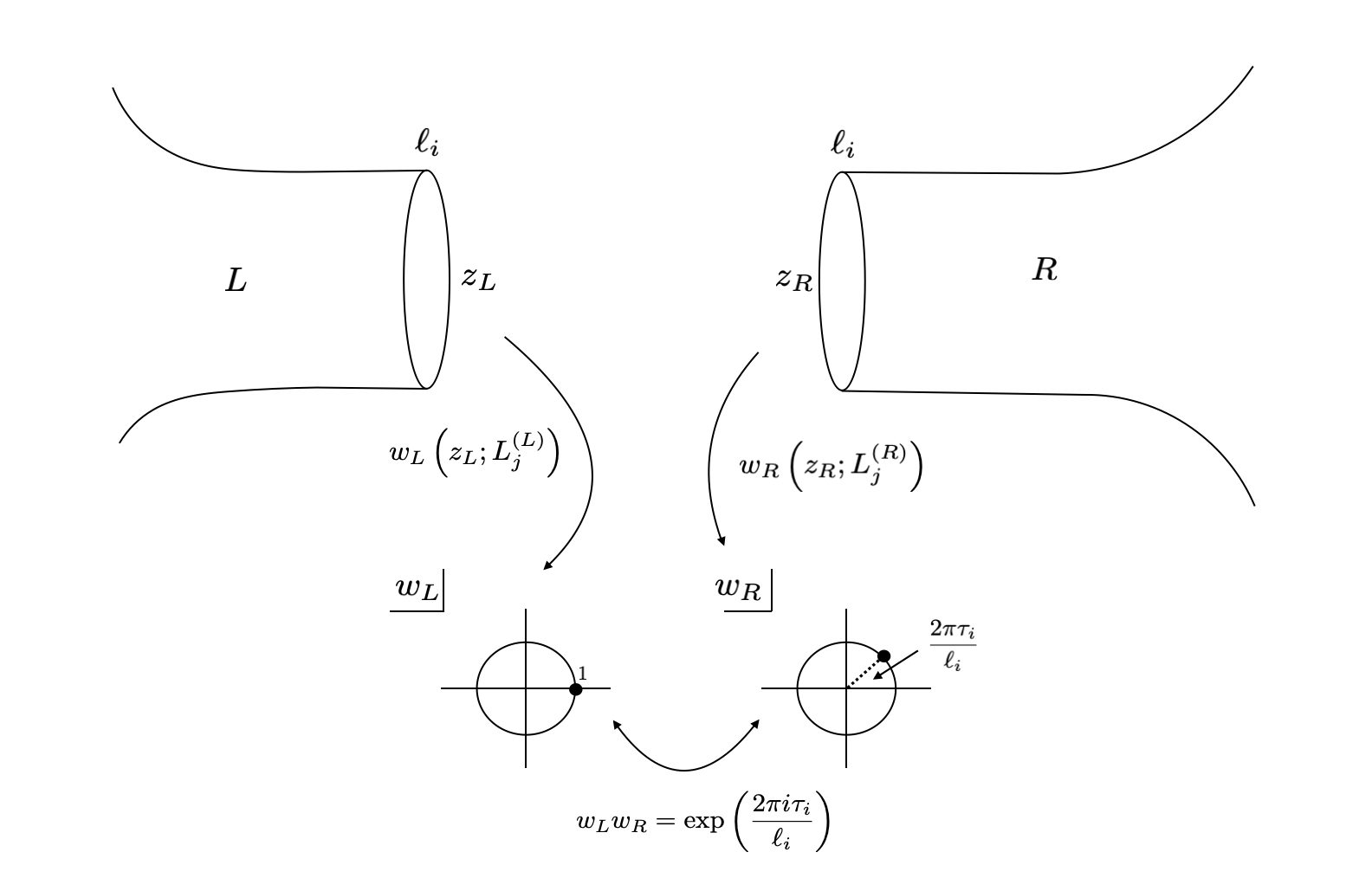}
	\caption{The geometry of the identification~\eqref{eq:4.3}.}\label{fig:coord-trans}
\end{figure}

Let $z_{L}, z_{R}$ be the coordinate patches on the left and right pants $L,R$ for the semi-infinite flat cylinder grafted $i$-th seam whose length is $\ell_i$. They are related to each other by (observe figure~\ref{fig:coord-trans})
\begin{align} \label{eq:4.3}
	w_L\left(z_{L}; L^{(L)}_j\right) \, &w_R\left(z_{R}; L^{(R)}_j\right) = e^{2 \pi i\tau_i  / \ell_i} 
	\\ \nonumber 
	&\implies z_{L} = w_L^{-1} \left( { e^{2 \pi i\tau_i  / \ell_i}  \over w_R\left(z_{R} ; L^{(R)}_j\right)} ; L^{(L)}_j
	\right) = f_{LR} (z_{R} ; L^{(L)}_j, L^{(R)}_j)
	\, ,
\end{align}
where $w_L, w_R$ are the local coordinates for the generalized hyperbolic three-string vertex around the puncture at $z=0$, whose expression is given in appendix~\ref{app:H3V}. The formula above is just a consequence of performing the sewing fixture~\eqref{eq:TS} for the $i$-th seam. We importantly highlight that the definition of coordinates depends on the lengths of the borders of the left ($L^{(L)}_j$) and right ($L^{(R)}_j$) pairs of pants with $j=1,2,3$. However it doesn't depend on the relative twist $\tau_i$ between them. We take $L^{(L)}_1 = L^{(R)}_1 = \ell_i$.\footnote{If the surface is a one-bordered torus, we take $z = z_L = z_R$ and $w_R= w_R(z)$ to be the local coordinate around $z=1$ instead. We also specify the border lengths by $L^{(L)}_j = L^{(R)}_j= (\ell_i, \ell_i, L_3)$.} 

The aim is finding the Schiffer vectors $v_i$ and $u_i$ and their associated $b$-ghost insertions based on~\eqref{eq:4.3} now. Begin by investigating the twist deformation $\tau_i \to \tau_i + \delta \tau_i$ while keeping all other coordinates the same. Using~\eqref{eq:2.15} we simply find
\begin{align}
	u_i(w_R) = - {2 \pi i \over \ell_i } w_R \, .
\end{align}
The corresponding $b$-ghost insertion according to~\eqref{eq:2.29} is then given by
\begin{align} \label{eq:4.5}
	b(u_i) = - {2 \pi i \over \ell_i} \left[
	\oint dw_R \, w_R \, b(w_R) - \oint  d\overline{w}_R \, \overline{w}_R \, \overline{b}(\overline{w}_R)
	\right] = - {2 \pi i \over \ell_i} \, b_0^- \, ,
\end{align}
applied to the states inserted in the local coordinates $w_R$ around the puncture $z_R=0$. It has the same expression in $w_L$ by $L \leftrightarrow R$ symmetry. This is a well-known result from the ordinary factorization analysis~\cite{Erbin:2021smf}.

Now we return our attention to the variation $\ell_i \to \ell_i + \delta \ell_i$. This is slightly more involved compared to the twist deformations since the definition of the local coordinates themselves depends on the pair of pants. Nonetheless we can express the associated Schiffer vector $v_i$ as a sum of the following vectors using~\eqref{eq:2.15}
\begin{align} \label{eq:4.6}
	v_i =
	v^{(L)}_i(w_{L}) + v^{(R)}_i(w_{R})+ {2 \pi i \tau_i \over  \ell_i^2} \, w_R \, ,
\end{align}
where
\begin{align}
	v^{(L)}_i (w_{L}) = - {\p w_L \left(z_{L} \left(w_L \right)  ; L^{(L)}_j \right)  \over \p \ell_i} \, .
\end{align}
This holds similarly for $L \leftrightarrow R$ and their anti-holomorphic counterparts.

We can understand the reasoning behind $v_i$~\eqref{eq:4.6} as follows. When we vary the length of the seam $\ell_i$, the local coordinate $w_L$ ($w_R$), as a function of $z_{L}$ ($z_{R}$) changes, see~\eqref{eq:A.3}. Keeping the coordinates $w_L$, $w_R$ and their identification~\eqref{eq:4.3} fixed, the vector $v^{(L)}_i$ ($v^{(R)}_i$) is then simply one part of the Schiffer vector $v_i$. Clearly, the parts associated with $L$ and $R$ add up since we consider infinitesimal changes. Finally we have another part associated with changing $\ell_i$ in the identification~\eqref{eq:4.3}. However, this produces a $b_0^-$ insertion like in~\eqref{eq:4.5}, which would vanish upon multiplying the twist's insertion $b(u_i) $~\eqref{eq:4.5}. These together produce the Schiffer vector $v_i$.

The nonvanishing part of the $b$-ghost insertion for $\ell_i \to \ell_i + \delta \ell_i$ is then given by
\begin{align} \label{eq:4.8}
	b(v_i) = b\left(v^{(L)}_i\right) + b\left(v^{(R)}_i\right) \, ,
\end{align}
where
\begin{align} \label{eq:4.9}
	b(v^{(L)}_i) 
	&= - \oint\limits d w_L \,  {\p w_L \left(z_L (w_L ) ; L^{(L)}_j\right)  \over \p \ell_i} \, b(w_L)  
	 - \oint\limits d \overline{w}_L \,  {\p \overline{w}_L \left(\overline{z}_L (\overline{w}_L ) ; L^{(L)}_j\right)  \over \p \ell_i} \, \overline{b} (\overline{w}_L) 
	 \\ \nonumber
	 &= {1 \over 2 \pi} \left[
	 {1 \over \rho_1} {\partial \rho_1 \over \partial \lambda_1} \left( b_0 + \overline{b}_0 \right)
	 + {\lambda_1 \left( \lambda_2^2 - \lambda_3^2\right) \over (1+ \lambda_1^2)^2} \rho_1 
	 \left( b_1 + \overline{b}_1 \right)
	 + \cdots
	 \right] \, ,
\end{align}
for which we take
\begin{align}
	2 \pi \lambda_j  = L_j^{(L)} \, , \quad \quad
	2 \pi \lambda_1 = \ell_i \, ,
\end{align}
to simplify the presentation. Here $\rho_1 = \rho_1 (\lambda_j)$ is the mapping radius associated with the local coordinate around the puncture $z_L =0$, see~\eqref{eq:R}. This $b$-ghost insertion~\eqref{eq:4.9} acts on the left pair of pants but  similar considerations apply after exchanging $L \leftrightarrow R$ for the right one. 

The same $b$-ghost insertion applies to changing the border length of any pair of pants. This has already been presented in~\eqref{eq:1.7}, which we report here again for $L_i > 0$
\begin{align} \label{eq:1.7a}
	&\langle \Sigma_{0,3}( L_1, L_2 , L_3 )| \left(   \mathfrak{B} \otimes \id \otimes \id \right)
	\\
	&\hspace{0.5in}= 
	\langle \Sigma_{0,3}( L_1, L_2 , L_3 )|
	\left[ {1 \over 2 \pi}  \left(
	{1 \over \rho_1} {\partial \rho_1 \over \partial \lambda_1} \, \left( b_0 + \overline{b}_0 \right)
	+{\lambda_1  \, (\lambda_2^2  - \lambda_3^2) \over (1+ \lambda_1^2)^2} \, \rho_1 
	\left( b_1+ \overline{b}_1 \right)
	+ \cdots
	\right) \otimes \id \otimes \id 
	\right]
	\, .\nonumber
\end{align}
In these general situations we denote $b(v_i)$ by $\mathfrak{B}$. There may be multiple applications of $\mathfrak{B}$ to the surface states if we consider the deformation of lengths of multiple borders simultaneously. In such cases $1,2,3$ labels in $\mathfrak{B}$ should be permuted accordingly and $\mathfrak{B}$ should act on the border whose length is deformed. 

We emphasize strongly that the $\mathfrak{B}$ insertion depends on~\emph{all} of the border lengths of a given pair of pants---not just to the border whose length has shifted. This is quite unorthodox, especially when contrasted to the ordinary factorization analysis for off-shell amplitudes~\cite{Erbin:2021smf}. For the latter the associated $b$-ghosts insertion $b_0^- b_0^+$ doesn't depend on the details of the surface they are acting on whatsoever. They take a universal form. The $b$-ghost insertion due to the Fenchel-Nielsen deformation $\mathfrak{B} b_0^-$, on the other hand, naturally depends on the particular pants decomposition.

One may worry that such dependence may obstruct factorizing generic off-shell amplitudes by requiring ``global'' knowledge of the surface. However, in the view of~\eqref{eq:4.8} and~\eqref{eq:4.9}, we can decompose the $b$-ghost insertions into two disjoint parts where each part~\emph{exclusively} depends on either the left or right pant---but not both of them simultaneously. So after cutting the surface along the $i$-th seam any dependence on the opposite pants disappears. While they still depend on the pants decomposition itself, the parts of the $b$-ghost insertions can be separated in a way that the knowledge of the opposite part of the surface is irrelevant at least. This will be sufficient for our purposes.

\subsection{The recursion for hyperbolic vertices} \label{sec:4.2}

In the previous subsection we have derived the $b$-ghost insertions associated with the Fenchel-Nielsen deformations~\eqref{eq:1.7a}. Using such insertions we can factorize the string measure $\langle \Omega_{g,n} (L_i) |$ according to the different types of pants excisions shown in figure~\ref{Mirz-gluing-figure}. They take the form (twists are taken respect to the $L$ pants):
\begin{subequations} \label{eq:d}
\begin{enumerate}
	\item \underline{$R$-term for the $i$-th border}:
	\begin{align}
		\langle \Omega_{g,n}(L_i)| =  d \ell \wedge {d \tau \over \ell}  \, 
		\sum_{\epsilon = \pm}\,
		&\bigg[ 
		\langle \mathcal{V}_{0,3}(L_1, L_i, \epsilon \ell) | 
		\otimes \langle \Omega_{g, n - 1}(- \epsilon \ell, \mathbf{L} \setminus \{ L_i \}) |
		\bigg ] | \mathfrak{S} (\tau) \rangle \, .
	\end{align}
	
	\item \underline{Nonseparating $D$-term}:
	\begin{align}
	\langle \Omega_{g,n}(L_i)| &= d \ell_1 \wedge {d \ell_2} \wedge { d \tau_1 \over \ell_1 }\wedge {d \tau_2 \over \ell_2}
	\\ \nonumber
	&\hspace{-0.2in}\sum_{\epsilon_1,\epsilon_2 = \pm}\,
	\bigg[ 
	\langle \mathcal{V}_{0,3}(L_1, \epsilon_1 \ell_1, \epsilon_2  \ell_2) | 
	\otimes \langle \Omega_{g-1, n + 1}(- \epsilon_1 \ell_1, - \epsilon_2 \ell_2, \mathbf{L} ) |
	\bigg ] \, | \mathfrak{S} (\tau_1)  \rangle_1 \, | \mathfrak{S} (\tau_2)  \rangle_2  \, .
	\end{align}
	
	\item \underline{Separating $D$-term}:
		\begin{align}
		\langle \Omega_{g,n}(L_i)| &= d \ell_1 \wedge  d \ell_2 \wedge {d \tau_1 \over \ell_1}  \wedge {d \tau_2 \over \ell_2}
		\\ \nonumber
		&\hspace{-1.3in}\sum_{\epsilon_1,\epsilon_2 = \pm}
		\bigg[ 
		\langle \mathcal{V}_{0,3}(L_1, \epsilon_1 \ell_1, \epsilon_2  \ell_2) | 
		\otimes
		\sum_{\text{stable}} \langle \Omega_{g_1,n_1}(-\epsilon_1 \ell_1, \mathbf{L_1}) | \otimes
		\langle \Omega_{g_2,n_2}(-\epsilon_2 \ell_2,  \mathbf{L_2}) |
		\bigg ] | \mathfrak{S} (\tau_1)  \rangle_1  | \mathfrak{S} (\tau_2) \rangle_2  \, .
	\end{align}
\end{enumerate}
\end{subequations}

These results can be established following a similar analysis to~\cite{Ishibashi:2022qcz} while being mindful about various signs, for example~\eqref{eq:2.20}. We point out this decomposition only holds~\emph{locally} on $\mathcal{M}_{g,n} (L_i)$. Above we have adopted the conventions stated in~\eqref{eq:1.5} and~\eqref{eq:A1.6}
\begin{subequations} \label{eq:4.15}
\begin{align} 
	&\langle \mathcal{V}_{0,3} ( L_1, L_2, L_3) |  \equiv  \langle \Sigma_{0,3}( |L_1|, |L_2| , |L_3| )|  
	\, \left(  \mathfrak{B}^{\otimes \, \theta(-L_1)} \otimes \mathfrak{B}^{\otimes \, \theta(-L_2)} \otimes \mathfrak{B}^{\otimes \, \theta(-L_3)} \right) \, , \\
	&\langle \Omega_{g,n} ( L_1, \cdots, L_n) | \equiv
	\langle \Omega_{g,n} (|L_1|, \cdots, |L_n|)  | \, 
	\left(\mathfrak{B}^{\otimes \, \theta(-L_1)} \otimes \cdots \otimes \mathfrak{B}^{\otimes \, \theta(-L_n)} \right) \, ,
\end{align}
\end{subequations}
of having negative lengths. Since the $\mathfrak{B}$ insertion either acts on the left or right of the seam (but not both at the same time), having a negative length interpretation is quite natural in this context and is associated with the orientation of the pants with respect to each other.

Observe that we insert the non-level-matched bivector $| \mathfrak{S} (\tau) \rangle$~\eqref{eq:2.42} to the entries of the bras in~\eqref{eq:d}, which are supposed to act only on the level-matched states~\eqref{eq:2.18}. This means that factorizations above come with global phase ambiguities. As we shall see shortly, however, this ambiguity is going to disappear for the final expression.\footnote{The same ambiguity also presents itself in the ordinary factorization analysis of the off-shell amplitudes with the string propagator before summing over the twists and disappear only afterward. We observe an analogous behavior. For the recent attempts of relaxing level-matching condition~\eqref{eq:2.17} in CSFT, see~\cite{Erbin:2022cyb,Okawa:2022mos}.}

Now, we can promote~\eqref{eq:3.29} to the partition of the hyperbolic string measure~\eqref{eq:4.2} as
\begin{align}  \label{eq:4.16}
	|L_1| \cdot \langle \omega_{g,n} (L_i) | &= 
	\sum_{i=2}^n  \sum_{\gamma \in \mathcal{P}_i} \sum_{\epsilon_\gamma = \pm} \,
	{d \gamma} \wedge {d \tau_\gamma \over \gamma } \, 
	\bigg[ \langle \mathfrak{R}( L_1, L_i, \epsilon_\gamma \gamma) | \, 
	\otimes \langle \omega_{g, n - 1}(- \epsilon_\gamma \gamma, \mathbf{L} \setminus \{ L_i \}) |
	\bigg] | \mathfrak{S}(\tau_\gamma) \rangle_\gamma
	\nonumber \\ 
	&\hspace{-1in}+
	\sum_{\{\gamma, \delta\}\in \mathcal{P}_1} \sum_{\epsilon_\gamma, \epsilon_\delta = \pm}
	d \gamma \wedge d \delta \wedge {d \tau_\gamma \over \gamma} \wedge {d \tau_\delta \over \delta} 
	\, 
	\bigg[
	\langle \mathfrak{D}( {L_1, \epsilon_\gamma \gamma,  \epsilon_\delta  \delta }) |
	\otimes 
	\bigg(
	\,   \langle \omega_{g-1, n + 1}(- \epsilon_\gamma \gamma, - \epsilon_\delta \delta, \mathbf{L} ) |
	\nonumber \\
	&+\sum_{\text{stable}} \langle \omega_{g_1,n_1}(-\epsilon_\gamma \gamma, \mathbf{L_1}) | \otimes
	\langle \omega_{g_2,n_2}(-\epsilon_\delta \delta,  \mathbf{L_2}) |
	\bigg)
	\bigg] | \mathfrak{S}(\tau_\gamma) \rangle_\gamma  \, | \mathfrak{S}(\tau_\delta) \rangle_\delta \, ,
\end{align}
in light of~\eqref{eq:d}. Above we have collected the cubic parts into the string kernels~\eqref{eq:1.7}
\begin{subequations} \label{eq:4.15a}
	\begin{align}
		\langle \mathfrak{R}(L_1,L_2,L_3) | &= \widetilde{R}_{|L_1| |L_2| |L_3|} \, \langle \mathcal{V}_{0,3}(L_1,L_2,L_3)| \, ,\\
		\langle \mathfrak{D}(L_1,L_2,L_3) | &= \widetilde{D}_{|L_1| |L_2| |L_3|}   \, \langle \mathcal{V}_{0,3}(L_1,L_2,L_3)| \, . \label{eq:4.15b}
	\end{align}
\end{subequations}
There is an implicit dependence on the threshold length $L$ in these objects.

Upon integrating~\eqref{eq:4.16} over the moduli space $\mathcal{M}_{g,n}(L_i)$ and repeating the unrolling trick~\eqref{eq:main} for evaluating such integrals, we obtain the desired recursion relation for $2g-2+n > 1$ and $n \geq 1$ (refer to figure~\ref{fig:sft})
\begin{align}  \label{eq:4.18}
	&|L_1| \cdot \langle \mathcal{V}_{g,n} (L_i) | = \sum_{i=2}^n \, \int\limits_{-\infty}^\infty \, d \ell 
	\, \bigg[ \langle \mathfrak{R} (L_1, L_i, \ell) | \otimes \, \langle \mathcal{V}_{g, n-1} (-\ell, \mathbf{L} \setminus \{L_i \}) |
	\bigg] |\omega^{-1} \rangle \\
	&\hspace{0.5in} + {1 \over 2} \int\limits_{-\infty}^\infty  \, d \ell_1  \, \int\limits_{-\infty}^\infty  
	\, d \ell_2 \,
	\bigg[\langle \mathfrak{D} (L_1, \ell_1, \ell_2) | \otimes \,
	\bigg(\langle \mathcal{V}_{g-1,n+1} (-\ell_1, -\ell_2, \mathbf{L} )| 
	\nonumber \\ 
	&\hspace{2in} +
	\sum_{\text{stable}} 
	\langle \mathcal{V}_{g_1,n_1} (-\ell_1, \mathbf{L_1} ) | \otimes \, \langle  \mathcal{V}_{g_2,n_2} (-\ell_2, \mathbf{L_2} ) |
	\bigg)
	\bigg] | \, \omega^{-1} \rangle_1  | \, \omega^{-1}  \rangle_2 \nonumber \, ,
\end{align}
given that the string amplitudes should be independent of the particular marking on the surface. Above we have evaluated the twist integrals, taken to be running from $0$ to $\ell$'s, using~\eqref{eq:2.43}. After such integration the global phase ambiguity mentioned below~\eqref{eq:4.15} disappears since only the level-matched states appear in the decomposition. This is the main result of this paper. Similar logic can be used to establish the following identity for the one-bordered torus (see~\eqref{eq:3.35})
\begin{align}
	|L_1| \cdot \langle \mathcal{V}_{1,1} (L_1) | = {1 \over 2} \int\limits_{-\infty}^\infty \,  d \ell \,
	 \bigg(
	 D_{|L_1| |\ell| |\ell|} - | L_1| \cdot \theta(L - |\ell|) 
	 \bigg)  \,
	 \langle \mathcal{V}_{0,3}(L_1, \ell,-\ell) |
	\,
	\bigg(\id \otimes | \omega^{-1} \rangle \bigg) \, .
\end{align}
We again included $1/2$ due to the nontrivial $\mathbb{Z}_2$ symmetry.

A few remarks are in order here. We first highlight that~\eqref{eq:4.18} relates hyperbolic vertices with each other iteratively in the negative Euler characteristic $-\chi_{g,n} = 2g - 2 +n$. This makes it a~\emph{topological recursion}. Furthermore, it contains more information on the nature of vertices compared to the ordinary loop $L_\infty$ relations satisfied by CSFT interactions~\cite{Erler:2019loq}. The latter follow from the constraint satisfied by the boundaries of string vertices $\partial \mathcal{V}$ as a consequence of the geometric master equation---they don't provide any information on the behavior of the interior of $\mathcal{V}$ whatsoever. On the other hand~\eqref{eq:4.18} also relates the interiors.

As mentioned earlier, a similar recursive formulation has been carried out by Ishibashi~\cite{Ishibashi:2022qcz,Ishibashi:2024kdv}. Despite its inspiring features, such as its connection with the Fokker-Planck formalism, Ishibashi's off-shell amplitudes don't factorize according to the prescription provided by covariant CSFT for which the Feynman diagrams contain flat cylinders. This obstructs their field theory interpretation and degenerating behavior of the amplitudes. Our recursion overcomes these particular issues by relating $\langle \mathcal{V}_{g,n} (L_i) | $, for which the string measure is integrated over the  systolic subsets $\mathcal{V}^L_{g,n}(L_i)$.

Precisely speaking, the difference mentioned above follows from the distinct choices of kernels for~\eqref{eq:4.15a}---contrast with the equation (3.20) in~\cite{Ishibashi:2022qcz}. The impact of this modification particularly presents itself when the length of the seams becomes smaller than $L \leq L_\ast$. For example
\begin{align} \label{eq:4.18ab}
	&R(L_1, L_2, \ell) = L_1 - 
	\frac{\sinh \left( {L_1 \over 2} \right)}{\sinh\left({L_1 \over 2}\right) + \cosh\left({L_2 \over 2}\right) } \, \ell + \mathcal{O} (\ell^2) \, ,
\end{align}
while in the twisted version the leading term $L_1$ disappears and the expansion holds for $0 \leq \ell \leq L$~\eqref{eq:3.6}. Similar expressions can be worked out for $D(L_1, L_2, L_3)$.

Thanks to the twist~\eqref{eq:218}, we subtract the Feynman region contributions while maintaining the recursive structure, resulting in a recursion for hyperbolic CSFT. However these subtracted contributions are~\emph{not} ordinary Feynman diagrams that contain flat propagators: rather they are hyperbolic amplitudes as well. As an example, consider $(g,n)=(0,4)$ in~\eqref{eq:4.18}. From the expansion~\eqref{eq:4.18ab} the subtracted term is
\begin{align}
	\int\limits_{-L}^L \, d \ell \, \bigg[  \langle \mathcal{V}_{0,3} (L_1, L_2, \ell) | 
	\otimes \langle \mathcal{V}_{0,3} (-\ell, L_3, L_4) | \bigg] | \omega^{-1} \rangle \, ,
\end{align}
for one of the channels. This is the hyperbolic contribution from this channel's associated Feynman region. Upon subtracting the other channel's contributions one is left with purely the vertex region as desired.\footnote{Even though the final result is finite, this is not necessarily the case for the individual subtracted terms, due to the $|\ell| \to 0$ regime. They may need analytic continuation.} The form of the subtraction for the higher-order vertices would be more complicated in a similar way to systolic volumes.

The recursion relation~\eqref{eq:4.18} works for any bordered hyperbolic surface, but it can also be used for obtaining the vacuum vertices $\langle \mathcal{V}_{g,0} | \in  \widehat{\mathcal{H}}^{\otimes 0}$ for $g \geq 2$ using the dilaton theorem~\cite{Bergman:1994qq,Rahman:1995ee}. The dilaton theorem in the context of hyperbolic CSFT states
\begin{align} \label{eq:4.21a}
	\langle \mathcal{V}_{g,n+1} (L_i, L_{n+1} = 2 \pi i) | 
	 \, \bigg( \underbrace{\id \otimes \cdots \otimes \id}_{n \text{ times}} \,
	\otimes \, | D \rangle \bigg)
	= (2g-2+n) \, \langle \mathcal{V}_{g,n} (L_i) | \, ,
\end{align}
where
\begin{align}
	|D \rangle = \left( c_1 \, c_{-1} - \overline{c}_1 \, \overline{c}_{-1} \right) | 0 \rangle \, ,
\end{align}
is the ghost-dilation. Taking $n=0$ in~\eqref{eq:4.21a} and using~\eqref{eq:4.18} for $\langle A_{g,1} (2 \pi i) | D \rangle$ leads to the desired recursion. 

Observe, however, $D$ is not a $(0,0)$-primary and~\eqref{eq:4.21a} implicitly contains a specific prescription for its insertions. In~\eqref{eq:4.21a} the ghost-dilaton is grafted to the border of formal length $L_{k+1} = 2 \pi i$.\footnote{This corresponds to the border degenerating to a cone point with the opening angle $2 \pi$ and becoming a regular point on a surface with one less border~\cite{do2011moduli}.} For this, the choice of local coordinates is made according to section (2.2) of~\cite{Bergman:1994qq} using the regular hyperbolic metric on the surface. Since the other borders are geodesics, there are no contributions from the integrals of $c_0^-$, see~\cite{Bergman:1994qq}.

Clearly~\eqref{eq:4.18} can be restricted to apply  for only genus $0$ surfaces like in~\eqref{eq:3.25}. We report this case for the completeness ($n>3$)
\begin{align}  \label{eq:4.18a}
	&|L_1| \cdot \langle \mathcal{V}_{0,n} (L_i) | = \sum_{i=2}^n \, \int\limits_{-\infty}^\infty \, d \ell 
	\, \bigg[ \langle \mathfrak{R} (L_1, L_i, \ell) | \otimes \, \langle \mathcal{V}_{0,n-1} (-\ell, \mathbf{L} \setminus \{L_i \}) |
	\bigg] |\omega^{-1} \rangle \\
	&\hspace{-0.25in} + {1 \over 2} \int\limits_{-\infty}^\infty  \, d \ell_1  \, \int\limits_{-\infty}^\infty  
	\, d \ell_2 \,
	\bigg[\langle \mathfrak{D} (L_1, \ell_1, \ell_2) | \otimes \,
	\sum_{\text{stable}} 
	\langle \mathcal{V}_{0,n_1} (-\ell_1, \mathbf{L_1} ) | \otimes \, \langle \mathcal{V}_{0,n_2} (-\ell_2, \mathbf{L_2} ) |
	\bigg] | \, \omega^{-1} \rangle_1  | \, \omega^{-1}  \rangle_2 \nonumber \, .
\end{align}

We finally highlight that~\eqref{eq:4.18} holds for every quantum hyperbolic CSFT given the former works for $L \leq L_\ast$ which is the regime for which the quantum hyperbolic vertices $\mathcal{V}^L$ obey the geometric master equation~\eqref{eq:2.2}~\cite{Costello:2019fuh}. The condition on $L$ can be relaxed slightly for~\eqref{eq:4.18a}, but it can't be entirely disposed of as in~\eqref{eq:3.39}. But recall the classical CSFT is consistent for any choice of $L$ and it is in principle possible to relate different choices of $L$ through field redefinitions. A question is what is the fate of~\eqref{eq:4.18a} for generic classical CSFT.

In a wider context, it is possible to modify string vertices $\mathcal{V}$ while maintaining that they solve the relevant geometric master equation, which is equivalent to performing field redefinitions in CSFT~\cite{Hata:1993gf}. Under generic field redefinitions the form of the recursion~\eqref{eq:4.18} or~\eqref{eq:4.18a} gets highly obstructed. Nevertheless, its observable consequences should remain the same. This relates back to our emphasis on using the ``correct'' string vertices: we would like to manifest these special structures as much as possible in order to ease the extraction of physics.

\subsection{Recursion as a differential constraint}

An alternative and more succinct way to encode~\eqref{eq:4.18} and~\eqref{eq:4.18a} is through a second-order differential constraint on an appropriate generating function. So imagine the string field
\begin{align} \label{eq:4.23}
	\Psi(\ell) \in \widehat{\mathcal{H}} \otimes \mathbb{R} \, , \quad \quad
	\Psi (\ell)  = \Psi (-\ell)  \, ,
\end{align}
that has a dependence on a real parameter and introduce the generating function $Z[\Psi (\ell)]$ and its associated free energy $W[\Psi (\ell)]$ as a functional of $\Psi (\ell) $\footnote{This partition function is formal since the integral may diverge around $\ell = 0$.}
\begin{align} \label{eq:4.20}
	Z[\Psi (\ell)] &\equiv \exp \left[ {W[\Psi (\ell)]} \right]
	\\ \nonumber
	&\equiv \exp \left[
	\, \sum_{g,n} \, {1 \over n!} \, \hbar^{g-1} \, \kappa^{2g-2+n}
	\int\limits_{-\infty}^\infty d \ell_1 \cdots  	\int\limits_{-\infty}^\infty d \ell_n \,
	\langle \mathcal{V}_{g,n} (\ell_i) |\, \bigg(   \Psi (\ell_1)  \otimes \cdots \otimes  \Psi  (\ell_n) \bigg)
	\, \right] \, .
\end{align}
We always consider $\Psi (\ell)$ to have even statistics (but arbitrary ghost number) in order to interpret $Z[\Psi (\ell)] $ as a ``partition function''. This is in the same vein of taking the string field even in the BV master action~\eqref{eq:2.37}. As we shall see this will be sufficient for our purposes.

We further introduce the $2$-products $A,B,C: \widehat{\mathcal{H}}^{\otimes 2} \to \widehat{\mathcal{H}}$
\begin{subequations} \label{eq:4.24}
\begin{align} 
	&A_{L_1 L_2 L_3} =
	\bigg[ \id \otimes \langle \mathcal{V}_{0,3} (L_1, L_2, L_3) | \bigg] 
	\bigg[ | \omega^{-1} \rangle \otimes  \id \otimes \id \bigg] \, , \\
	&B_{L_1 L_2 L_3} =
	\bigg[ \id \otimes \langle \mathfrak{R} (L_1, L_2, L_3) | \bigg] 
	\bigg[ | \omega^{-1} \rangle \otimes  \id \otimes \id \bigg] \, , \\
	&C_{L_1 L_2 L_3} =
	\bigg[ \id \otimes \langle \mathfrak{D} (L_1, L_2, L_3) | \bigg] 
	\bigg[ | \omega^{-1} \rangle \otimes  \id \otimes \id \bigg] \, ,
\end{align}
\end{subequations}
and the $0$-product $D:  \widehat{\mathcal{H}}^{\otimes 0} \to \widehat{\mathcal{H}}$
\begin{align}
	D_{L_1} =
	\bigg[ \id \otimes \langle \mathcal{V}_{1,1} (L_1) | \bigg] | \omega^{-1} \rangle \, .
\end{align}
We point out these products depend on the length of the borders so they are~\emph{not} graded-symmetric. However $A,B$ are symmetric if we also exchange $L_2 \leftrightarrow L_3$ thanks to the symmetry~\eqref{eq:27}. When $L_1 = L_2 = L_3 = L$, the products $A$ and $D$ become part of the ordinary $L_\infty$ products of CSFT, see~\eqref{eq:2.47}.

The final object we introduce is ``differentiation'' with respect to the level-matched string fields
\begin{align}
	 {\delta \Psi'(\ell') \over \delta \Psi (\ell) } \, 
	 \equiv {\delta | \Psi'(\ell') \rangle  \over \delta \left( \langle  \Psi (\ell) | c_0^- \right)  } \, 
	 &= {\delta \over \delta \left( \langle  \Psi (\ell) | c_0^- \right)   }
	 \bigg( \id \otimes \langle \Psi'(\ell') | \bigg) | \Sigma_{0,2} \rangle \\
	 &= {\delta \over \delta \left( \langle  \Psi (\ell) | c_0^- \right)   }
	 \bigg( \id \otimes \langle \Psi'(\ell') | \, c_0^- \delta(L_0^-)  b_0^-  \bigg) | \Sigma_{0,2} \rangle 
	 \nonumber \\
	 &=\bigg( \id \otimes \delta [ \Psi'(\ell') - \Psi (\ell)] \bigg) \, | \omega^{-1} \rangle 
	 \in \widehat{\mathcal{H}}^{\otimes 2} \, , \nonumber
\end{align}
where we have used in order,~\eqref{eq:2.22};~\eqref{eq:2.42a};~\eqref{eq:2.40} and the fact that $c_0^-$ doesn't support a cohomology. Notice the right-hand side contains the delta functional in the space $\widehat{\mathcal{H}}\otimes \mathbb{R}$. This derivative obeys the Leibniz rule. The effect of acting the derivative on~\eqref{eq:4.20} is
\begin{align}
	 {\delta Z \over \delta \Psi (\ell) }
	 &= { \delta W  \over \delta \Psi(\ell) } \cdot Z 
\end{align}
where
\begin{align} \label{eq:4.27}
	{ \delta W  \over \delta \Psi(\ell )} = 
	\sum_{g,n} \, {\hbar^{g-1} \, \kappa^{2g-2+n} \over (n-1)!} \, 
	&\int\limits_{-\infty}^\infty d \ell_2 \cdots  	\int\limits_{-\infty}^\infty d \ell_{n} \,
	\bigg( \id \otimes \langle \mathcal{V}_{g,n} (\ell, \ell_2, \cdots \ell_n) \, \bigg) 
	\\ \nonumber
	&\hspace{1.75in}\bigg(|  \omega^{-1} \rangle \otimes
	 \Psi  (\ell_{2}) \otimes \cdots \otimes \Psi  (\ell_{n}) 
	\bigg) \, ,
\end{align}
after applying the Leibniz rule and symmetry of the vertices. Note that taking this derivative produces a state in $\widehat{\mathcal{H}}$.

One can similarly find the second derivative
\begin{align}
	{\delta^2 \, Z \over \delta \Psi' (\ell') \, \delta \Psi (\ell)}
	= \left[
	 { \delta^2 \, W  \over \delta  \Psi'(\ell') \, \delta \Psi(\ell) } + { \delta W  \over \delta \Psi'(\ell') } \, { \delta W  \over \delta \Psi(\ell) } 
	\right] \cdot Z \, ,
\end{align}
where
\begin{align}
	{ \delta^2 W  \over \delta  \Psi'(\ell') \, \delta \Psi(\ell) }  = 
	\sum_{g,n} \, &{\hbar^{g-1} \, \kappa^{2g-2+n} \over (n-2)!} \,
	\int\limits_{-\infty}^\infty d \ell_3 \cdots  	\int\limits_{-\infty}^\infty d \ell_{n} \,
	\\ \nonumber
	&\bigg( \langle \mathcal{V}_{g,n} (\ell', \ell,  \ell_3, \cdots \ell_n) \, | \bigg) 
	\, | \omega^{-1} \rangle_{\ell'}  \, | \omega^{-1} \rangle_\ell \,
	\bigg( \Psi  (\ell_{3})  \otimes \cdots \otimes \Psi  (\ell_{n}) 
	\bigg) \, .
\end{align}
This object belongs to $\widehat{\mathcal{H}}^{\otimes 2}$. Observe how the Poisson bivector acts here for lack of a better notation.

Upon defining the string field $| \Psi (\ell) \rangle$-valued operator
\begin{align} \label{eq:4.28}
	\mathcal{D}_{\Psi(L_1)} \equiv \bigg[ \hbar \, {\delta \over \delta \Psi (L_1)}
	&- {\kappa  \over 2} \, \int\limits_{-\infty}^\infty d \ell_1 \,  \int\limits_{-\infty}^\infty d \ell_2 \,
	A_{L_1 \ell_1 \ell_2 } \left(\Psi(\ell_1), \Psi(\ell_2)\right)
	\nonumber \\
	&- \hbar \kappa \int\limits_{-\infty}^\infty d \ell_1 \,  \int\limits_{-\infty}^\infty d \ell_2 \,
	B_{L_1 \ell_1 \ell_2} \left(\Psi(\ell_1), \, {\delta \over \delta \Psi(\ell_2)} \right)
	\\ \nonumber
	&- {\hbar^2 \kappa  \over 2} 
	\int\limits_{-\infty}^\infty d \ell_1 \,  \int\limits_{-\infty}^\infty d \ell_2 \,
	C_{L_1 \ell_1 \ell_2}  \left({\delta^2 \over \delta \Psi(\ell_1) \, \delta \Psi(\ell_2)} \right)
	- \hbar \kappa D_{L_1} \bigg]  \, ,
\end{align}
the differential constraint reads
\begin{align} \label{eq:4.29}
	\forall \, \Psi(L_1) \in \widehat{\mathcal{H}} \otimes \mathbb{R} 
	\quad \quad \quad
	\mathcal{D}_{\Psi(L_1)} \cdot Z[\Psi(\ell)] =0 \, .
\end{align}
It is not difficult to establish this is equivalent to the recursion~\eqref{eq:4.18} using the derivatives above. Observe that~\eqref{eq:4.29} admits a trivial solution $Z = 1$, on top of those provided by the hyperbolic amplitudes. This case corresponds to having no interactions in CSFT that trivially realizes~\eqref{eq:4.18}.

It is important to point out the uncanny resemblance between~\eqref{eq:4.28}-~\eqref{eq:4.29} and the differential operators that form~\emph{quantum Airy structures} and the unique function that is annihilated by them~\cite{kontsevich2017airy,Andersen:2017vyk}. This is not coincidental: quantum Airy structures can be used to encode topological recursion as differential equations and we have established the latter already. However the relation between quantum airy structures and CSFT can be more than what meets the eye initially. For example, turning the logic on its head, one may imagine CSFT as something that is constructed by solving~\eqref{eq:4.29} perturbatively. We are going to comment on these points at the end of the paper.
	
We can also state the recursion restricted to genus $0$ surfaces~\eqref{eq:4.18a} as a first-order differential constraint. This can be simply obtained by taking the classical $\hbar \to 0$ limit of~\eqref{eq:4.29}
\begin{align} \label{eq:4.32}
	0 = {\delta W \over \delta \Psi (L_1)}
	&- { \kappa  \over 2} 
	\int\limits_{-\infty}^\infty d \ell_1 \,  \int\limits_{-\infty}^\infty d \ell_2 \,
	A_{L_1 \ell_1 \ell_2} \left(\Psi(\ell_1), \Psi(\ell_2) \right)
	\\ \nonumber
	&- \kappa \int\limits_{-\infty}^\infty d \ell_1 \,  \int\limits_{-\infty}^\infty d \ell_2 \,
	B_{L_1 \ell_1 \ell_2} \left(\Psi(\ell_1), {\delta W \over \delta \Psi(\ell_2)} \right)
	\\ \nonumber
	&- { \kappa  \over 2} \int\limits_{-\infty}^\infty d \ell_1 \,  \int\limits_{-\infty}^\infty d \ell_2 \,
	C_{L_1 \ell_1 \ell_2} \left({\delta W \over \delta \Psi(\ell_1)} , {\delta W \over \delta \Psi(\ell_2) } \right)  \, ,
\end{align}
for all $ \Psi (L_1)  \in \widehat{\mathcal{H}} \otimes \mathbb{R} $. It is further possible to restrict the string fields to be ghost number $2$ in this equation. This particular form is going to be useful in section~\ref{sec:6}.

\section{Comparison with a stubbed theory} \label{sec:5}

Before we proceed to investigating the implications of the recursion relation described in the previous section, let us show that similar structures can be exhibited in theories with stubs~\cite{Chiaffrino:2021uyd,Schnabl:2023dbv,Schnabl:2024fdx, Erler:2023emp,Maccaferri:2024puc,Erbin:2023hcs} and investigate their consequences to get an intuition for the recursion~\eqref{eq:4.18}. We work with the classical stubbed cubic scalar field theory of~\cite{Erler:2023emp} for simplicity and adopt its conventions unless stated otherwise. This section is mostly self-contained.

\subsection{The stubbed cubic scalar field theory}

We consider the real scalar field theory in $D$ dimensions
\begin{align}
	I[\phi] = \int d^D x \left(- {1 \over 2} \, \partial_\mu \phi \, \partial^\mu \phi - V(\phi)\right) \, ,
\end{align}
with the cubic potential
\begin{align} \label{eq:6.2}
	V(\phi) = - {\mu^2 \over 2!} \phi^2 + {\kappa \over 3!} \phi^3 \, .
\end{align}
Here $\mu^2 > 0$ is the mass parameter and $\kappa $ is the coupling constant. This potential has an unstable perturbative vacuum at $\phi = 0$ and the stable nonperturbative tachyon vacuum $\phi = \phi_\ast$ residing at
\begin{align}
	\phi_\ast = {2 \mu^2 \over \kappa } \, , \quad \quad
	V(\phi_\ast) = - {2 \over 3} {\mu^6 \over \kappa ^2} \, .
\end{align}
The squared mass of the linear fluctuations around this vacuum are $V''(\phi_\ast) = \mu^2 > 0$.

We would like to deform this theory by stubs~\cite{Erler:2023emp}. Focusing on the zero-momentum sector of the theory for simplicity, i.e., spacetime-independent configurations, we find the potential of the theory deforms to
\begin{align} \label{eq:6.4}
	V(\phi; \Lambda)  &= - {1 \over 2!} \mu^2 \phi^2 
	+ {1 \over 3!} \kappa  \left(e^{\Lambda \mu^2 \over 2} \phi\right)^3
	- {1 \over 4!} 3 \kappa ^2 \left({e^{\Lambda \mu^2} - 1 \over \mu^2 }\right) \left(e^{\Lambda \mu^2 \over 2} \phi\right)^4
	+ \cdots \nonumber \\
	&= - {1 \over 2!} \mu^2 \phi^2  
	+ \sum_{n=3}^\infty {1 \over n!} b_{n-2} \kappa ^{n-2} \left({1 - e^{\Lambda \mu^2}  \over \mu^2 }\right)^{n-3} \left(e^{\Lambda \mu^2 \over 2} \phi\right)^n\, ,
\end{align}
after including stubs of length $\Lambda / 2$. The positive integers $b_n$ are the number of unordered rooted full binary trees with $n+1$ \emph{labeled} leaves, whose formula is given by a double factorial
\begin{align} \label{eq:5.5}
	b_{n} = (2 n - 1)!! = {(2n)! \over 2^n \, n !}= 1 \times 3 \times 5 \times \cdots \times (2 n - 1) \, ,
\end{align}
for $n \geq 1$ and it is $1$ when $n=0$. The first few terms are
\begin{align}
	b_n = 1, 1,3,15, 105, \cdots \, .
\end{align}
These numbers come from the combinatorics of the Feynman diagrams. Stubs instruct us to treat the Feynman diagrams whose propagator's proper time is shorter than $\Lambda$ as an elementary interaction. That means we need to include each topologically distinct Feynman diagrams with these ``partial propagators'' to the potential. Our conventions in \eqref{eq:6.2} and~\eqref{eq:6.4} suggests us to consider the~\emph{labeled} Feynman diagrams.\footnote{In other words we perform the homotopy transfer with the partial propagator in the context of the homotopy Lie algebras instead of their associative counterpart like in~\cite{Erler:2023emp}.}

We point out the form of this stubbed potential is slightly different from~\cite{Erler:2023emp}. Nevertheless they can be related to each other after setting
\begin{align}
	\kappa ^{\text{here}} = 2 g^{\text{there}} \, ,
\end{align}
compare~\eqref{eq:6.2} with equation (2.2) of~\cite{Erler:2023emp}, since
\begin{align}
	{1 \over n!} \, b_{n-2} \, 2^{n-2} 
	=   {1 \over n!} \,  {(2n - 4)! \over 2^{n-2} (n-2) !} \, 2^{n-2} 
	= {1 \over n} \, {(2n-4)! \over (n-1)! (n-2)!} = {1 \over n} \, C_{n-2} \, ,
\end{align}
where $C_n$ are the Catalan numbers. Plugging this into~\eqref{eq:6.4} one indeed obtains the stubbed potential of~\cite{Erler:2023emp}, see equation (2.24). Taking this slightly different convention would help us to draw parallels with the topological recursion presented in the previous section more directly.

Resumming the expansion~\eqref{eq:6.4} gives
\begin{align} \label{eq:5.9}
	V(\phi; \Lambda) = - {\mu^2 \over 2} \, \phi^2 \, { e^{\Lambda \mu^2 } f_3(x_3) - 1 \over e^{\Lambda \mu^2 } - 1} \, ,
\end{align}
where
\begin{align} \label{5.10}
	f_3(x) = {6 x - 1 + (1-4x)^{3/2} \over 6x^2 } \, ,
	\quad \quad
	x_3 = - {\kappa  \over 2} {e^{\Lambda \mu^2} -1 \over \mu^2} e^{\Lambda \mu^2 \over 2} \phi \, ,
\end{align}
and the nonperturbative  vacuum is shifted exponentially far away
\begin{align} \label{eq:5.11}
	\phi_\ast(\Lambda) = {2 \mu^2 \over \kappa } \exp \left( {\Lambda \mu^2 \over 2} \right) \, ,
\end{align}
in the stubbed theory while the depth of the potential remains the same. We also point out the expansion~\eqref{eq:6.4} in $\phi$ has the radius of convergence
\begin{align} \label{eq:5.12a}
	r(\Lambda) = 
	{\mu^2 \over 2 \kappa} \, {e^{- { 3\Lambda \mu^2  \over 2}} \over 1 - e^{- \Lambda \mu^2}} \, ,
\end{align}
due to the fractional power in~\eqref{5.10}.

\subsection{Topological recursion for the stubs}

Now we would like to obtain the nonperturbative solution~\eqref{eq:5.11} of the stubbed theory using an alternative perspective. From the form of the series in \eqref{eq:6.4} and the identity 
\begin{align}
	b_{n+1} = (2n+1) \, b_n \, ,
\end{align}
it shouldn't be surprising that there is a recursion among the elementary vertices of the stubbed theory. Introducing
\begin{align} \label{eq:5.12}
	W_n(s_i)  =
	b_{n-2} \left({1 - e^{\Lambda \mu^2}  \over \mu^2 }\right)^{n-3} 
	\prod_{i=1}^n \exp \left( {s_i \, \mu^2 \over 2} \right) \, ,
	\quad \quad
	s_i \geq 0 \, ,
\end{align}
one can form the topological recursion
\begin{align} \label{eq:5.13}
	W_n(s_i)  &=\sum_{i=2}^n \int\limits_0^\infty \, dt \, r_{s_1 s_i t} \, W_{n-1}( t, \mathbf{s} \setminus \{ s_i \} )  
	\\ \nonumber
	&\hspace{1in}+
	{1 \over 2} \, \int\limits_0^\infty \, dt_1 \, \int\limits_0^\infty \, dt_2 \,
	\, \sum_{\text {stable} } \, d_{s_1 t_1 t_2} \,
	W_{n_1} (t_1,  \mathbf{s_1}) \, W_{n_2} (t_2,  \mathbf{s_2}) \, ,
\end{align}
for $n > 3$, where the~\emph{stub kernels} are given by
\begin{subequations} \label{eq:5.14}
	\begin{align}
		r_{s_1 s_2 s_3} 
		&= \big(-1 + \theta(s_3 - \Lambda) \big) \, a_{s_1 s_2 s_3}  \, \\
		d_{s_1 s_2 s_3}  &= \big(1- \theta(s_2 - \Lambda)- \theta(s_3 - \Lambda) + \theta(s_2 - \Lambda) \, \theta(s_3 - \Lambda) \big) \, a_{s_1 s_2 s_3}  \, ,
	\end{align}
\end{subequations}
and
\begin{align}
	a_{s_1 s_2 s_3} = W_3(s_1, s_2, s_3) =  \prod_{i=1}^3 \exp \left( {s_i \, \mu^2 \over 2} \right) \, .
\end{align} 
Notice the functions $W_n$ are totally symmetric in their arguments, whereas the stub kernels $r$ and $d$ have the expected counterparts of the symmetry properties~\eqref{eq:27}. Comparing with the twisted kernels~\eqref{eq:218} we can roughly identify $\Lambda \sim 1/L$.

\begin{figure}[t]
	\centering
	\includegraphics[height=2.5in,trim={0.75cm 0.5cm 1.25cm 1cm}, clip]{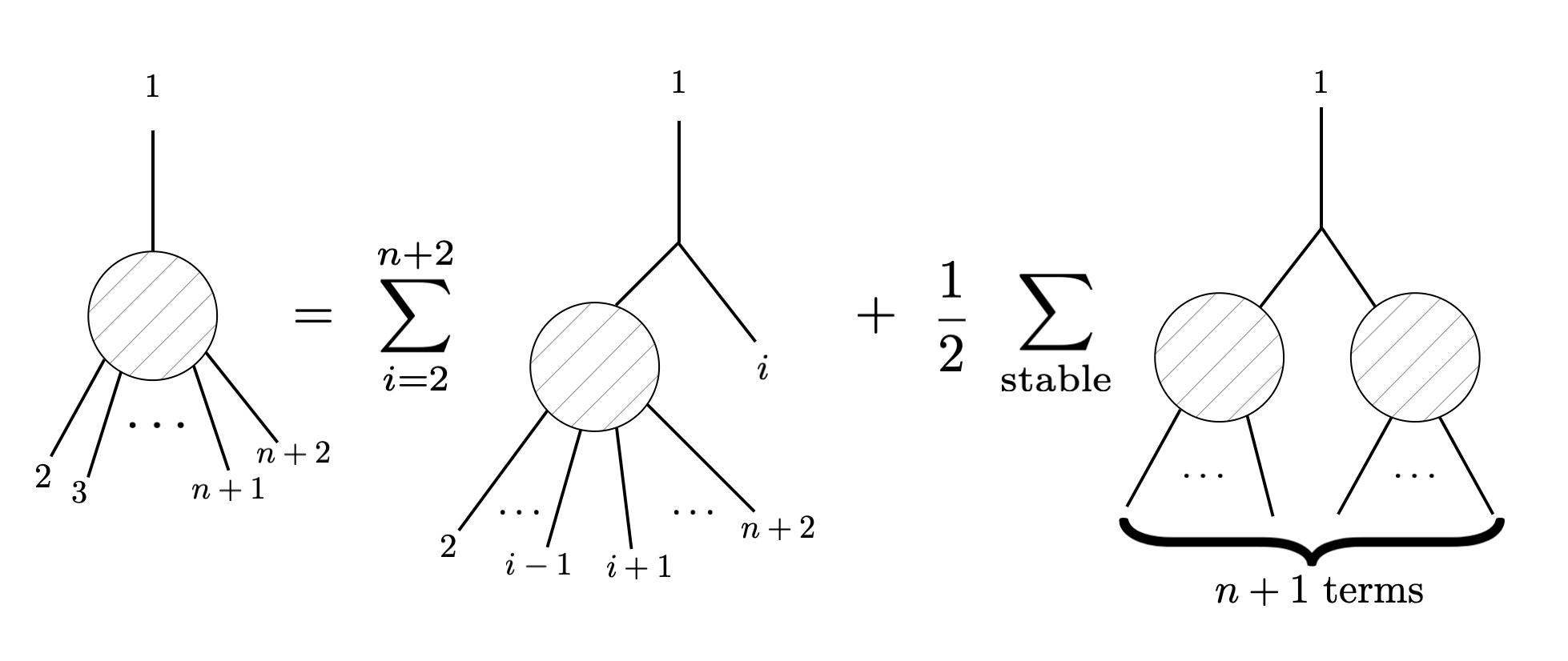}
	\caption{The recursion for the unordered rooted full binary trees with $n+1$ labeled leaves. Observe the similarity between the $R$-term and separating $D$-term excisions in figure~\ref{Mirz-gluing-figure}.}\label{fig:ct}
\end{figure}

The expression~\eqref{eq:5.13} can be derived by the recursive structure of the labeled binary trees in figure~\ref{fig:ct}. From the labeled binary trees we read the identity
\begin{align}
	b_n = {1 \over 2} \, \sum_{i=1}^n \binom{n+1}{i} \, b_{i-1} \, b_{n-i}
	= \sum_{i=2}^{n+2} b_{n-1} + {1 \over 2} \sum_{\text{stable}} b_{n_1} b_{n_2} \, ,
\end{align}
which can be also derived using~\eqref{eq:5.5}. Together with~\eqref{eq:5.12}, it is possible to include the rest of the terms in~\eqref{eq:5.12} through integrals like in~\eqref{eq:5.13}. We run the bounds of integration from $0$ to $\infty$, which requires introducing step functions as in~\eqref{eq:5.14}. So we see the stubbed scalar theory obeys a (classical) topological recursion~\eqref{eq:5.13} with the twisted kernels~\eqref{eq:5.14}.\footnote{Here we considered  the ``classical'' recursion for simplicity, however similar considerations can also be applied to the ``quantum'' recursion after one considers cubic graphs instead of trees.}

It is useful to look at two limiting cases of~\eqref{eq:5.14}. First, we see that $\Lambda = 0$ trivializes the recursion, $W_n(s_i) = 0$ for $n > 3$ and $W_3(s_i) = 1$, which corresponds to having the polynomial and manifestly local formulation. Such a limit doesn't exist in the stringy analog~\eqref{eq:4.18a} since there is an upper-bound bound on~\eqref{eq:3.39} $L \sim 1/ \Lambda$. This makes sense: the ordinary CSFT can't admit a polynomial and local formulation~\cite{Sonoda:1989sj}. On the opposite end, $\Lambda \to \infty$, we see~\eqref{eq:5.14} becomes well-defined upon analytic continuation. The stringy analog of this case is Ishibashi's recursion~\cite{Ishibashi:2022qcz}.

We can similarly cast the recursion~\eqref{eq:5.13} in the form of a second-order differential constraint. Introducing the following functionals of $\phi(s)$
\begin{align} \label{eq:5.17}
	z[\phi(s)] \equiv \exp \left( {w[\phi(s)]} \right) &\equiv \exp \left( \sum_{n=3}^\infty {\kappa^{n-2} \over n!} \int\limits_{0}^\infty d s_1 \cdots \int\limits_{0}^\infty  ds_n \,
	W_n(s_1, \cdots, s_n) \, \phi(s_1) \, \cdots \, \phi(s_n) 
	\right) \, ,
\end{align}
the relation~\eqref{eq:5.13} can be expressed compactly as
\begin{align} \label{eq:5.18}
	0 = {\delta w \over \delta \phi(t_1)} 
	&- {\kappa \over 2}  \,  \int\limits_{0}^\infty dt_2 \int\limits_{0}^\infty dt_3 \, a_{t_1 t_2 t_3} \, \phi(t_2) \, \phi(t_3) 
	-  \kappa \, \int\limits_{0}^\infty dt_2 \int\limits_{0}^\infty dt_3 \, r_{t_1 t_2 t_3} \, \phi(t_2) \,  {\delta w \over \delta \phi(t_3)} 
	\\ \nonumber 
	&\hspace{0.5in} -  {\kappa \over 2} \, \int\limits_{0}^\infty dt_2 \int\limits_{0}^\infty dt_3  \, d_{t_1 t_2 t_3} \left( {\delta w \over \delta \phi(t_2)} \right)
	\left( {\delta w \over \delta \phi(t_3)} \right) \, .
\end{align}
This is the analog of~\eqref{eq:4.32}. Note that we didn't need to place any constraint like~\eqref{eq:3.39} on the value of $\Lambda$ in contrast to its stringy counterpart---we can have any stub length. Note
\begin{align} \label{eq:5.21a}
	V(\phi; \Lambda ) = - {1 \over 2!} \, \mu^2 \phi^2 + w \left[\phi(s) = \phi \, \delta(s - \Lambda)\right]  \, ,
\end{align} 
see~\eqref{eq:6.4} and~\eqref{eq:5.17}.

We remark that~\eqref{eq:5.18} was a consequence of the geometric interpretation of stubs in terms of suitable binary trees. If we had used a different way of integrating out UV modes, which is choosing a different field parametrization, this geometric presentation would have been highly obstructed. This is the avatar of what we have discussed for the stringy case at the end of subsection~\ref{sec:4.2}. But again, the implications of the recursion are supposed to stay the same. 

\subsection{The implications}

Let us now investigate the consequences of~\eqref{eq:5.18}. First notice the functional derivative of the free energy $w[\phi(s)]$ evaluates to
\begin{align}
	{\delta w \over \delta \phi(t_1)} = 
	\sum_{n=3}^\infty \, {\kappa^{n-2} \over (n-1)!} \, \int\limits_{0}^\infty d s_1 \cdots \int\limits_{0}^\infty  ds_n \, W_n(t_1, \mathbf{s}) \, \phi(s_2) \, \cdots \, \phi(s_n) \, ,
\end{align}
after using the symmetry property of $W_n$. In particular consider taking $\phi(s_i) = \phi_\ast(\Lambda) \delta(s_i - \Lambda)$, where $\phi_\ast(\Lambda)$ is a solution to the stubbed theory of stub length $\Lambda/2$. We have
\begin{align} \label{eq:5.19}
	\varphi(t_1; \Lambda) 
	&\equiv {\delta w \over \delta \phi(t_1; \Lambda)}  \bigg|_{\phi \to \phi_\ast}
	 = \sum_{n=3}^\infty {\kappa^{n-2} \over (n-1)!} \, W_n(t_1, \mathbf{\Lambda}) \, \phi_{\ast}(\Lambda)^{n-1}\, ,
\end{align}
where we introduced $\varphi(t_1) $ for convenience. Here $\mathbf{\Lambda}$ indicates that all remaining entries are equal to $\Lambda$ here. We often suppress the $\Lambda$ dependence to simplify the presentation of the expressions.

\begin{figure}[t]
	\centering
	\includegraphics[height=1.8in,trim={0.5cm 0.5cm 0.5cm 1cm}, clip]{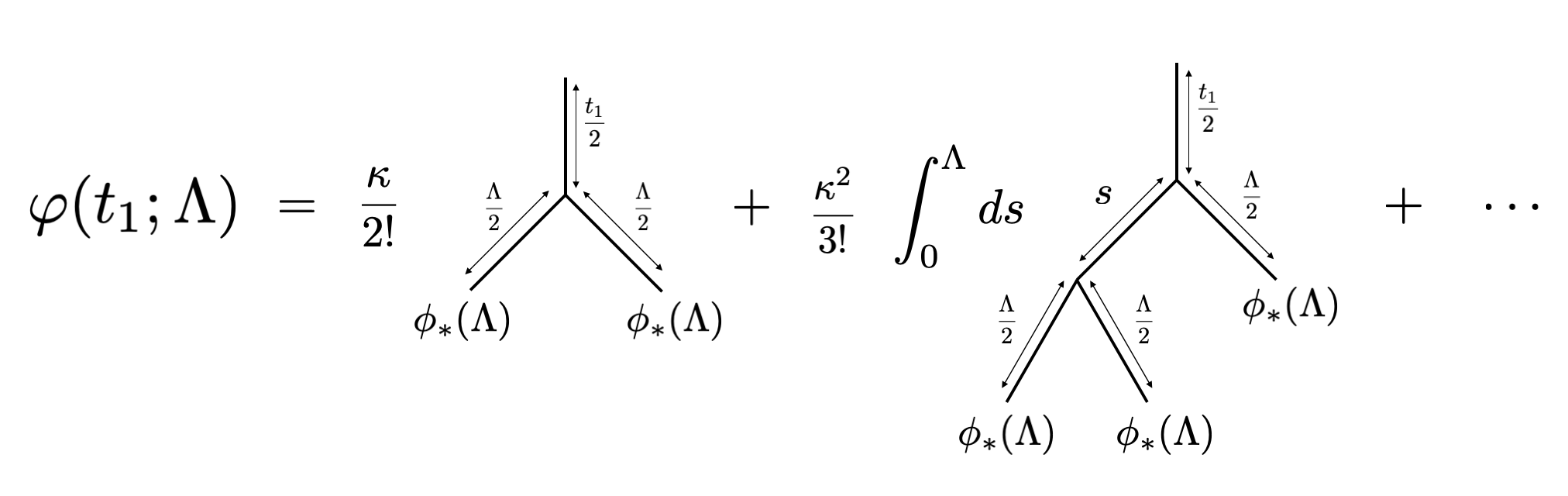}
	\caption{The graphical representation of the definition of the function $\varphi(t_1; \Lambda)$~\eqref{eq:5.19}.}\label{fig:def}
\end{figure}

Importantly, when $t_1= \Lambda$~\eqref{eq:5.19} evaluates to (see~\eqref{eq:5.21a})
\begin{align} \label{eq:5.21}
	\varphi(t_1 = \Lambda) 
	&=\sum_{n=3}^\infty {\kappa^{n-2} \over (n-1)!} \, W_n(s_i = \Lambda) \, \phi_{\ast}^{n-1}
	 \\ \nonumber
	&= {\partial V \over \partial \phi} \bigg|_{\phi = \phi_\ast}  + \mu^2 \phi_{\ast}
	= \mu^2 \phi_{\ast}
	\, ,
\end{align}
since $\phi_\ast$ is a solution and it extremizes the potential~\eqref{eq:6.4} by construction. Therefore one can think of $\varphi(t_1) $ as a function that captures the solution when $t_1 = \Lambda$, but generalizes it otherwise. The definition of this function is schematically shown in figure~\ref{fig:def}. 

\begin{figure}[t]
	\includegraphics[height=3in]{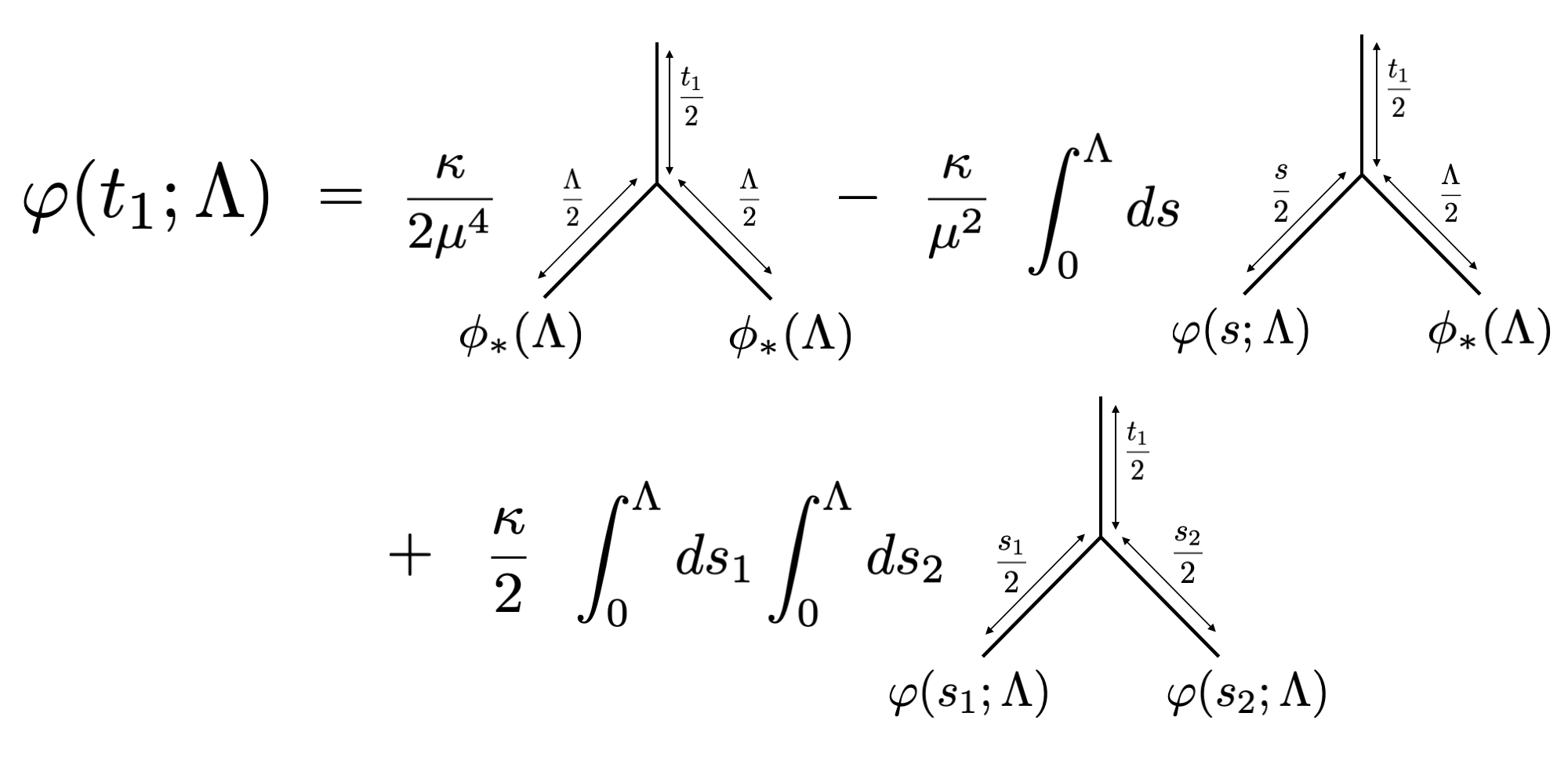}
	\caption{The graphical representation of~\eqref{eq:5.23}. This representation makes apparent that~\eqref{eq:5.23} is a statement of the recursion that goes into the definition of $\varphi(t_1; \Lambda)$, refer to figure~\ref{fig:def}.}\label{fig:def2}
\end{figure}

Taking $\phi(s_i) = \phi_\ast \, \delta(s_i - \Lambda) =( \varphi(\Lambda) / \mu^2 )\, \delta(s_i - \Lambda) $ in~\eqref{eq:5.18} then implies 
\begin{align}
	0 = \varphi(t_1)
	- {\kappa \over 2 \mu^4} \, a_{t_1 \Lambda\Lambda} \,  \varphi(\Lambda)^2
	&- {\kappa \over \mu^2}  \,  \int\limits_{0}^\infty dt_3  \, r_{t_1 \Lambda t_3} \, \varphi(\Lambda) \, \varphi(t_3)
	\\ \nonumber
	&\hspace{0.75in} -{\kappa  \over 2 }  \, \int\limits_{0}^\infty dt_2 \int\limits_{0}^\infty dt_3  \, d_{t_1 t_2 t_3} \, \varphi(t_2) \,  \varphi(t_3) \, ,
\end{align}
and we obtain a quadratic integral equation for $\varphi(t_1)$
\begin{align} \label{eq:5.23}
	0 = \varphi(t_1)
	&- {\kappa \over 2 \mu^4} \, \exp\left({ t_1 \mu^2 \over 2 }+ {\Lambda \mu^2 }\right) \, \varphi(\Lambda)^2
	\\ \nonumber
	&+ {\kappa\over \mu^2} \, \exp\left( {t_1 \mu^2 \over 2} + {\Lambda \mu^2 \over 2} \right)
	\, \varphi(\Lambda) \,
	\left[ \int\limits_{0}^\Lambda dt' \, \exp\left({t' \mu^2 \over 2}\right) \, \varphi(t') \right]
	\\
	&-{\kappa \over 2} \, \exp\left({t_1 \mu^2 \over 2}\right) \,  
	\left[ \int\limits_{0}^\Lambda dt' \, \exp\left( {t' \mu^2 \over 2}  \right)
	\varphi(t') \right]^2
	\nonumber
	\, ,
\end{align}
after substituting the stub kernels~\eqref{eq:5.14}. The schematic representation of this quadratic integral equation is shown in figure~\ref{fig:def2}.

Observe that setting $t_1 = \Lambda = 0$ above produces $V'(\phi) = 0$, see~\eqref{eq:6.2}. This suggests~\eqref{eq:5.23} can be understood as a resummation of the equation of motion $V'(\phi; \Lambda) = 0$ of the stubbed theory. We can iteratively insert $\varphi(t_1)$ into itself and expand in $\kappa$ to find 
\begin{align} \label{eq:5.26}
	0 
	&= \varphi(t_1) 
	- {\kappa \over 2} \, \exp\left({ t_1 \mu^2 \over 2 }+ {\Lambda \mu^2}\right) \, \phi_\ast^2
	- {\kappa^2 \over 2} \, \exp\left( {t_1 \mu^2 \over 2} + {3 \Lambda \mu^2 \over 2} \right)
	\, 
	\left( {1- e^{\Lambda \mu^2} \over \mu^2} \right)
	\, \phi_\ast^3 
	+ \cdots \, .
\end{align}
Upon taking $t_1 = \Lambda$ and using~\eqref{eq:5.21} one indeed obtains $V'(\phi; \Lambda) = 0$ in the expanded form, see~\eqref{eq:6.4}. The argument here can be generalized to higher orders in $\kappa$.

In fact, upon close inspection we see
\begin{align} \label{eq:5.27}
	\varphi(t_1) = {2 \mu^4 \over \kappa} \exp\left( {t_1 \mu^2 \over 2} \right) \, ,
\end{align}
satisfies~\eqref{eq:5.23} and we again obtain the solution~\eqref{eq:5.11} for the stubbed theory
\begin{align} \label{eq:5.28}
	\phi_\ast = {1 \over \mu^2} \, \varphi(t_1 = \Lambda ) 
	= {2 \mu^2 \over \kappa} \exp\left( {\Lambda \mu^2 \over 2} \right) \, .
\end{align}
This demonstrates the differential constraints encoding the topological recursion~\eqref{eq:5.18} in a theory can be used to find its solutions. Of course, we have found $\varphi(t_1)$ by guesswork, while it can be quite challenging to do so for more complicated situations. We highlight resumming the potential $V(\phi; \Lambda)$~\eqref{eq:6.4} as in~\cite{Erler:2023emp} wasn't necessary to obtain this solution: we were able to access beyond the radius of converges $\phi > r (\Lambda)$~\eqref{eq:5.12a} directly.

This approach, however, does not give a novel method for finding the energy difference between the solutions since this calculation still has to involve the resummed potential. On the other hand, it is possible to find the mass of the linear fluctuations around the vacuum~\eqref{eq:5.28} based on the recursion alone. In order to do that we vary $\varphi \to \varphi + \delta \varphi$ in~\eqref{eq:5.23} for which we find
\begin{align} \label{eq:5.29}
	- {1 \over \mu^2} M^2 (t_1) \, \delta \varphi(t_1)  =
	\delta \varphi(t_1) 
	&- 2 \, \exp \left( {t_1 \mu^2 \over 2} + {\Lambda \mu^2 \over 2} \right)
	\delta \varphi(\Lambda)
	\\ \nonumber
	&\hspace{0.5in}+ 2 \, \mu^2 \exp\left({t_1 \mu^2 \over 2}\right) \, \int\limits_{0}^\Lambda dt' \, 
	\exp\left({t' \mu^2 \over 2}\right) \, \delta \varphi(t') \, ,
\end{align}
using~\eqref{eq:5.27}. We suggestively named this variation $M^2 (t_1) \delta \varphi(t_1) /\mu^2 $ given that~\eqref{eq:5.23} was essentially the resummed $V'(\phi; \Lambda) = 0$. Its variation can be understood as the resummed version of $-V''(\phi; \Lambda)$ and it should encode the spectrum of the linear fluctuations around $\phi = \phi_\ast$. The overall sign follows from the choice of signs in~\eqref{eq:5.19} and the division by $\mu^2$ is to get the correct units for $M^2$. We have $\delta \varphi(\Lambda) = \mu^2 \delta \phi$ for the genuine fluctuations.

Indeed, by taking 
\begin{align} \label{eq:5.30}
	\delta \varphi(t_1) = 
	\exp \left( {t_1 \mu^2 \over 2}  \right) \epsilon \, ,
\end{align}
with infinitesimal $\epsilon$ without dependence on $t_1$, we see
\begin{align} \label{eq:5.31}
	- {1 \over \mu^2} M^2 \, \delta \varphi(\Lambda ) = - \delta \varphi(\Lambda) 
	\quad \implies \quad
	M^2 \, \delta \phi = \mu^2 \, \delta \phi
	\, ,
\end{align}
so the squared mass of the fluctuations around the nonperturbative vacuum is $M^2 = \mu^2 > 0$---as it should be. We guess the particular form of~\eqref{eq:5.30} for $\delta \varphi(t_1)$ as a function of $t_1$ by getting inspired from~\eqref{eq:5.27}, but the linearity in $\delta \varphi(t_1)$ of~\eqref{eq:5.31} justifies it a posteriori in any case. We again emphasize that we didn't need to resum the potential to derive~\eqref{eq:5.31}. In fact, obtaining this result from the resummed form~\eqref{eq:5.9} would have been quite intricate due to the complicated higher derivative structure of the stubbed theory.

Given that the integral equation~\eqref{eq:5.23} is a consequence of the differential constraint and topological recursion, it is natural to wonder whether there is an analog of the integral equation in hyperbolic CSFT as a result of~\eqref{eq:4.32}. Indeed, we are going to see there is such an equation in the upcoming section.

\section{Quadratic integral equation for hyperbolic CSFT} \label{sec:6}

Now we apply the reasoning from the last section to derive a quadratic integral equation for hyperbolic CSFT following from~\eqref{eq:4.32}. We adapt the conventions from before and we are going to be brief in our exposition since the manipulations are very similar to the previous section. 

We begin with the derivative of $W[\Psi(\ell)]$. Upon substituting
\begin{align} \label{eq:6.1}
	\Psi(\ell_i)  =  \Psi_\ast  \, \delta(\ell_i - L) \, ,
\end{align}
into~\eqref{eq:4.27} we get
\begin{align} \label{eq:6.2a}
		\Phi (L_1)
		\equiv {\delta W \over \delta \Psi(L_1)}  \bigg|_{\Psi  \to \Psi_\ast}
		&= \sum_{n=3}^\infty {\kappa^{n-2} \over (n-1)!} \, 
		\bigg( \id \otimes \langle \mathcal{V}_{0,n} (L_1, \mathbf{L}) | \bigg) \,
		\bigg( | \omega^{-1} \rangle \otimes | \Psi_{\ast} \rangle^{\otimes (n-1)} \bigg) 
		\\ \nonumber
		&= \sum_{n=2}^\infty {\kappa^{n-1} \over n !} \, L_{0,n} \left( \Psi_\ast^{n}; L_1 \right) \,  .
\end{align}
Here $ \Psi_\ast = \Psi_\ast(L) $ is a critical point of the classical CSFT action and we defined the string field $\Phi (L_1) = \Phi (L_1; L)$ accordingly. These string fields are even and assumed to have ghost number 2.

We also introduce a new set of (graded-symmetric) string products $L_{g,n-1}(L_1): \widehat{\mathcal{H}}^{\otimes (n-1)} \to \widehat{\mathcal{H}}$
\begin{align}
	L_{g, n-1} (L_1) =
	\bigg( \id \otimes \langle \mathcal{V}_{g,n} (L_1, \mathbf{L}) | \bigg) \,
	\bigg( | \omega^{-1} \rangle \otimes \id^{\otimes (n-1)} \bigg)  \, ,
\end{align}
that generalize the ordinary string products $L_{g,n}$~\eqref{eq:2.47} such that the border length of the output is $L_1$ instead of $L$. They are equivalent when $L_1 = L$. Note that we have
\begin{align} \label{eq:6.4b}
	\Phi (L_1 = L)   = - Q_B \, \Psi_\ast \, ,
\end{align}
by the equation of motion. We therefore see $\Phi (L_1 = L) $ is related to $Q_B$ acting on the solutions. This definition can also be schematically understood like in figure~\ref{fig:def}.

We remind the reader there is a gauge redundancy for the solutions $\Psi_\ast $~\eqref{eq:2.48b}, which indicates there should be a redundancy in the choice of $\Phi (L_1)$ as well. We demand this is given by
\begin{align} \label{eq:6.5}
	\delta_\Lambda \, \Phi (L_1)  =
	\sum_{n=2}^\infty {\kappa^{n-1} \over (n-1)!} \, L_{0,n} \left( \delta_\Lambda \Psi_\ast, \Psi_\ast^{n-1}; L_1 \right) \, ,
\end{align}
so that the definition~\eqref{eq:6.2a} is invariant under~\eqref{eq:2.48b}. Here $\delta_\Lambda \Psi_\ast $ is given by~\eqref{eq:2.48b}. Upon taking $L_1 = L$ and using the $L_\infty$ relations~\cite{Erler:2019loq} this gauge transformation simplifies further and can be seen to be consistent with~\eqref{eq:6.4b}.

We need to fix this gauge symmetry in order to invert~\eqref{eq:6.4b} and write down an integral equation. We do this by~\emph{assuming} a grassmann odd operator $\beta (L_1) = \beta(L_1; L) $, possibly to be constructed with $b$-ghost modes and depending on $L_1$, satisfying\footnote{We take $\beta = \beta(L_1)$ so that it indicates $\beta$ is to be used for fixing the gauge redundancy~\eqref{eq:6.5} for all $L_1$.} 
\begin{align} \label{eq:6.6}
	\left\{ Q_B, \beta(L_1 = L) \right\} = -P \,  ,
\end{align}
where $P$ is the projector to the complement of the cohomology of $Q_B$. We impose the gauge
\begin{align} \label{eq:6.7}
	\beta(L_1 = L) \, \Psi_\ast  = 0 \, ,
\end{align}
on the CSFT solutions. As a result the equation~\eqref{eq:6.4b} gets inverted to
\begin{align}
	\beta \,  \Phi (L_1 = L)  =  \Psi_\ast \, .
\end{align}
Note that we have $\beta \, \Phi (L_1) \neq 0 $ and $\{ Q_B, \beta(L_1) \} \neq - 1$ for $L_1 \neq L$ in general. We have taken $P \to 1$ above and ignored the subtleties associated with the on-shell modes.

Taking~\eqref{eq:6.1} in the differential constraint~\eqref{eq:4.32} then implies 
\begin{align} \label{eq:6.4a}
	0 = \Phi (L_1) 
	- {\kappa \over 2} \, 
	A_{L_1 L L } \big( \beta \, \Phi (L) , \beta \, \Phi (L) \big)
	&- \kappa \, \int\limits_{-\infty}^\infty d \ell \, 
	B_{L_1 L \ell}\left( \beta \, \Phi (L), \Phi(\ell)\right)
	\\ \nonumber
	&\hspace{0.75in}- {\kappa \over 2} \, \int\limits_{-\infty}^\infty d \ell_1 \, \int\limits_{-\infty}^\infty d \ell_2 \,
	C_{L_1 \ell_1 \ell_2}\left(\Phi(\ell_1), \Phi(\ell_2) \right) \, ,
\end{align}
where the $2$-products $A,B,C$ are already defined in~\eqref{eq:4.24}. The similarity to~\eqref{eq:5.23} is apparent, also observe the structure in figure~\ref{fig:def2}. This is the quadratic integral equation and it makes the cubic nature of CSFT manifest.

Like earlier, iteratively substituting $\Phi (L_1) $ into this equation, taking $L_1 = L$, and expanding in $\kappa$ we obtain the equation of motion for CSFT~\eqref{eq:2.48a} using~\eqref{eq:4.18a}.  A similar phenomenon occurs for the stubs~\eqref{eq:5.26} and it is associated with resummation as discussed. So we again interpret~\eqref{eq:6.4a} as the resummation of the CSFT equation of motion. Having such a form is reassuring: trying to resum the action~\eqref{eq:A1.11} as in~\eqref{eq:5.9} would have been quite a daunting task, bordering on the impossible. 

Being able to recast the CSFT equation of motion to a quadratic integral equation~\eqref{eq:6.4a} is encouraging. It is explicitly~\emph{and} exclusively given  in terms of the generalized hyperbolic three-vertex of~\cite{Firat:2021ukc} and the twisted Mirzakhani kernels~\eqref{eq:218}. It may be viable to attempt to solve this integral equation in the future after identifying a convenient gauge-fixing operator $\beta(L_1)$. Furthermore, it should be possible to read the spectrum of the linear excitations (i.e., cohomology) around the solution by varying~\eqref{eq:6.4a} like in~\eqref{eq:5.29} and repeating a similar analysis.\footnote{On the other hand we don't have a new way to compute the on-shell value of the action as before. This quantity vanishes up to boundary terms in CSFT~\cite{Erler:2022agw}.} However, we expect the form of solutions and their variations would be more complicated than the stub counterparts~\eqref{eq:5.27} and~\eqref{eq:5.30}. We comment on this approach more in the next section.

\section{Discussion} \label{sec:disc}

In this paper:
\begin{enumerate}
	
	\item We identified a background-independent topological recursion relation satisfied by the elementary vertices of hyperbolic closed string field theory, refer to~\eqref{eq:4.18} or~\eqref{eq:4.29}. This explicitly demonstrates closed string field theory has an underlying cubic structure and makes its features manifest. The moduli integrations are simplified considerably. The recursion relation satisfied among the volumes of the systolic subsets~\eqref{eq:218A} provides the essential ingredients for its closed string field theory counterpart.
		
	\item We showed a suitable generalization of the classical solutions to closed string field theory obey a quadratic integral equation~\eqref{eq:6.4a} as a consequence of the aforementioned topological recursion. We discussed how this equation can be used to construct analytic solutions in principle. 
	
\end{enumerate}

\noindent There are numerous interesting directions left to be investigated in future work. We list the ones that we find the most appealing and exciting:

\begin{enumerate}
	
	\item Developing a systematic approach for solving the quadratic integral equation~\eqref{eq:6.4a} is the natural next step. Even though this equation appears to contain sufficient information to construct solutions based on our investigations in the stubbed scalar theory, it is far from clear how to obtain them beyond guessing judiciously. Identifying a convenient gauge choice~\eqref{eq:6.7} and the smallest subspace for which a solution to~\eqref{eq:6.4a} exists are possibly among the first prerequisites for making any further progress. Understanding the underlying algebraic structure to our reformulation may also help.
	
	\item Even if a solution is constructed, it is somewhat unclear how to probe its physics. For example, it has been established that the on-shell action vanishes up to boundary terms in closed string field theory~\cite{Erler:2022agw}, thanks to the dilaton theorem~\cite{Bergman:1994qq}. We don't have a novel way to test this. However, investigating the spectrum around the background may still be doable with a correct approach. One may try to establish (or rule out) the closed string version of Sen's third conjecture~\cite{Sen:1999mh,Sen:1999xm,Ellwood:2006ba} for the vanishing cohomology around the tachyon vacuum for instance.
	
	\item The form of the differential constraint encoding the topological recursion~\eqref{eq:4.29} is suggestive. We have already pointed out its possible interpretation as a quantum Airy structure~\cite{kontsevich2017airy,Andersen:2017vyk}. It would be interesting to investigate this further to see whether a deeper principle and/or theory is lurking behind here and how it relates to the nonperturbative structure of closed string field theory. There may be connections to the ideas of~\cite{costello2007topological} for example. 
	
	\item It may also be useful to illuminate how our approach compares to the themes in the relation between matrix models~\cite{Eynard:2007fi} and Jackiw-Teitelboim gravity~\cite{Saad:2019lba,Stanford:2019vob,Post:2022dfi,Altland:2022xqx}. Here, the initial point of investigation would be understanding and clarifying the matrix model interpretation of the systolic volumes and their recursion~\eqref{eq:218A}.
	
	\item We have seen the string fields are naturally endowed with an additional positive real parameter: the length of the geodesic border $L_i$~\eqref{eq:4.23}. A similar feature also appears in the lightcone SFT~\cite{Kaku:1974zz, Kaku:1974xu, hata1986covariant, Hata:1986kj, Kugo:1992md, Erler:2020beb, Bernardes:2024ncs}, where the lightcone momentum $k_-$ relates to the border length. It may be useful to understand how similar their roles are in the interactions.
	
	\item Physics following from the recursion shouldn't change under field redefinition even though a particular form of the relations will be no longer manifest. Nevertheless, it is still rather counter-intuitive that our derivation falls short of establishing a recursion for all classical hyperbolic vertices---especially for the polyhedral vertices based on Strebel quadratic differentials~\cite{Saadi:1989tb,Firat:2023glo,Moeller:2004yy,Moeller:2006cw,Moeller:2007mu,Erbin:2022rgx}. Such a recursion may still exist considering the work~\cite{Ishibashi:2024kdv} and it may be possible to find a generalization to our argument to cover these scenarios as well. Relatedly, it is desirable to establish a recursion relation for other hyperbolic theories~\cite{Cho:2019anu,Firat:2023gfn} and/or their supersymmetric versions~\cite{Pius:2018pqr}.
	
\end{enumerate}

\noindent  We hope these results help construct closed string field theory solutions some day.

\section*{Acknowledgments}

We are grateful to Scott Collier, Harold Erbin, Ted Erler, Daniel Harlow, David Kolchmeyer, Raji Mamade, and Edward Mazenc for many enlightening discussions; Barton Zwiebach for his comments on the early draft and discussions; and Nobuyuki Ishibashi for the correspondence. NVM in particular thanks Daniel Harlow for the freedom and encouragement to explore. 

The work of AHF is supported by the U.S. Department of Energy, Office of Science, Office of High Energy Physics of U.S. Department of Energy under grant Contract Number DE-SC0012567, DE-SC0009999, and the funds from the University of California. NVM is supported by the Hertz Foundation and the MIT Dean of Science Fellowship.

\appendix

\section{Hyperbolic three-string vertex} \label{app:H3V}

In this appendix we summarize the local coordinates for the generalized hyperbolic string vertex $\langle \Sigma_{0,3} (L_1, L_2, L_3)|$ for $L_1,L_2,L_3 \geq 0$. The reader can refer to~\cite{Firat:2021ukc,Hadasz:2003he} for the details of the derivation using the connection between hyperbolic geometry on the three-bordered sphere and the hypergeometric equation. We also comment on various branch choices in this appendix.

We begin with a three-punctured sphere whose punctures are placed at $z=0,1,\infty$ and there are coordinate patches
\begin{align} \label{eq:A.1}
	D_i = \big\{ z \in \mathbb{C} \; \; \big| \; \; 0 < |w_i(z; L_1, L_2, L_3)| \leq 1 \big\} \,  ,
\end{align}
around them such that the surface
\begin{align}
	\Sigma_{0,3}(L_1, L_2, L_3) = \mathbb{C} \setminus \bigcup_{i=1}^3 D_i \, , 
\end{align}
endows a regular hyperbolic metric with geodesic borders of length $L_i = 2 \pi \lambda_i$. The local coordinate maps $w_1(z; L_1, L_2, L_3)$ in~\eqref{eq:A.1} are given by
\begin{align} \label{eq:A.3}
	w_1(z; L_1, L_2, L_3) &= 
	\, {z \, (1-z)^{-{\lambda_2 / \lambda_1}} \over \rho(L_1, L_2, L_3 )}\, \left[
	{
		_2F_1\left( { 1+ i \lambda_1 - i \lambda_2 + i \lambda_3 \over 2}, 
		{ 1+ i \lambda_1 - i \lambda_2 - i \lambda_3 \over 2};
		1 + i \lambda_1; z \right) \over 
		_2F_1\left( { 1- i \lambda_1 + i \lambda_2 - i \lambda_3 \over 2}, 
		{ 1- i \lambda_1 + i \lambda_2 + i \lambda_3 \over 2};
		1 - i \lambda_1; z \right) 
	}
	\right]^{1 / i \lambda_1} \nonumber \\
	&= {1 \over \rho_1} \, \left[ z + 
	{1 + \lambda_1^2 + \lambda_2^2 - \lambda_3^2 \over 2 \, (1+\lambda_1^2)} \,  z^2 + \cdots
	\right] \, ,
\end{align}
where the expression for the mapping radius $\rho_1 = \rho (L_1, L_2, L_3 )$ is
\begin{align} \label{eq:R}
	\rho (L_1, L_2, L_3 ) = e^{-{\pi / 2 \lambda_1} } \left[
	{\Gamma(1-i \lambda_1)^2 \over \Gamma(1+i \lambda_1)^2 } 
	{ \gamma\left({1 + i\lambda_ 1 + i \lambda_2 + i \lambda_3 \over 2}  \right)
		\gamma\left({ 1 + i\lambda_ 1 - i \lambda_2 + i \lambda_3 \over 2} \right)
		\over 
		\gamma\left( {1 - i\lambda_ 1 - i \lambda_2 + i \lambda_3 \over 2}  \right)
		\gamma\left({ 1 - i\lambda_ 1 + i \lambda_2 + i \lambda_3) \over 2 } \right)  } 
	\right]^{ i / 2 \lambda_1} \, .
\end{align}
Here $_2F_1(a,b;c;z)$ is the hypergeometric function and $\Gamma(x)$ is the gamma function. The latter also goes into the definition of
\begin{align}
	\gamma(x)  = {\Gamma(x) \over \Gamma(1-x)} \, .
\end{align}
Take note the mapping radius $\rho_1$ diverges as $\sim e^{- 1 / L_1}$ as $L_1 \to 0$. In the inverted form, the expansion~\eqref{eq:A.3} is given by
\begin{align}
	z(w_1; L_1, L_2, L_3) = 
	\rho_1  w
	- {1 + \lambda_1^2 + \lambda_2^2 - \lambda_3^2 \over 2 \, (1+\lambda_1^2)}  \,
	(\rho_1 w )^2 
	+ \cdots \, .
\end{align}
Unfortunately the closed-form expressions for this inverse function is not known except for $L_i \to 0$. The expressions for the local coordinates $w_2$ and $w_3$ around $z=1$ and $z=\infty$ are given by
\begin{subequations}
\begin{align}
	&w_2(z; L_1, L_2, L_3) = w_1 (1-z \, ; L_2, L_1, L_3) \, , 
	\\
	&w_3(z; L_1, L_2, L_3) = w_1 \left( {1 \over z} \, ; L_3, L_2, L_1 \right) \, ,
\end{align}
\end{subequations}
up to possible global phases. 

It is important to be mindful about the various branch choices for the imaginary exponents in the mapping radius~\eqref{eq:R}. This has been partially investigated in~\cite{Firat:2021ukc}, however we briefly comment on them again here. First notice there is an overall ambiguity in the mapping radii
\begin{align}
	\rho(L_1,L_2, L_3) \simeq 
	\exp\left( { n \, \pi \over \lambda_1 } \right) \,
	\rho(L_1,L_2, L_3)  \, 
	\quad \quad
	n \in \mathbb{Z} \, .
\end{align}
as a result of the imaginary exponent. If one would like to work in a particular branch for the local coordinates then one has to find the correct integer $n = n (L_1,L_2,L_3)$. It has been argued that the correct choice is $n=0$ when $L_1=L_2 = L_3$ together with the choice of principal branch for the branches~\cite{Firat:2021ukc}. However there may be branch crossings for generic border lengths. Since the mapping radius should be continuous in $L_i$ the integer $n$ should be adjusted accordingly when a branch cut is crossed. In the expressions~\eqref{eq:R} we assume such adjustments are implicitly present.

\section{Sample computations for the systolic volumes} \label{app:B}

In this appendix we perform sample computations for the systolic volumes using the recursion~\eqref{eq:218A}. We have already computed $(g,n)=(0,4),(1,1)$ in the main text, see~\eqref{eq:321} and~\eqref{eq:322}. Using them we can compute the systolic volumes for the surfaces with $\chi_{g,n} = -3$ and $n \geq 1$:
\begin{enumerate}
	\item \underline{Five-bordered sphere $g=0, n=5$.} Writing the recursion~\eqref{eq:218A} explicitly we have
	\begin{align} 
		L_1 \cdot V \mathcal{V}_{0,5} (L_i) &= 
		\sum_{i=2}^5 \int\limits_{0}^\infty \ell d \ell \, \widetilde{R}_{L_1L_i\ell} \, V \mathcal{V}^L_{0,4}\left(\ell, \mathbf{L} \setminus \{L_i \} \right) 
		+ {3} \int\limits_{0}^\infty \ell_1 d \ell_1 \, \int\limits_{0}^\infty \ell_2 d \ell_2 \,
		\widetilde{D}_{L_1 \ell_1 \ell_2} 
		\, ,
	\end{align}
	after taking $V \mathcal{V}^L_{0, 3} (L_i) =1$.  This can be expressed as
	\begin{align}
		L_1 \cdot V \mathcal{V}^L_{0,5} (L_i)  &= L_1 \cdot V \mathcal{M}_{0,5} (L_i) 
		- {3 \over 2}\, L^2 \, \sum_{i=2}^5 \, \int\limits_{0}^\infty \ell d \ell \, R_{L_1L_i\ell} 
		- L_1 \, \sum_{i=2}^5 \, \int\limits_{0}^L \, \ell d \ell \, \mathcal{V}^L_{0,4}\left(\ell, \mathbf{L} \setminus \{L_i \} \right) \nonumber \\
		& \hspace{0.5in} - 6 \, \int\limits_{0}^L \ell_1 d \ell_1 \, \int\limits_{0}^\infty \ell_2 d \ell_2 \, R_{L_1 \ell_1 \ell_2}
		+  3 L_1 \, \int\limits_{0}^L \ell_1 d \ell_1 \, \int\limits_{0}^L \ell_2 d \ell_2 \, ,
	\end{align}
	using the twisted Mirzakhani kernels~\eqref{eq:218}. Focus on the integral
	\begin{align} \label{eq:B.3}
		I_{L_1 L_i} \equiv \int\limits_{0}^\infty \ell d \ell \, R_{L_1 L_i \ell} 
		= \int\limits_{0}^\infty \ell d \ell
		\left[
		L_1 -
		\log \left({\cosh\left({L_i \over 2}\right) + \cosh\left({L_1 + \ell \over 2}\right) \over 
		\cosh\left({L_i \over 2}\right) + \cosh\left({L_1 - \ell \over 2}\right)}\right)
		\right] \, .
	\end{align}
	Clearly this integral converges. It can be evaluated by taking the derivative of both sides with respect to $L_1$ and using the identity~\eqref{eq:38}
	\begin{align}
		{\p  I_{L_1 L_i}  \over \p L_1} &= {1 \over 2}\int\limits_{0}^\infty \ell d \ell  \sum_{\epsilon_1, \epsilon_i = \pm} 
		\left[ 1 + \exp\left({1 \over 2} \left({\ell + \epsilon_1 L_1+ \epsilon_i L_i }\right)\right) \right]^{-1}
		 = {1 \over 2} \left( L_1^2 + L_i^2\right) + {2 \pi^2 \over 3} \\
		&\implies I_{L_1 L_i} = {1\over 6} L_1^3  + {1 \over 2}L_1 L_i^2 + {2 \pi^2 \over 3} L_1
		\, . \nonumber
	\end{align}
	The integration constant fixed by noticing $I_{L_1=0,L_i} =0$, see~\eqref{eq:B.3}. Then we have
	\begin{align}
		\int\limits_{0}^L \ell_1 d \ell_1 \,  I_{L_1 \ell_1} = 
		{1 \over 12} L_1^3 L^2 + {1 \over 8}L_1 L^4  + {\pi^2 \over 3}  L_1 L^2 \, ,
	\end{align}
	and
	\begin{align}
		\int\limits_{0}^L \ell d \ell \, \mathcal{V}^L_{0,4}\left(\ell, \mathbf{L} \setminus \{L_i \} \right)
		&= 	\int\limits_{0}^L \ell d \ell \bigg[
		2 \pi^2 +{1 \over 2} \ell^2 +  {1 \over 2} \sum_{\substack{j=2 \\  j \neq i}}^5 L_j^2 - {3 \over 2} L^2
		\bigg]  \\
		&= \pi^2 L^2 +  {1 \over 4} L^2 \sum_{\substack{j=2 \\  j \neq i}}^5 L_j^2 - {5 \over 8} L^4 \, . \nonumber
	\end{align}
	Combining these integrals we obtain
	\begin{align}
		L_1 \cdot V \mathcal{V}^L_{0,5} (L_i)  &= L_1 \cdot V \mathcal{M}_{0,5} (L_i) 
		- {3\over 2} L^2 \sum_{i=2}^5 \left[ {1\over 6} L_1^3  + {1 \over 2} L_i^2 L_1 + {2 \pi^2 \over 3} L_1\right]
		\\
		&\hspace{-1in} - L_1 \sum_{i=2}^5 \bigg[
		\pi^2 L^2 +  {1 \over 4} L^2 \sum_{\substack{j=2 \\  j \neq i}}^5 L_j^2 - {5 \over 8} L^4
		\bigg] 
		- 6 \left[ {1 \over 12} L_1^3 L^2 + {1 \over 8}L_1 L^4  + {\pi^2 \over 3}  L_1 L^2 \right] 
		+ {3  \over 4} L_1 L^4
		\nonumber \\
		&= L_1 \cdot V \mathcal{M}_{0,5} (L_i) - {L_1^3 L^2} - {3 \over 4} L_1 L^2 \sum_{i=2}^5 L_i^2 - 4 \pi^2 L_1 L^2
		- 4 \pi^2 L_1 L^2 
		\nonumber \\
		&\hspace{-1in} - {3 \over 4} L_1 L^2 \sum_{i= 2}^5 L_i^2 + {5 \over 2} L_1 L^4
		- {1 \over 2} L_1^3 L^2 - {3 \over 4} L_1 L^4 - 2\pi^2 L_1 L^2 + {3 \over 4} L_1 L^4
		\nonumber \\
		&= L_1 \cdot V \mathcal{M}_{0,5} (L_i)  - {3 \over 2}  {L_1^3 L^2} - {3 \over 2} L_1 L^2 \sum_{i=2}^5 L_i^2 - 10\pi^2 L_1 L^2 + {5 \over 2} L_1 L^4 \, , \nonumber
	\end{align}
	and the final result is
	\begin{align} \label{eq:B.8}
		V \mathcal{V}^L_{0,5} (L_i)   =
		V \mathcal{M}_{0,5} (L_i)  - L^2 \, \left[ {3 \over 2} \sum_{i=1}^5 L_i^2  + 10 \pi^2 \right]
		+ {5 \over 2} L^4 \, .
	\end{align}
	This expression is symmetric under permutations of $L_i$, as it should be. Observe that the term subtracted from $V V \mathcal{M}_{0,5} (L_i)$ to get $V \mathcal{V}^L_{0,5} (L_i) $ is quite nontrivial unlike~\eqref{eq:321} and~\eqref{eq:322}.
	
	\item \underline{Two-bordered torus $g=1, n=2$}. We begin by writing~\eqref{eq:218A} specific to this case
	\begin{align}  
		L_1 \cdot V \mathcal{V}^L_{1,2} (L_i) &= 
		\int\limits_{0}^\infty \ell d \ell \, \widetilde{R}_{L_1L_2\ell} \, V \mathcal{V}^L_{1,1}\left(L_1 \right) 
		+ {1 \over 2} \int\limits_{0}^\infty \ell_1 d \ell_1 \, \int\limits_{0}^\infty \ell_2 d \ell_2 \,
		\widetilde{D}_{L_1 \ell_1 \ell_2}
		\, .
	\end{align}
	after using $V \mathcal{V}^L(\ell_1, \ell_2,L_i) = 0$. The twisted Mirzakhani kernels~\eqref{eq:218} produce
	\begin{align}  
		L_1  \cdot V \mathcal{V}^L_{1,2} (L_i) &= 
		L_1 \cdot V \mathcal{M}_{1,2} (L_i)
		- {1 \over 4} L^2 \int\limits_{0}^\infty \ell d\ell R_{L_1 L_2 \ell}
		- L_1 \int\limits_{0}^L \ell d \ell \, V \mathcal{V}^L_{1,1}\left(\ell \right) 
		\\ \nonumber
		&\hspace{0.75in}-\int\limits_{0}^L \ell_1 d \ell_1 \, \int\limits_{0}^\infty \ell_2 d \ell_2 \,
		R_{L_1 \ell_1 \ell_2}
		+  {1 \over 2 } L_1 \int\limits_{0}^L \ell_1 d \ell_1 \, \int\limits_{0}^L \ell_2 d \ell_2
		\, .
	\end{align}
	This is almost the same as the case considered above, the only difference being that we have to use $V \mathcal{V}^L_{1,1}\left(\ell \right) = \pi^2/12 + \ell^2/48 - L^2 /4$~\eqref{eq:322} now. We perform the integral
	\begin{align}
		\int\limits_{0}^L \ell d \ell \, V \mathcal{V}^L_{1,1}\left(\ell \right) =
		\int\limits_{0}^L \ell d \ell \left[ {\pi^2 \over 12} +{ \ell^2 \over 48} -{L^2 \over 4} \right]
		= {\pi^2 \over 24} L^2 - {23 \over 192} L^4\, ,
	\end{align}
	and obtain
	\begin{align}
		L_1 \cdot V \mathcal{V}^L_{1,2} (L_i) &=
		L_1 \cdot V \mathcal{M}_{1,2} (L_i) 
		- {1 \over 4} L^2 \left[ {1\over 6} L_1^3  + {1 \over 2} L_1 L_2^2 + {2 \pi^2 \over 3} L_1\right]
		- L_1 \left[{\pi^2 \over 24} L^2  - {23 \over 192} L^4 \right]
		\nonumber \\
		&\hspace{0,5in}
		- \left[{1 \over 12} L_1^3 L^2 + {1 \over 8}L_1 L^4  + {\pi^2 \over 3}  L_1 L^2\right] 
		+  {1\over 8} L_1 L^4
		\nonumber \\
		&= L_1 \cdot V \mathcal{M}_{1,2} (L_i) - {1 \over 24} L_1^3 L^2 - {1 \over 8} L_1 L_2^2 L^2 - {\pi^2 \over 6} L_1 L^2
		- {\pi^2 \over 24} L_1 L^2  + {23 \over 192}L_1 L^4
		\nonumber \\
		&\hspace{0,5in}
		- {1 \over 12} L_1^3 L^2 - {1 \over 8} L_1 L^4 - {\pi^2 \over 3} L_1 L^2 + {1 \over 8} L_1 L^4
		\nonumber \\
		&= L_1 \cdot V \mathcal{M}_{1,2} (L_i) - {1 \over 8} L_1 L^2 \sum_{i=1}^2 L_i^2  + {23 \over 192} L_1 L^4
		- {13 \pi^2 \over 24} L_1 L^2 \, .
	\end{align}
	Then the final result is
	\begin{align}
		V \mathcal{V}^L_{1,2} (L_i) = V \mathcal{M}_{1,2} (L_i) - 
		L^2 \left[ {1 \over 8} \sum_{i=1}^2 L_i^2 +  {13 \pi^2 \over 24} \right]
		+ {23 \over 192} L^4  \, .
	\end{align}
	This is also invariant under exchanging $L_1 \leftrightarrow L_2$. Defining
	\begin{align}
		x \equiv L_1^2 + L_2^2  \, ,
	\end{align}
	we can alternatively express~\eqref{eq:B.8} as 
	\begin{align} \label{eq:B.15}
		V \mathcal{V}^L_{1,2} (x)  = {1 \over 192} (x + 4 \pi^2) (x + 12 \pi^2) - {1 \over 8} L^2 x 
		- {13 \pi^2 \over 24} L^2
		+  {23 \over 192} L^4  \, .
	\end{align}
	which can be plotted as a function of the threshold length $L$ and $x$, see figure~\ref{fig:V12}. Take note that the quantity is positive for all $0 \leq L \leq 2 \sinh^{-1} 1$. 
	\begin{figure}[t]
		\centering
		\includegraphics[height=3in]{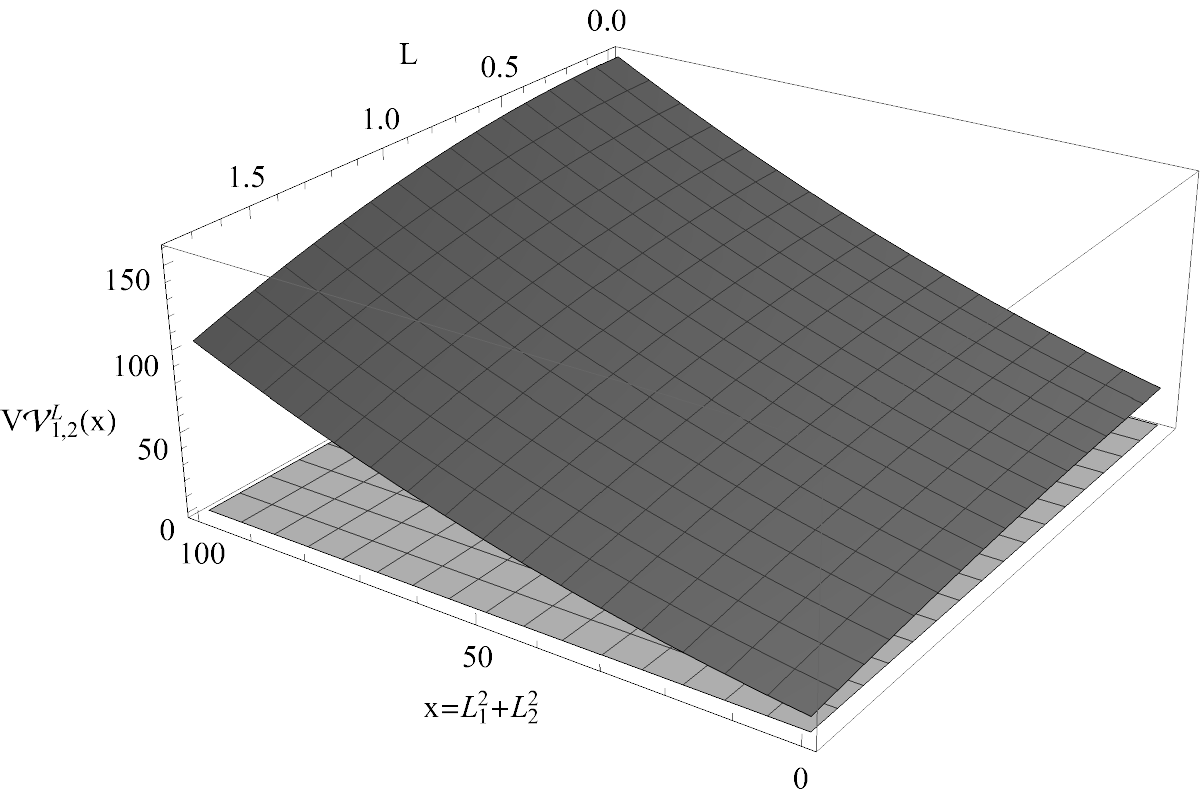}
		\caption{The systolic volume $V \mathcal{V}^L_{1,2} (L_i)$ as a function of $x=L_1^2 +L_2^2$ and the threshold length $L$~\eqref{eq:B.15}. We have $V \mathcal{V}^L_{1,2} (L_i) \geq 0$ when $0 \leq L \leq L_\ast$.}\label{fig:V12}
	\end{figure} 
	
\end{enumerate}

\section{Numerical evaluation of $V \mathcal{V}^L_{1,1}(L_1)$} \label{sec:A}

Following~\cite{zograf1987liouville, Firat:2023glo, Firat:2023suh, Artemev:2023bqj,zograf1988uniformization}, it is possible to work out the WP geometry of $\mathcal{M}_{0,4}(L_i)$ and $\mathcal{M}_{1,1}(L_1)$ explicitly using~\emph{classical conformal blocks}~\cite{Zamolodchikov:1995aa,Hadasz:2005gk,Hadasz:2006rb,Hadasz:2009db,Piatek:2013ifa}.\footnote{Also see the works~\cite{Matone:1993tj,Matone:1994pz,Matone:1995rx,Bertoldi:2004cc}.} Despite the fact that this procedure doesn't have a mathematically rigorous basis in general, there is overwhelming evidence that it holds true by the aforementioned works. We can use this approach to numerically compute the volumes $\mathcal{V}^L_{1,1}(L_1)$ to provide numerical evidence for the systolic volumes~\eqref{eq:322}. The reader can refer to~\cite{Firat:2023suh} for a deeper exposition. 

We begin by reminding the WP metric on the moduli space $\mathcal{M}_{1,1}(L_1)$, parameterized by the moduli $\tau \in \mathbb{H} / PSL(2,\mathbb{Z})$, is given by
\begin{align}
	g^{(1,1)}_{\tau \overline{\tau}}(L_1) = - 4 \pi \p_\tau  \p_{\overline{\tau}} S_{HJ}^{(1,1)} (\tau, \overline{\tau}; L_1) \, .
\end{align}
Here the K\"ahler potential $ S_{HJ}^{(1,1)} (\tau, \overline{\tau}; L_1)$ is a suitably-regularized on-shell Liouville action that can be found as an expansion in $q = e^{2 \pi i \tau}$, see equation (3.25) of~\cite{Firat:2023suh}.  The K\"ahler form associated with the metric $g^{(1,1)}(L_1)$ is related to the WP form on $\mathcal{M}_{1,1}(L_1)$. We can use this to evaluate $V \mathcal{V}^L_{1,1}(L_1)$ by performing the integral
\begin{align} \label{eq:A2}
	\mathcal{V}^L_{1,1}(L_1) = {i \over 4} \int\limits_{\mathcal{V}^L_{1,1}(L_1)} g_{\tau \overline{\tau}}(L_1) \, d \tau \wedge d \overline{\tau} \, .
\end{align}
Note the factor of $2$ difference between the conventions for the volumes here and in~\cite{Firat:2023suh}. This is a result of our conventions explained below~\eqref{eq:3.17}.

The integral for $L=0$ has been numerically evaluated in~\cite{Firat:2023suh} and a good agreement with the analytic results has been observed. Here we repeat an analogous computation after placing a cutoff on the moduli space to find the volumes of the systolic subsets. This first requires us to find the demarcation curve $ \partial \mathcal{V}^{2 \pi \lambda}_{1,1}(L_1) $, as a function of the moduli $\tau$, that separates the systolic subset from the region whose volume we subtract. This curve is
\begin{align} \label{eq:A3}
	\rho(2 \pi \lambda, 2 \pi \lambda_1, -2 \pi \lambda)^\lambda = \bigg| 
	\exp {\p f_{\lambda'}^\lambda \over \partial \lambda' } (q) 
	\bigg|_{\lambda' = \lambda} 
	\quad \quad \text{where} \quad \quad
	q = e^{2 \pi i \tau} \, ,
\end{align}
and the left-hand side of~\eqref{eq:A3} is given by the classical torus conformal blocks~\cite{Hadasz:2009db}
\begin{align} \label{eq:A5n}
	f_{\lambda'}^\lambda (q) = {\lambda'^2 \over 4} \log q + { (1+\lambda^2)^2 \over 8 \, (1+ \lambda'^2)} q + \mathcal{O}(q^2) \, ,
\end{align}
and the right-hand side is given by the mapping radius~\eqref{eq:R}.

Now it is possible to approximate the integral~\eqref{eq:A2} numerically, for which we use Monte-Carlo (MC) integration and consider the representative cases $L_1 = 0, \pi/2$. We approximate the curve~\eqref{eq:A3} by keeping only the $\log q$ term, as the shape of the curve is observed to remain almost the same including the higher order terms in $q$. To this order $ \partial \mathcal{V}^{2 \pi \lambda}_{1,1}(L_1) $ is simply described by the line
\begin{align} \label{eq:c5}
	\mathrm{Im} \, \tau = {1 \over \pi} \, {\log \rho(2\pi\lambda, 2\pi\lambda_1, -2\pi\lambda)} \, .
\end{align}
In our evaluation, we sampled points between this line and $\mathrm{Im} \, \tau = 40$ with the restriction $-1/2 \leq \mathrm{Re} \tau \leq 1/2$ and evaluated the volume of this region. We then subtracted it from the total volume of the moduli space $V \mathcal{M}_{1,1}(L_1)$ to find $V \mathcal{V}^{L}_{1,1}(L_1)$. We repeated the MC integration five times for each value of $L$.  The results are already shown in figure~\ref{fig:vol} and it is consistent with the analytical result~\eqref{eq:322}. We can also repeat a similar numerical evaluation to argue $V \mathcal{V}^{L}_{0,4}(L_i)$ is given by~\eqref{eq:321} after constructing the WP metric on the four-bordered sphere~\cite{Firat:2023glo}.

We emphasize the restriction $L \leq L_\ast$ is necessary to cover the region $\mathcal{M}_{1,1} (L_1) \setminus \mathcal{V}^L_{1,1} (L_1)$ once and only once. This can be seen from setting $L=L_\ast$. The symmetric punctured-torus ($L_1 = 0, \tau =i$) saturates the systolic condition in this case~\cite{maskit1989parameters} and any increase in the threshold length $L$ would overcount the region $\mathcal{M}_{1,1} (L_1) \setminus \mathcal{V}^L_{1,1} (L_1)$ as a result~\cite{Firat:2023suh}. This implies the formula~\eqref{eq:322} no longer holds and the form of the subtraction term should be modified from $L^2/4$ appropriately. Hence the condition $L \leq L_\ast$ is necessary. Having $L_1 > 0$ doesn't lead to a stricter condition. Notice this argument doesn't inform whether $L \leq L_\ast$ is sufficient or not. This follows from the collar lemma instead.

%\bibliography{all}{}
%\bibliographystyle{utphys}
%Copy bbl file if you don't want to deal with bib!

\providecommand{\href}[2]{#2}\begingroup\raggedright\endgroup

\end{document}